\documentclass{aa}

\usepackage{graphicx}
\usepackage[svgnames]{xcolor}
\usepackage[pdftex,colorlinks,citecolor=DarkBlue]{hyperref}

\usepackage[varg]{txfonts}

\usepackage{natbib}
\bibpunct{(}{)}{;}{a}{}{,} % to follow the A&A style

\newcommand{\ad}{{\mathrm{ad}}}

\begin{document}

%   \title{How the formation and characteristics of protoplanetary discs depend on dense core properties  and  on accretion}

   \title{What determines the formation and characteristics of protoplanetary discs?}

%   \titlerunning{How the formation of protoplanetary discs depends on dense core properties  and  on accretion}

   \author{Patrick Hennebelle \inst{\ref{inst1}}
	\and Benoit Commer{\c c}on \inst{\ref{inst2}}
      \and
      Yueh-Ning Lee \inst{\ref{inst4},\ref{inst3},\ref{inst1}}
	\and S\'ebastien Charnoz \inst{\ref{inst3}} }

   \institute{AIM, CEA, CNRS, Université Paris-Saclay, Université Paris Diderot, Sorbonne Paris Cité, F-91191 Gif-sur-Yvette, France,
      \label{inst1}
      \and
       Institut de Physique du Globe de Paris, Sorbonne Paris Cit\'e, Universirt\'e  Paris Diderot, UMR 7154 CNRS, F-75005 Paris, France
      \label{inst3}
      \and
 Univ Lyon, Ens de Lyon, Univ Lyon1, CNRS, Centre de Recherche Astrophysique de Lyon UMR5574, F-69007, Lyon, France
      \label{inst2}
      \and
      Department of Earth Sciences, National Taiwan Normal University, 88, Sec. 4, Ting-Chou Road, Taipei 11677, Taiwan
      \label{inst4}
      }

   \date{Received ---; accepted ---}

% \abstract{}{}{}{}{}
% 5 {} token are mandatory

   \abstract%
      % context heading (optional), leave it empty if necessary
      {Planets form in protoplanetary discs. Their masses, distribution, and orbits  sensitively depend 
on the structure of the protoplanetary discs. However, what sets the initial structure of the discs 
in terms of mass, radius and accretion rate is still unknown. 
}
      % methods heading (mandatory)
      {It is therefore of great importance to understand exactly how protoplanetary discs form and 
what determine their physical properties. We aim at quantifying the role of the initial dense core magnetisation, rotation, turbulence 
and misalignment between rotation and magnetic field axis as well as the role of the accretion  scheme onto 
the central object. }
{We perform non-ideal MHD numerical simulations using the adaptive mesh refinement code Ramses, of a collapsing, one solar mass,
 molecular 
core to study the disc formation and early, up to 100 kyr, evolution, paying great attention to the impact of 
numerical resolution and accretion scheme. }
      { We found that while the mass of the central object is
almost independent of the numerical parameters such as the resolution and the accretion scheme onto the 
sink particle,  the disc mass, and to a lower extent its size, heavily depend on the accretion scheme,
 which we found, is itself resolution dependent. This implies that the accretion onto the star and 
through the disc are  largely decoupled.
 For a relatively large domain of initial conditions 
(except at low magnetisation), we found 
that the properties of the disc do not change too  significantly. In particular both the 
level of initial rotation and turbulence do not influence the disc properties provide the 
core is sufficiently magnetized.
After a short relaxation phase, 
the disc settles in a stationary state.  
It then slowly grows in size but not in mass.
The disc itself is weakly magnetized but its immediate surrounding is on the contrary highly magnetized.  }
      {Our results show that the disc properties directly depend on the inner boundary condition, i.e. the accretion
scheme onto the central object, 
suggesting that the disc mass is eventually controlled by the small scale accretion process, 
possibly the star-disc interaction.  
Because of ambipolar diffusion and its significant resistivity, the disc diversity 
remains limited and except for low magnetisation, their properties are weakly sensitive to initial conditions 
such as rotation and turbulence. }

%For a  choice of the accretion scheme inferred
%from a simple analytical model, the disc has  a size on the order of 20 AU and  
%a mass of about 10$^{-2}$ $M_\odot$. 

   \maketitle

%________________________________________________________________

\section{Introduction}

It is well established that planets form within circumstellar discs, 
and therefore knowing the conditions under which these protoplanetary discs
form as well as their physical characteristics such as mass and size 
is of fundamental importance \citep{testi2014,dutrey2014}.
While most discs observed so far are located around objects that are not 
any more embedded, i.e. T-Tauri stars,  
there are increasing evidences that the discs form early during 
the class-0 stage \citep{jorgensen2009,maury2010,murillo2013,ohashi2014,tobin2016,maury2019}. 
While the statistics remain hampered by large uncertainties, it seems that 
there is a clear trend for the discs at this early stage to be limited in size with most discs 
having a radii below 60 AU, and possibly even smaller.
Understanding this initial phase is of primordial importance to get the complete 
history of the process, but also because the mass of the discs observed around 
T-Tauri stars may not be massive enough to explain the mass of the observed planets
\citep{greaves2010,najita2014,manara2018}, suggesting that planet formation 
may start early after the protostar formation during the infall envelope while the 
disc is still fed by fresh ISM materials.

From a theoretical point of view, much progress has been made in the  last decade
regarding the formation of protoplanetary discs during the collapse of prestellar cores. 
One of the major outcome of the first calculations of magnetized collapse 
\citep[e.g.][]{allen2003,galli2006,price2007,hf2008,li2014,wursterli2018,hennebelle2019}, is that magnetic 
braking could be so efficient that disc formation may be entirely prevented, 
a process named as catastrophic braking. 
Further studies have shown that the aligned configuration assumed in these calculations was
however a significant oversimplification and that discs should form in magnetized clouds, although the discs are 
smaller in size and fragment less than in the absence of magnetic field. Three reasons have been 
proposed to explain how magnetic braking is reduced  i) the magnetic field and the rotation axis are misaligned 
\citep{joos2012,li2013,gray2018}, ii)  turbulence diffuses the magnetic flux outwards  
\citep{santos2012,joos2013}, iii) tangled magnetic  field is inefficient to transport momentum
 \citep{seifrid2013}. Note that \citet{gray2018} carried out calculations with a turbulent velocity field in which 
the angular momentum is aligned with the magnetic field. They found  that discs do not form or are much smaller 
than when there is no particular alignment. This led them to conclude that misalignment is  the main effect.

The second over-simplification which has been made in the first series of studies, is the 
assumption of ideal MHD.
Non-ideal MHD  has been taken into account  by 
numerous groups \citep{dapp2010,dapp2012,inutsuka2012,li2014,hetal2016,masson2016,machida2016,wurster2016,wurster2019a,wurster2019b,zhao2018}.
These works found that at least small discs always form.
When non-ideal MHD is accounted for, turbulence and  misalignment tend to be relatively less important.
The resulting disc however depends on the magnetic resistivities \citep{wurster2018,zhao2018} which 
depend on the grain distribution and cosmic-ray ionisation \citep{padovani2013}.
It has also been found that the Hall effect may introduce a dependence on the respective 
direction between the magnetic field and the rotation axis \citep{tsukamoto2017} but 
large uncertainties both on the resistivities \citep{zhao2016} and the numerical scheme \citep{marchand2018} remain. 

Another important issue in the collapse calculations are the sink particles used to mimic 
the stars. These Lagrangian particles accrete the surrounding gas and interact through self-gravity 
with the whole flow in the simulations. Sink particle parameters such as 
their accretion radius or the criteria prevailing to their introduction, are not easily 
justified and remain largely empirical. However previous authors 
\citep{machida2014,vorobyov2019} report strong dependence of the resulting disc on the
parameters which control the sink particles, particularly the parameters 
that control the gas accretion.

In an attempt to provide a generic description of disc size in non-ideal 
MHD calculations, \citet{hetal2016} proposed an analytical approach to infer 
the disc radius. The calculation is based on simple timescale estimates 
at the edge of the cloud. Essentially, it is required that the braking and rotation times
should be comparable as well as  the diffusion and magnetic generation times.
 This leads to a simple prediction that is recalled later in the paper (Eq.~(\ref{radius_ad})), 
which has been found to be in reasonable agreement with a set of 3D simulations. 
A particularly important prediction of this study is that the 
disc radius, is not expected to vary much with the cloud initial parameters such 
as its magnetization and rotation (provide they are not too low). In the present paper, 
we pursue our investigation of disc formation along two lines. First we include 
sink particles as they were not considered in our previous studies 
\citep{hetal2016,masson2016} and second we vary more extensively the 
initial conditions. Importantly, we look in details into the disc 
properties and show that the accretion rate on the central 
region is an important parameter which largely contribute to determine the mass, and to a lower 
extent the radius, of the disc.

In the second section the numerical method as well as the initial conditions 
used in this work are presented. 
In the third section, we present in details the result of  a run that we 
use as a reference through the paper and with which other runs will be 
compared. In the fourth section, we perform a series of runs to carefully
investigate the influence of the numerical parameters such as the spatial 
resolution and the accretion scheme onto the central object. 
Fifth section presents a systematic investigation of the role of the 
initial conditions. The influence of the core initial rotation, 
turbulence, misalignment and magnetisation are all discussed. 
Finally section six concludes the paper.

\section{Setup and simulations}
 The simulations presented here used the same physical and numerical setup 
than the ones presented in \citet{hetal2016} and \citet{masson2016} but for 
the usage of the sink particles. They also explore a broader range 
of parameters.

\subsection{Code, physics and resolution}
To perform our simulations, we used the adaptive mesh refinement code, Ramses
\citep{teyssier02}, which solves the MHD equations using the finite volume 
method \citep{fromang06} and maintain the nullity of $\nabla B$ within 
machine accuracy thanks to the usage of the constraint transport method. 
 Periodic boundary conditions are employed.
The ambipolar diffusion is treated following the implementation of 
\citet{masson2012}, which is a simple explicit scheme. The main difficulty 
is that the timestep associated to the ambipolar diffusion can become very small, 
particularly in the diffuse gas. To limit this effect, we proceed as 
in \citet{masson2016}, that is to say we adjust, when required, the ionisation 
to force the timesteps due to the ambipolar diffusion term to be at least 1/3 
of the timestep due to the ideal MHD equations. The method has been tested and results
do not sensitively depend on the choice of the threshold \citep{vaytet2018}. 

For the non-ideal MHD coefficient we used the approach presented in \citet{marchand2016}.
Only the ambipolar diffusion is being considered. The Ohmic resistivity 
only becomes important at densities higher than the regime of interest of the present paper 
and correspondingly at spatial scale smaller that the typical scale described here ($\simeq 10$ AU).
The Hall effect is also not considered. 
A chemical network is solved and the species abundances are tabulated as a function 
of density and temperature allowing to fastly compute the resistivities within all cells at 
all timesteps. The assumed ionisation rate is $3 \times 10^{-17}$ s$^{-1}$.

To  mimic the radiative transfer and the fact that the dust becomes opaque to its own 
radiation at high density, we use an equation of state (eos) given by
\begin{eqnarray}
T = T_0 \left\{1 \!+\! { (n / n _{\ad})^{(\gamma_1-1)} \over 
1+( n / n _{\ad,2})^{(\gamma_1-\gamma_2)}  }  \right\}, 
\label{eq_full_eos}
\end{eqnarray}
where $T_0=10$ K, $n _{\ad}=10 ^{10}$ cm$^{-3}$, $n _{\ad,2}= 3 \times 10 ^{11}$ cm$^{-3}$, 
$\gamma_1=5/3$ and $\gamma _2 = 7/5$. Let us remind that this equation of states 
mimics the thermal behaviour of the gas, $n_{ad}$ is the density at which the gas becomes opaque  
to its radiation while $n_{\ad,2}$ corresponds to the density at which the H$_2$ molecules
is warm enough for the rotation level to be excited \citep[e.g.][]{Masunaga98}.

We used the sink particle algorithm developed by \citet{Bleuler14} .
The sinks are introduced at a density $n_{\rm thres}$ (our standard value is $n_{\rm thres} = 3 \times 10^{13}$ cm$^{-3}$). 
Sink particles are placed at the highest refinement level at the peak of clumps whose density 
is larger than $n_{\rm thres}/10$.

Further gas is being accreted when the density becomes higher than $n_{\rm acc} = n_{\rm thres}/3$ 
 within cells located 
at a distance less than $4 dx$, where $dx$ is the size of the most refined grid, from the sink particle. 
At each timestep a fraction $C_{\rm acc}$  of this gas ($C_{\rm acc}=0.1$ except for one run)
is removed from the grid and attributed to the sink.  To clarify, at each time step
the sink received a mass equal to  $C_{\rm acc} \int (n - n_{\rm acc}) m_p dV$, where the integral runs 
over the sink volume. The magnetic field inside the sink particles is not modified and evolves as
in the absence of the sink. 
 While this is very simple, this may not be so inaccurate because 
at small scales, i.e. below 1 AU or so, the magnetic field becomes poorly coupled 
to the gas and therefore it is not dragged along with the collapsing gas. 
 The sinks interact with the gas and also between them through their gravitational 
forces which are directly added to the force exerted by the gas.  Inside the sink particles 
gravitational softening is applied and the gravitational acceleration is given by
$G M_* / (r^2 + (2dx) ^2)$, where $dx$ is the cell size. Finally, in this work the sinks cannot merge.

The initial resolution is about 256 AU, implying that the  cloud diameter is described 
with 128 cells. 
Each level of refinement subdivides the cells into two in each of the three dimensions. 
 As the collapse proceeds 7-9 AMR levels  are being added to the initial grid which 
contains $(2^3)^6$ cells and starts at level 6. 
 This provides a 
 maximum resolution in this series of simulations of about  0.5-2 AU. 
The refinement 
criterion is the local Jeans length, which is resolved by at least 20 cells.

\subsection{Initial condition and runs performed}

Two series of runs have been performed. The first one is devoted to exploring the 
influence of the numerical scheme, in particular the sink particle algorithm, while the 
second series investigate the influence of initial conditions. 

 Unlike many works, the simulations 
are integrated during a relatively long time. This goes from at least 
10 kyr after the sink formation to almost 100 kyr for run $R1$ and $R2$, for which about 10$^7$ timesteps have been performed. 
We can therefore 
test the evolution of the disc up to the end of the class I phase. Given the small time steps induced by 
the ambipolar diffusion, the simulations take a lot of time. They have been computed on 40 cores and last several 
months (2-8). 

All cores have a mass equal to 1 $M_\odot$, and they have initially a uniform density 
equal to about 5 10$^5$ cm$^{-3}$  and their initial thermal over 
gravitational energy is equal to 0.4. Outside the core, the density is 
uniform and equal to this value divided by hundred.  This implies that the pressure outside is rather low 
and therefore the external part of the core undergoes an expansion. 
 The radius of the core is 0.019 pc and the computational box size is four times this value.
The cores are in solid-body 
rotation, which is quantified by the ratio of rotation over gravitational energy, $\beta_{\rm rot}$.
The magnetic field is initially uniform. Its intensity inside the cylinder that contains 
the cloud, is constant while its intensity outside this cylinder is also uniform  but reduced by a factor  100$^{2/3}$.  
An $m=2$ density perturbation perpendicular to the rotation axis and equal to 10$\%$ has been added to break axisymmetry.

\subsubsection{Numerically motivated runs}

      \begin{table}
         \begin{center}
            \begin{tabular}{lcccccccc}
               \hline\hline
               Name & lmax  & $n_{\rm thres} = 3 n_{\rm acc}$  & $R_{sink}$ (AU) &  $C_{\rm acc}$   &  rest  \\
               \hline
               $R2ns$ & 14  & NA  & NA & NA  &  no \\ %DC_2_nosink
               $R2$ & 14  & 3 10$^{13}$ & 4   & 0.1  &   no \\ %DC_2
               $R2hsink$ & 14  & 1.2 10$^{14}$ & 4 & 0.1  &   no \\ %DC_2_hsink
               $R2lowr$ & 13  & 3 10$^{13}$ & 8 & 0.1  &    no  \\ %DC_2_lowres
               $R2lowrlsink$ & 13  & 3 10$^{12}$ & 8  & 0.1  &    no \\ %DC_2_lowres_early
               $R2r$ & 15  & 3 10$^{13}$  &  2 & 0.1 &    no \\ %DC_2_res
               $R2rlowacc$ & 15  & 3 10$^{13}$  & 2 &  0.01 &    no  \\ %DC_2_res_lowacc
               $R2rr$ & 15  & 3 10$^{13}$  & 2 &  0.1 &   yes  \\ %DC_2_rest2_test
               $R2rrhsink$ & 15  & 3 10$^{14}$ & 2  &  0.1 &  yes  \\ %DC_2_rest2_test_hsink
%               $R2cad$ & 14  & 10$^{13}$ & 0.6  & 0.1 & baro & yes  & yes \\ %DC_2_test_coefAD
%               $R2fld$ & 14  & 10$^{13}$ & 0.3  & 0.1 & fld & yes  & no \\ %DC_2_fld
%               $R2fldr$ & 15  & 10$^{14}$ & 0.3  & 0.1 & fld & yes  & yes \\ %DC_2_fld_res
%               $R2fldhr$ & 17  & 10$^{15}$ & 0.3  & 0.1 & fld & yes  & yes \\ %DC_2_fld_rest2_hres_hsink
%               $R2fldvhr$ & 18  & 10$^{15}$ & 0.3  & 0.1 & fld & yes  & yes \\ %DC_2_fld_rest2_vhres_hsink
%               $R2fldvhrcad$ & 18  & $5 \times 10^{15}$ & 0.15  & 0.1 & fld & yes  & yes \\ %DC_2_fld_rest2_vhres_hsink_lowdt
               \hline
            \end{tabular}
         \end{center}
         \caption{Summary of the numerically motivated runs performed.  lmax is the maximum level of grid used,
$lmax=14$ corresponds to about 1 AU of resolution, $lmax=13$ to 2 AU and $lmax=15$ to 0.5 AU. $n_{thres}$ is the density at which the sink particles are introduced while $n_{\rm acc}$ is
the density above which the gas is being accreted in the sink, $R_{sink}$ is the sink radius,
 $C_{\rm acc}$ is the fraction of the gas mass above this density threshold and inside the sink radii which is 
accreted in a timestep. The parameter rest indicates whether the run uses an output of run $R2$ as starting point or is performed from $t=0$.}
\label{table_param_num}
      \end{table}

To understand the effect of the sink particle algorithm and the numerical resolution,
we perform 9 runs which  are summarized in Table~\ref{table_param_num}.
Run $R2$, which will serve as a reference model and is presented in great details below,
 has an initial mass-to-flux 
over critical mass-to-flux ratio, $\mu^{-1}=0.3$ (as in \citet{masson2016}, we follow the 
definition of \citet{mous1976}), implying that it is significantly 
magnetized but clearly supercritical. The magnetic field axis is along the z-axis and the 
rotation axis is initially tilted by an angle of 30$^\circ$ with respect to it. 
It has  $\beta_{\rm rot}=0.04$. This run is integrated 
during 100 kyr after the formation of the sink particle.

All runs start with the same initial conditions than run $R2$.
Run $R2ns$ has no sink particle, while run $R2hsink$ uses a density threshold, $n_{\rm thres}= 1.2 \times 10^{14}$ cm$^{-3}$
implying that the sink forms later and accretes gas at higher densities than  run $R2$. This allows us 
to test the influence of $n_{\rm thres}$ and $n_{\rm acc}$, which as we see later are crucial.
Runs $R2lowr$ and $R2lowrlsink$ have a lower resolution and two different values of $n_{\rm thres}$, namely 
$3 \times 10^{13}$ and $3 \times 10^{12}$ cm$^{-3}$.
Runs $R2r$ and $R2rlowacc$ have a higher spatial resolution and two values of $C_{\rm acc}$, namely 0.1 and 0.01.
With these two runs we can determine whether another accretion scheme influences the disc properties.
Finally runs $R2rr$ and $R2rrhsink$ restart from run $R2$ output. They have higher resolution and two 
different values of $n_{\rm thres}$. With these two runs we can distinguish the influence of the sink 
creation and the  accretion scheme. 

\subsubsection{Physically motivated runs}

      \begin{table}
         \begin{center}
            \begin{tabular}{lccccc}
               \hline\hline
               Name & $\mu^{-1}$  & $\beta_{\rm rot}$  &  $\theta$ & ${\mathcal M}$  \\
               \hline
               $R1$ & 0.3  & 0.01  &  30$^\circ$ & 0  \\
               $R2$ & 0.3  & 0.04  &  30$^\circ$ & 0  \\
               $R3$ & 0.1  & 0.04  &  30$^\circ$ & 0  \\
               $R4$ & 0.15  & 0.04  &  30$^\circ$ & 0  \\
               $R5$ & 0.07  & 0.04  &  30$^\circ$ & 0  \\
               $R6$ & 0.3  & 0.01  &  30$^\circ$ & 1  \\
               $R7$ & 0.3  & 0.04  &  90$^\circ$ & 0  \\
               $R8$ & 0.15  & 0.01  &  30$^\circ$ & 1  \\
               $R9$ & 0.1  & 0.04  &  90$^\circ$ & 0  \\
               $R10$ & 0.3  & 0.0025  &  30$^\circ$ & 0  \\
               \hline
            \end{tabular}
         \end{center}
         \caption{Summary of the physically motivated runs performed.
$\mu$ is the mass to flux ratio normalised to the critical value,  $\beta_{\rm rot}$ is the ratio of rotation over gravitational energy, 
$\theta$ is the initial angle between the magnetic field and the rotation axis, while ${\mathcal M}$
is the initial Mach number.}
\label{table_param}
      \end{table}

We performed 10 runs in total with the aim of sufficiently exploring the 
influence of initial conditions. They are summarized in Table~\ref{table_param}.
For all these runs, 8 levels of AMR are used and $n_{\rm thres}= 3 n_{\rm acc} = 3 \times 10^{13}$ cm$^{-3}$.

We remind  that run $R2$ is the reference run.
Runs $R1$ and $R10$ have a low rotation with  respectively $\beta_{\rm rot}=0.01$ and $\beta_{\rm rot}=0.0025$. 
They aim at testing the importance of rotation, which is found to 
play a weak role. 
With runs $R3$, $R4$ and $R5$ we study the influence of the magnetisation, which is the most critical parameter, 
when the magnetic field is weak.
Runs $R6$ and $R8$ present on top of a weak rotation a turbulent velocity field which leads to a rms 
Mach number of 1 initially. 
Finally, for runs $R7$ and $R9$ the magnetic field and rotation axis are initially perpendicular. 

\subsection{Disc identification}
\label{disc_iden}
We are primarily interested here by the disc formation and evolution and it is therefore essential 
to define it. We proceed in several steps. 
First, we apply a density criterion selecting only gas denser than $10^8$ cm$^{-3}$. We then select the cells 
for which the radial velocity with respect to the sink particle is smaller than half the azimuthal velocity.
This defines an ensemble of cells whose mass is well defined. Determining the disc radius is however more 
difficult because, the discs while reasonably well defined, present nevertheless large fluctuations and generally 
are not fully axisymmetric. 
We proceed as follows. First, we determine the mean angular momentum direction of the disc. We then 
project the  disc cell coordinates in the disc plane. We then subdivide the disc plane in 50 
angular bins.  For each of them we determine the most distant cell, that gives the radius 
of this angular bin. We then calculate the mean radius, 
$R_{\rm mean}$ by taking the mean value of the 50 radius obtained for each angular bin and the maximum 
radius, $R_{\rm max}$, of the 50 angular bins.

 Note that in the study of \citet{joos2012}, more criteria were used. The reason is that in their 
calculations, ideal MHD was employed and no sink particle was used. Indeed, both tend to 
produce smoother discs, which are easier to define. Ambipolar diffusion dissipates
motions and sink particles make the discs much less massive and therefore less prone to
gravitational instabilities. Another difference comes from the outflow cavities and 
their boundaries which are denser in the ideal MHD case and tend to be picked by simple 
disc criteria.

\setlength{\unitlength}{1cm}
\begin{figure*}%[h!]
%\centering
\begin{picture} (0,20)
\put(0,15){\includegraphics[width=7cm]{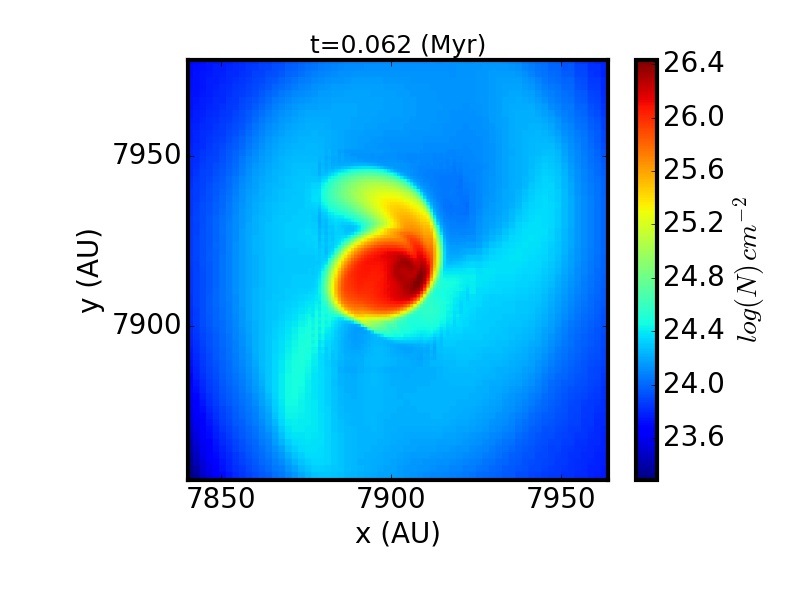}}  
\put(0,10){\includegraphics[width=7cm]{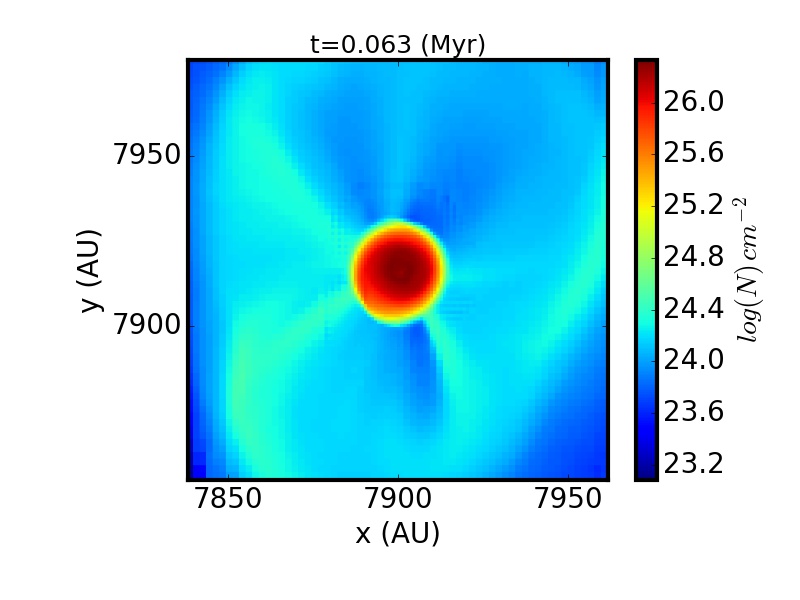}}  
\put(0,5){\includegraphics[width=7cm]{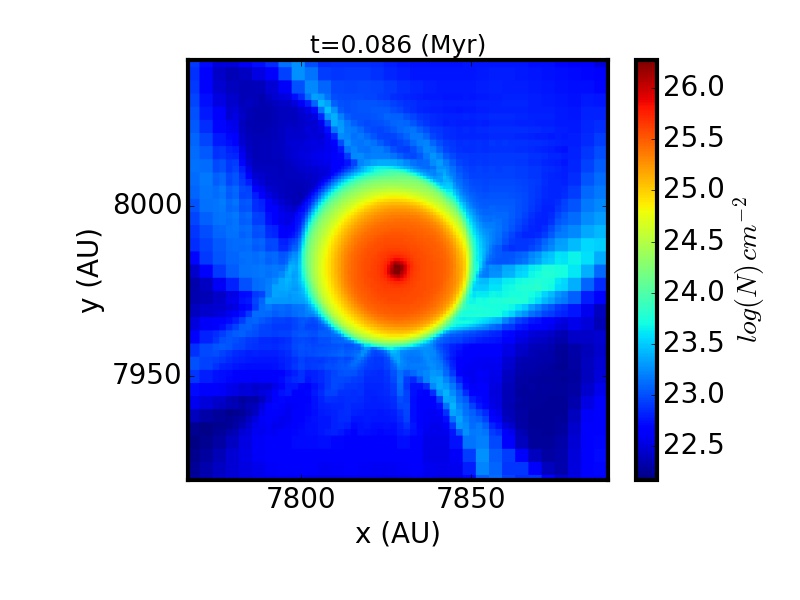}}  
\put(0,0){\includegraphics[width=7cm]{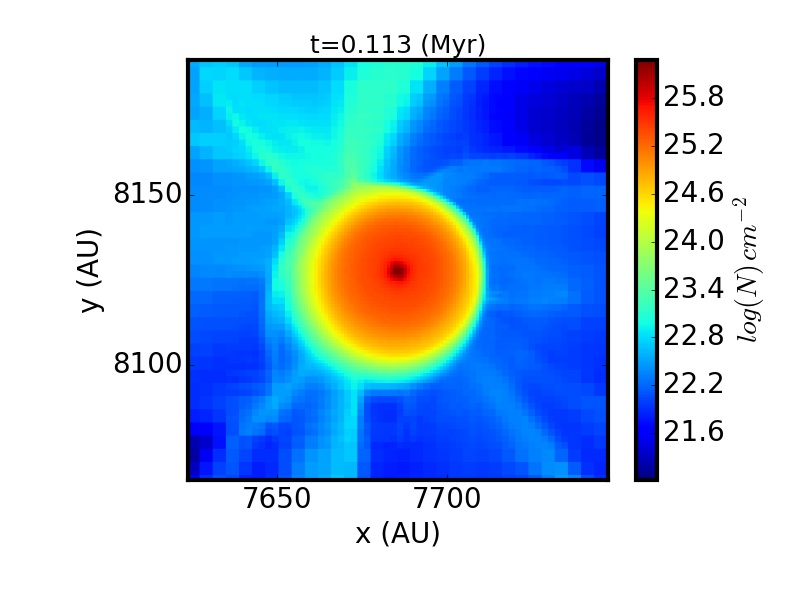}}  
\put(7,15){\includegraphics[width=7cm]{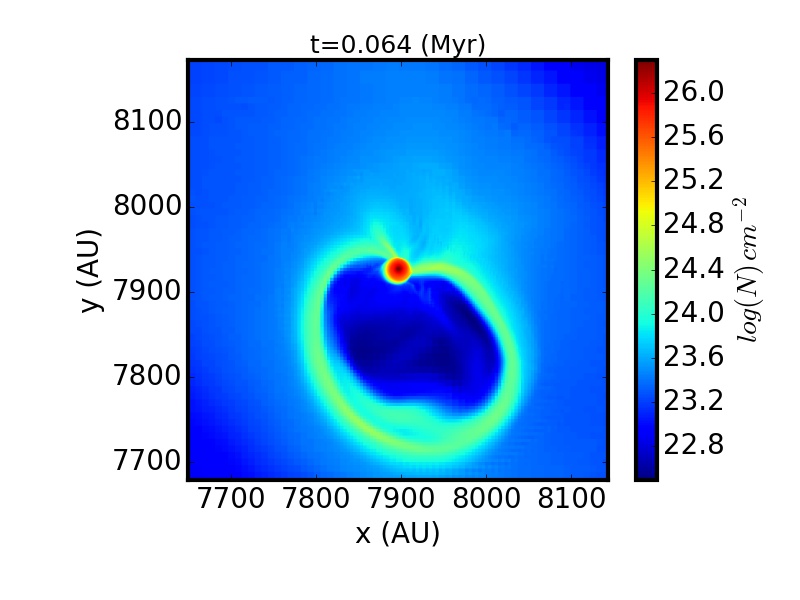}}  
\put(7,10){\includegraphics[width=7cm]{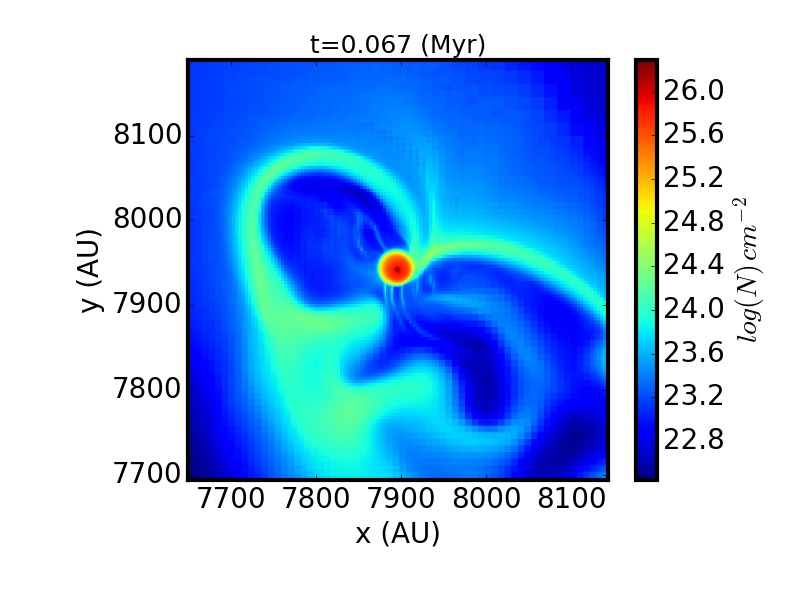}}  
\put(7,5){\includegraphics[width=7cm]{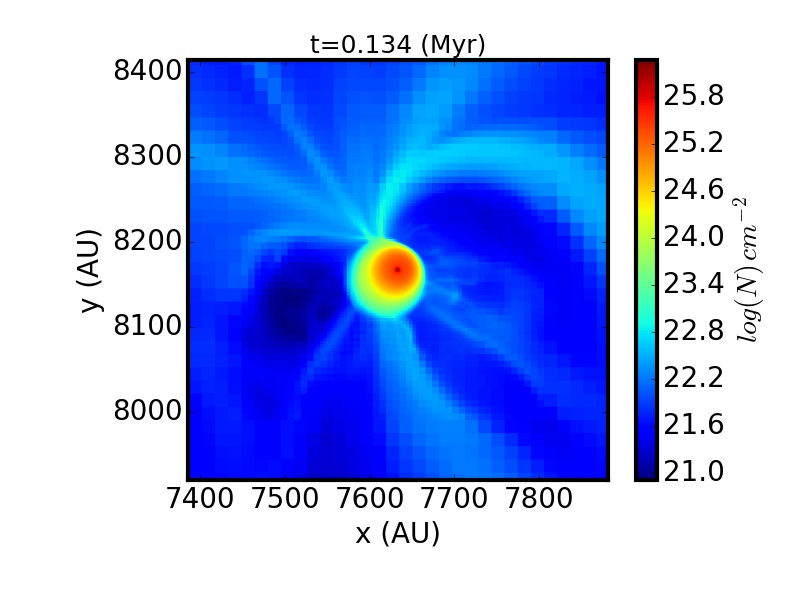}}  
\put(7,0){\includegraphics[width=7cm]{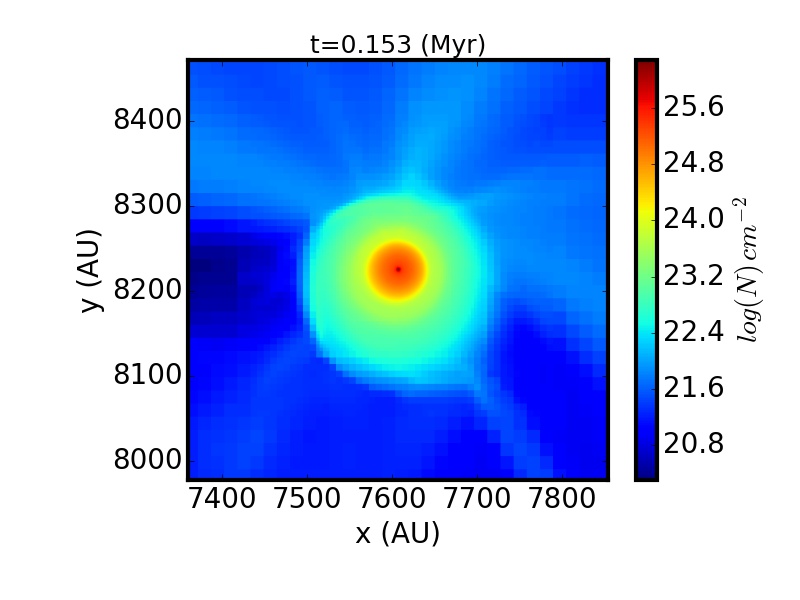}}  
\end{picture}
\caption{Column density of run $R2$ at several snapshots and two scales. Left: box length is about 100 AU, 
right: box length is about 500 AU. }
\label{coldens_refmod}
\end{figure*}

\section{Detailed description of  a reference model}
We start with a detailed description of run $R2$.

\subsection{Qualitative description}
Figure~\ref{coldens_refmod} displays the column density for a series of snapshots.
To show the disc structure and its close environment, in the left column, the box size is about 100 AU.
 In the right column, the box size  is about 500 AU, allowing to display the larger scales surrounding the disc.
 Note that the sink particle forms right in the middle of the box.  The origin of the 
x and y coordinates is taken at the left bottom corner of the computational box.

The first snapshot of the left column 
reveals that initially the disc is highly non-axisymmetric with a prominent spiral arm.
It however quickly (i.e. in a few kyr) relaxes in a nearly symmetric object whose radius is on the order 
of 10 AU. The third and fourth panels show that at later times, the disc grows in size. At time 0.113 Myr, 
its radius is about 25 AU while at time 0.153 Myr (fourth panel of right column), it 
is about 100 AU. At all times, dense filaments are connected to the disc 
and bring further material from the envelope. This anisotropic accretion maintains a certain degree 
of asymmetry within the  disc, that as we can see is never perfectly circular. 
At some snapshots, see for example time 0.134 (third snapshot of right column), the disc
is highly excentric.  
Another important aspect is that 
the disc position changes with time. Between the first and the fourth snapshot, it has moved 
of about 200 AU. Since the initial conditions of run $R2$ are axisymmetric, this implies that 
symmetry breaking occurs.

A symmetry breaking is indeed clearly visible on the first panel of right column 
where a loop of dense material has developed 
in the vicinity of the disc. Inside the loop, the density is much lower.  This is a consequence of the 
interchange instability as described in \citet{krasno2012} \citep[see also][]{joos2012}, which is 
essentially a Rayleigh-Taylor instability produced by the excess of magnetic flux that has been 
accumulated in the central region due to the collapse. It typically happens when the flux trapped 
within the dense gas diffuses out and prevent further accretion to occur. The interchange instability leads to 
strong instability as seen in the first and second panel of the right column. Not only the loop itself 
develops in one direction and not isotropically but also because of gravity, Rayleigh-Taylor fingers develop
and fall back onto the sink.

\subsection{Disc global evolution}

\setlength{\unitlength}{1cm}
\begin{figure}%[h!]
%\centering
\begin{picture} (0,22)
\put(0,16.5){\includegraphics[width=8cm]{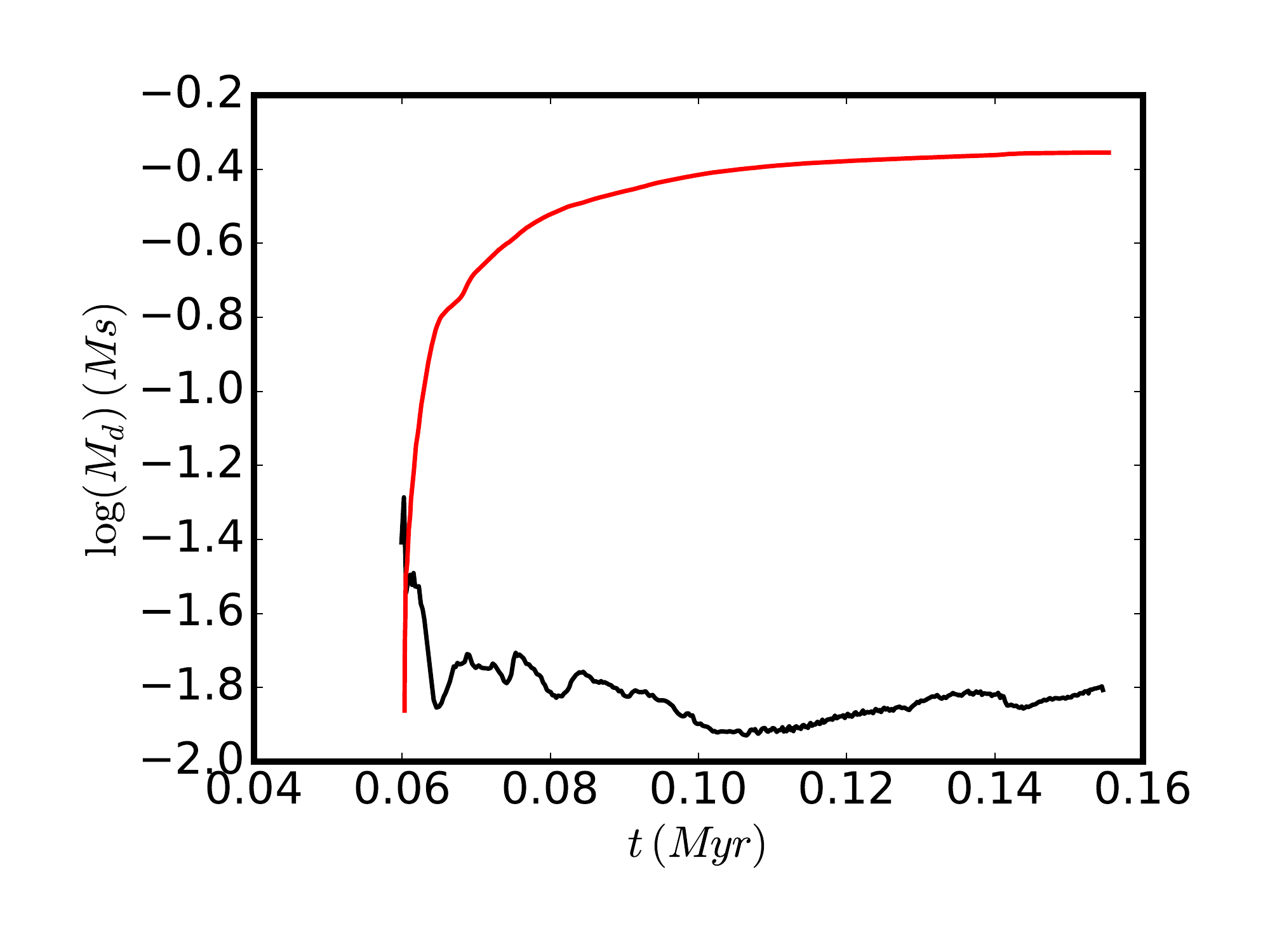}}  
\put(0,11){\includegraphics[width=8cm]{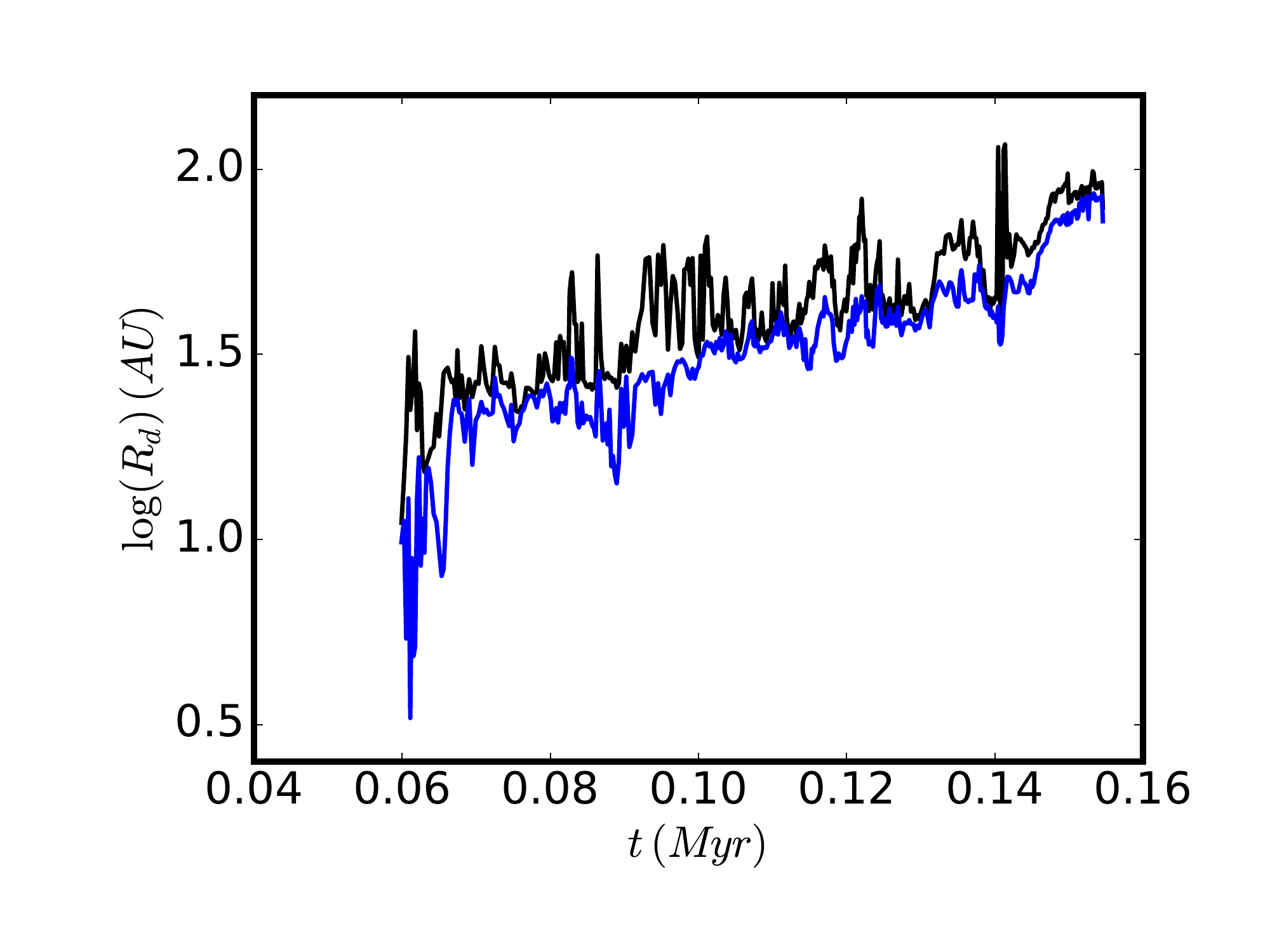}}  
\put(0,5.5){\includegraphics[width=8cm]{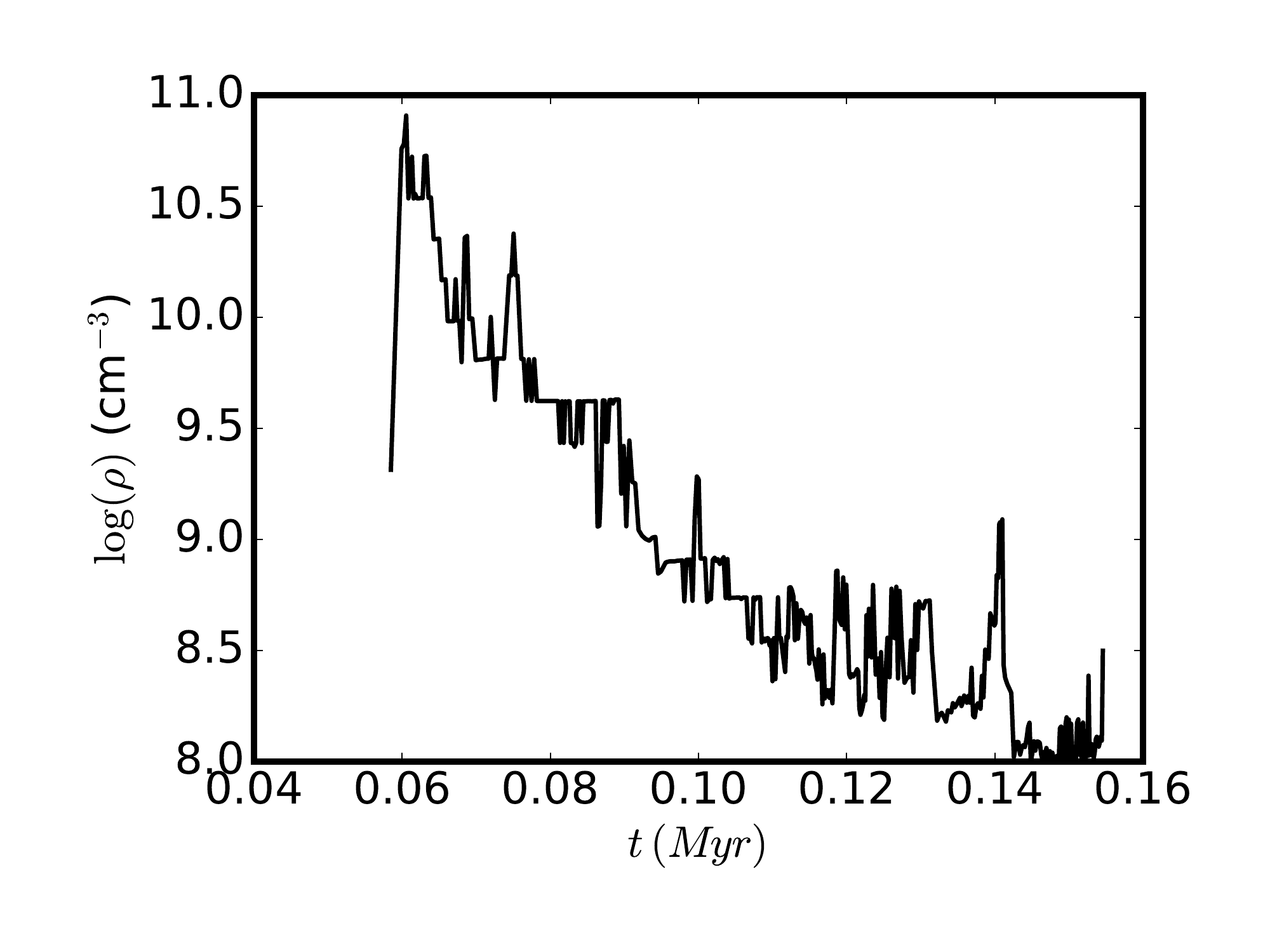}}  
\put(0,0){\includegraphics[width=8cm]{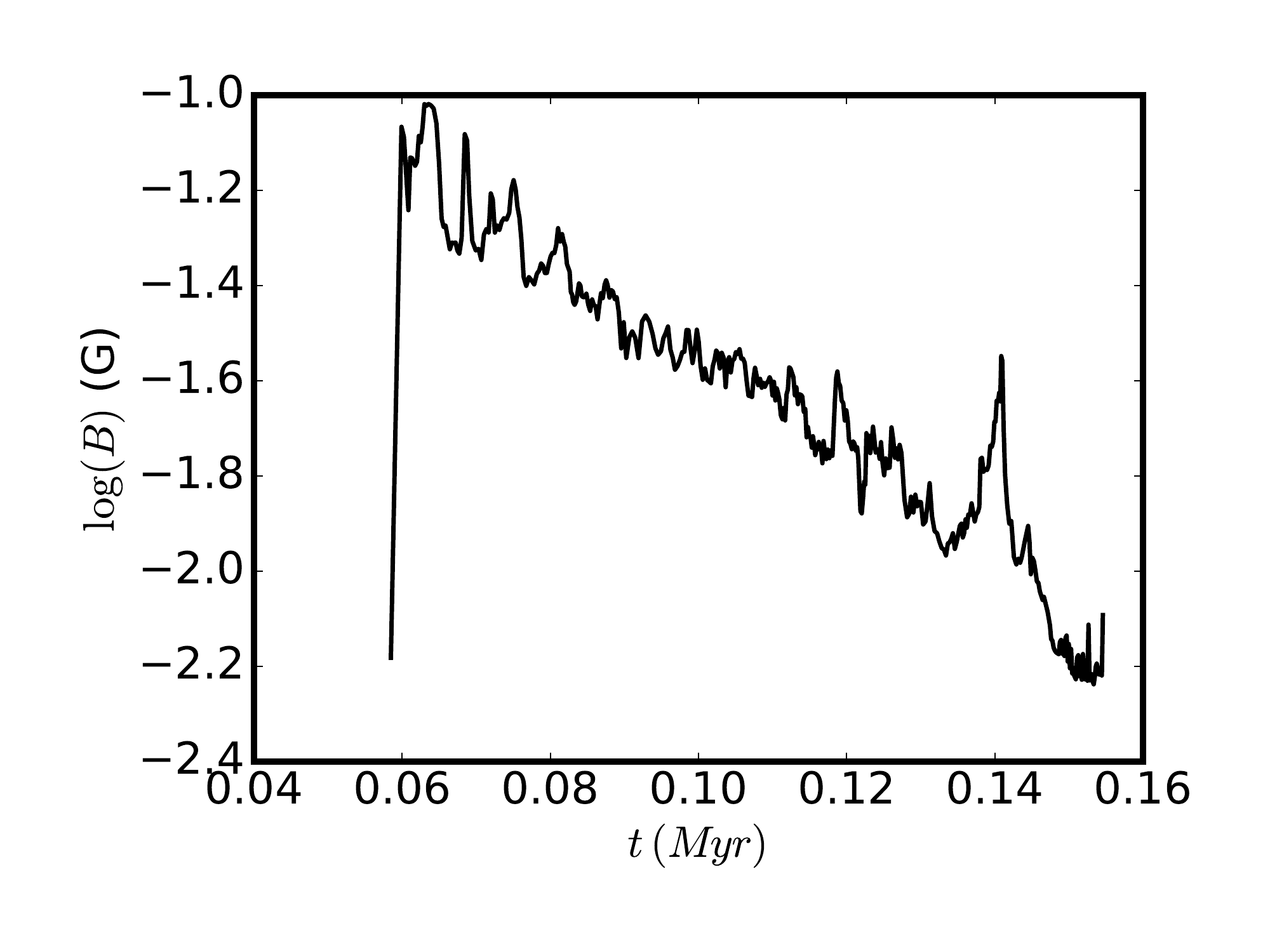}}  
\end{picture}
\caption{Global properties of the disc of run $R2$ as a function of time. 
Top panel: disc mass (black line) and red line (sink mass). Second panel: disc radius (maximum radius:
black line, mean radius: blue line, see Section~\ref{disc_iden}). Third panel: mean gas
 density within the disc. Fourth panel: mean magnetic intensity of the disc.}
\label{time_evol}
\end{figure}

Figure~\ref{time_evol} shows various relevant quantities integrated over the disc, i.e. 
its mass, radius, mean density and mean magnetic intensity as a function of time. 

In the top panel, the sink mass, $M_*$, is also given (red curve). The sink is introduced at 0.06 Myr
and in about 20 kyr its mass grows to about 0.3 $M_\odot$. Apart from a steep drop at early 
time where the disc is not well defined, $M_{\rm d}$ stays remarkably constant along time to a 
value of about $1.5 \times 10^{-2}$ $M_\odot$. This implies that the disc mass over star 
mass keeps decreasing and typically goes from about 1/10 to 1/30.  

Second panel displays the disc radius, blue curve is the mean value, $R_{\rm mean}$ (see Section~\ref{disc_iden}) 
while black is the maximum value, $R_{\rm max}$. Both values follow the same trend except for some fluctuations of 
$R_{\rm max}$, which are due to the difficulty to exclude the filaments that connect to the disc (see Fig.\ref{coldens_refmod}).
 Therefore we adopt $R_{\rm mean}$ as the disc radius. 
At sink age 10 kyr, i.e. at time 0.07 Myr, its value is about 20 AU. It then grows with time and after 100 kyr reaches roughly 
100 AU in size (see right bottom panel). 
As discussed in Section~\ref{compar_analyt}, this behaviour turns out to be in good agreement 
with the analytical prediction of \citet{hetal2016}. In particular the growth of the disc radius is 
proportional to $(M_*+M_{\rm d})^{1/3} \approx M_*^{1/3}$.

Third and fourth panels show the mean density and  volume averaged magnetic field through the disc 
(the mean value of all disc cells is taken). 
Within 100 kyr, the density has decreased
by almost 2 orders of magnitude, while the magnetic intensity drops by about a factor of 10. This is broadly consistent 
with the fact that the mean magnetic field is given by the value in the disc outer part, which itself is close
to the envelope value where typically $B \propto n^{1/2}$.

\subsection{Disc spatial structure}

\setlength{\unitlength}{1cm}
\begin{figure*}%[h!]
%\centering
\begin{picture} (0,22)
\put(0,16.5){\includegraphics[width=8cm]{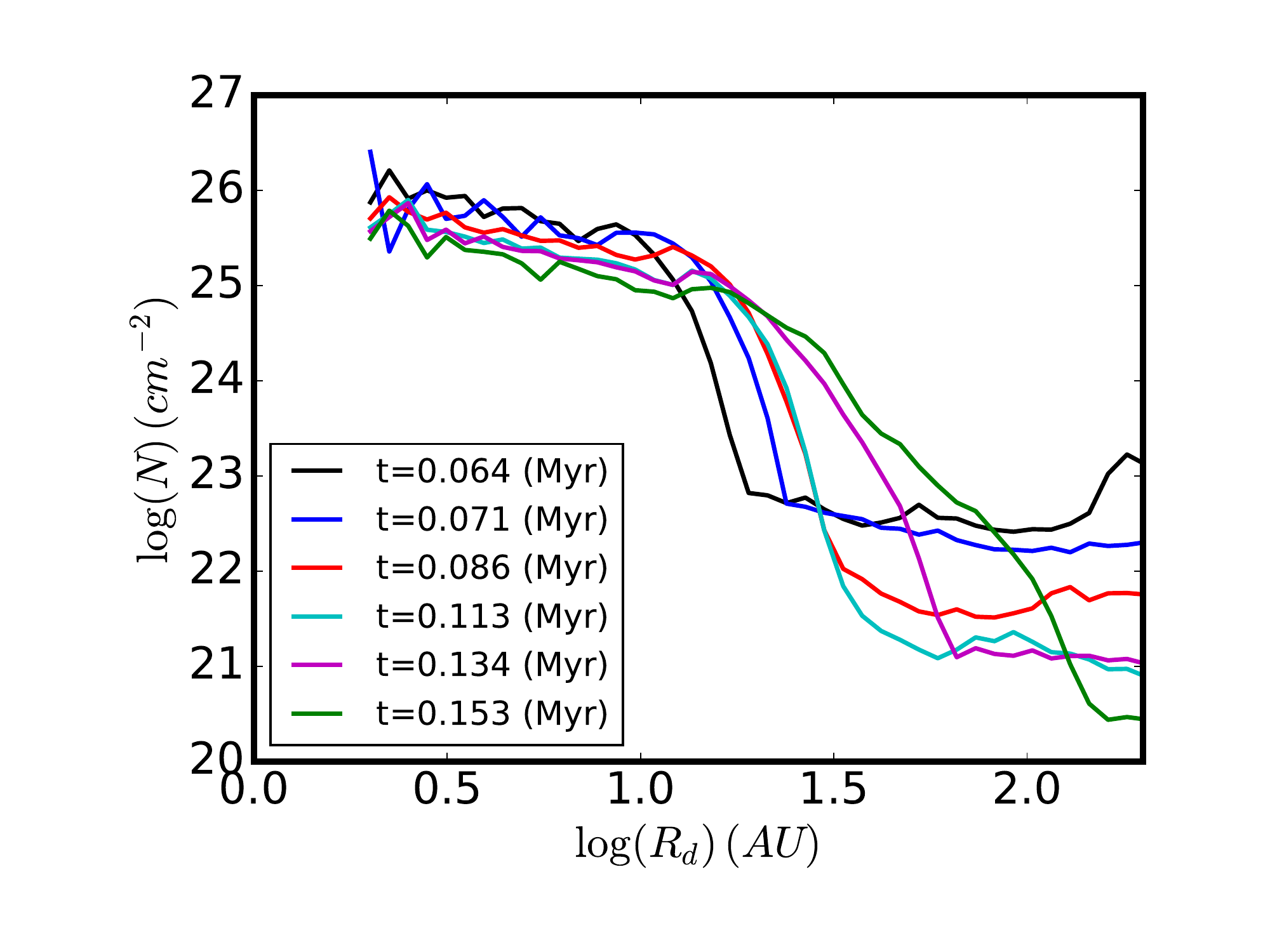}}  
\put(0,11){\includegraphics[width=8cm]{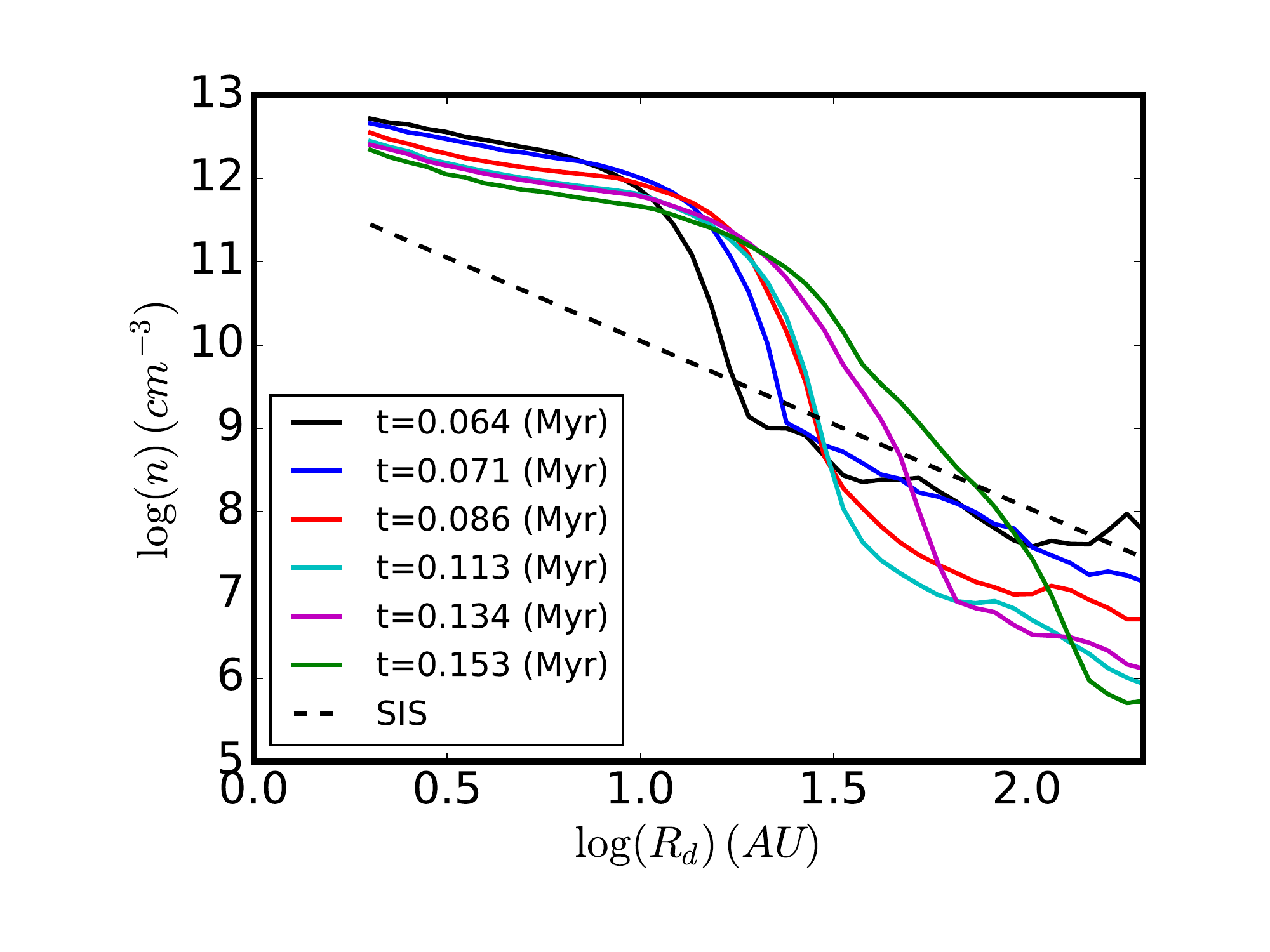}}  
\put(0,5.5){\includegraphics[width=8cm]{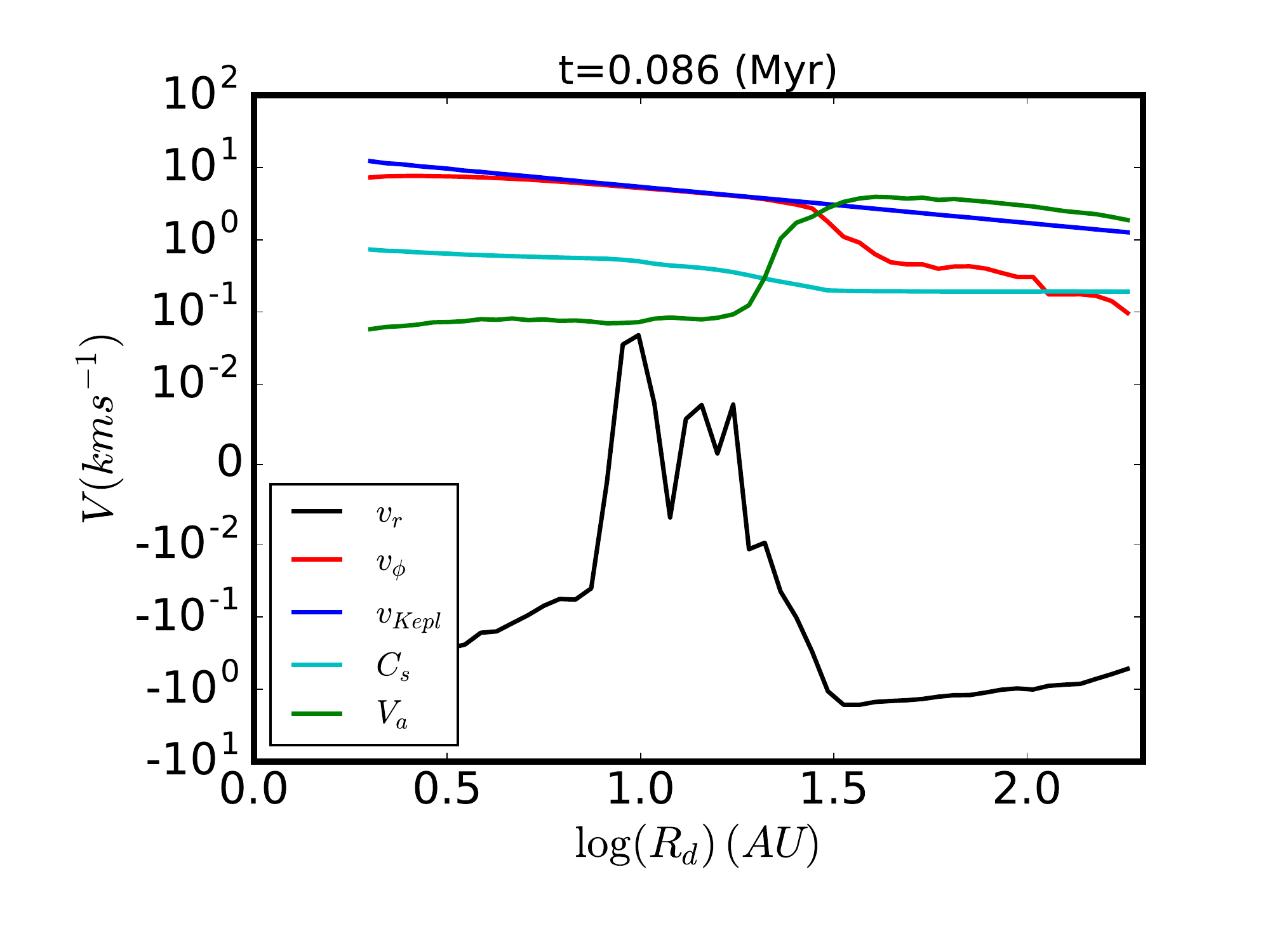}}  
\put(0,0){\includegraphics[width=8cm]{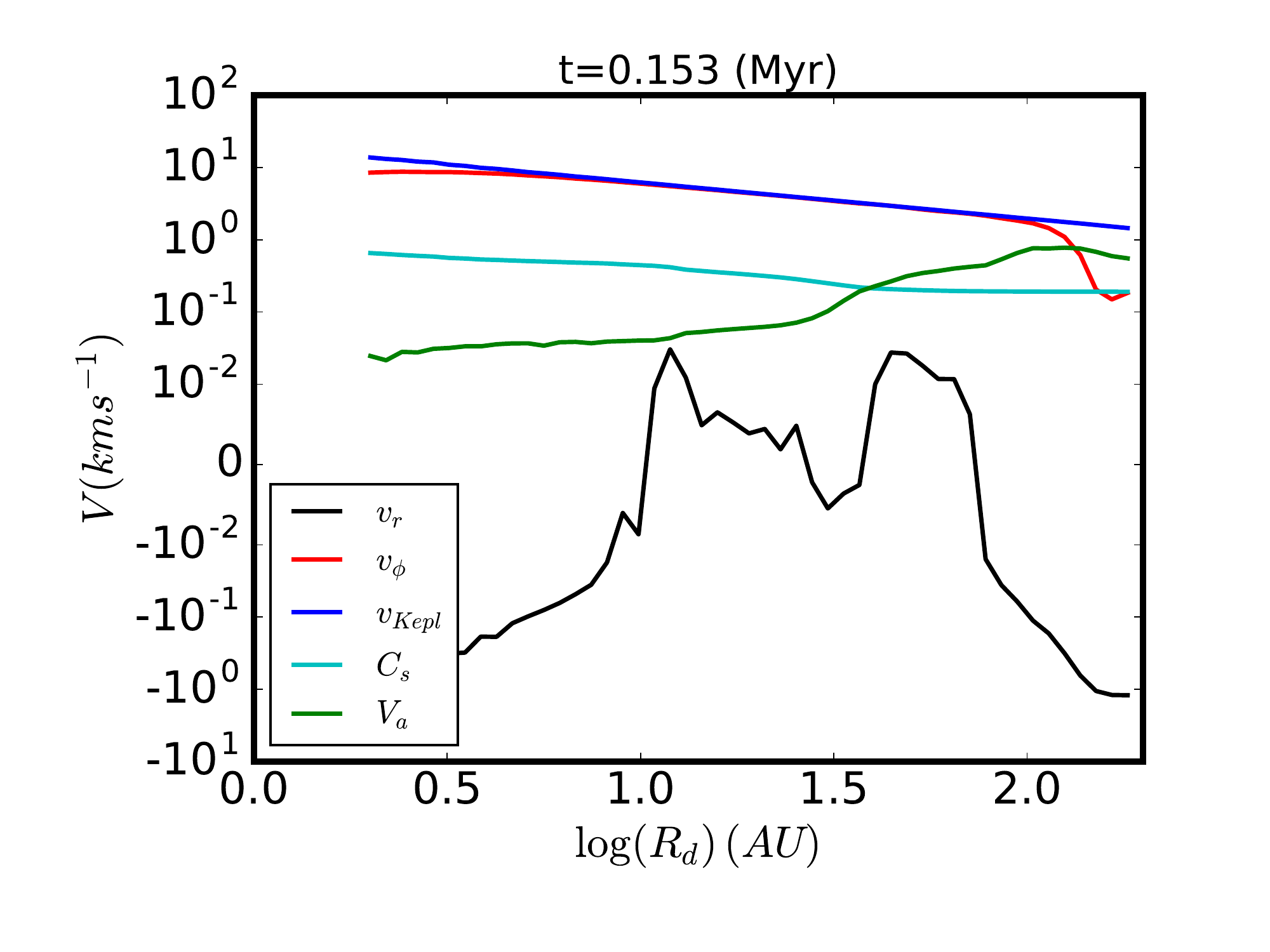}}  
\put(8,16.5){\includegraphics[width=8cm]{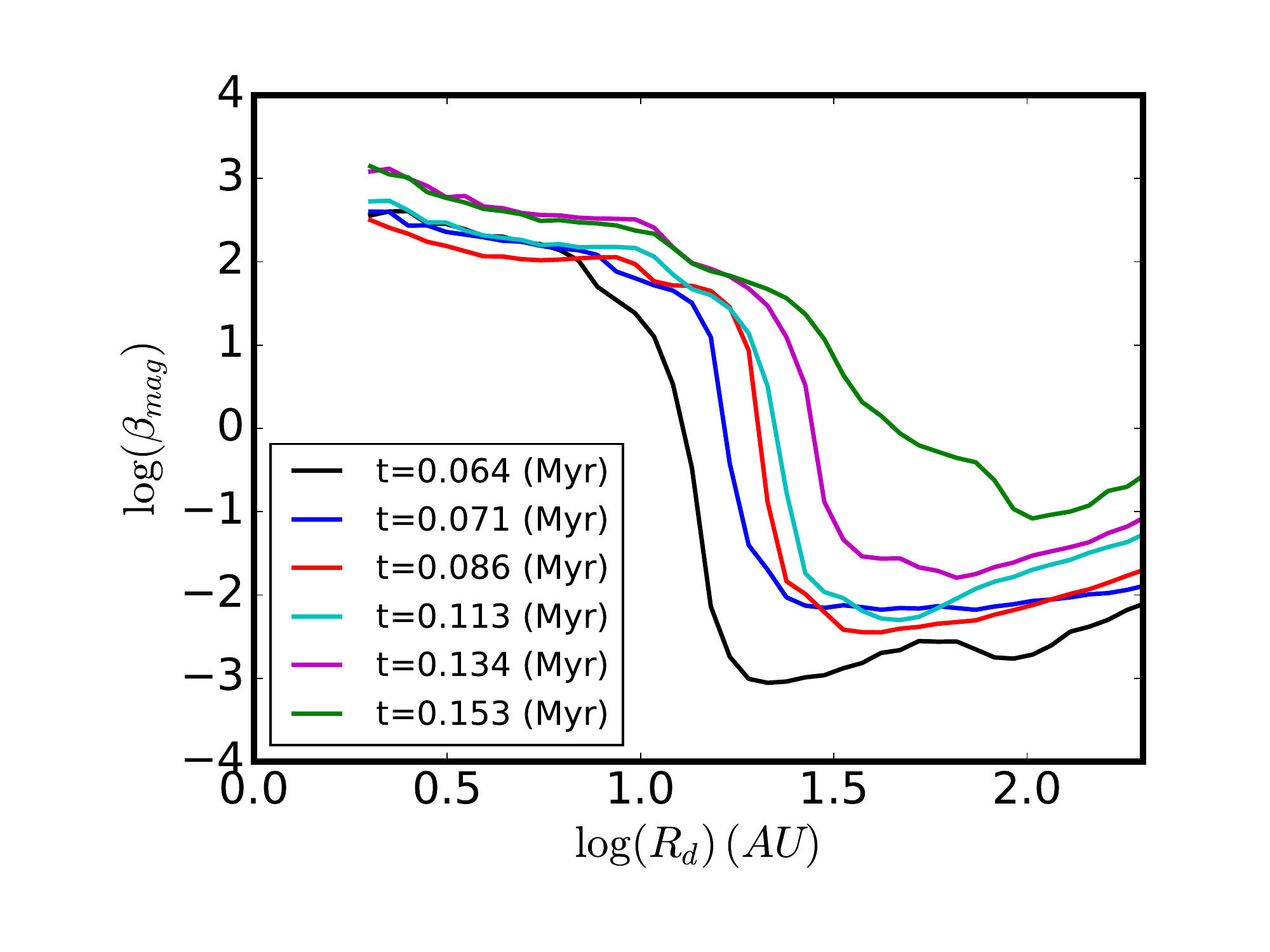}}  
\put(8,11){\includegraphics[width=8cm]{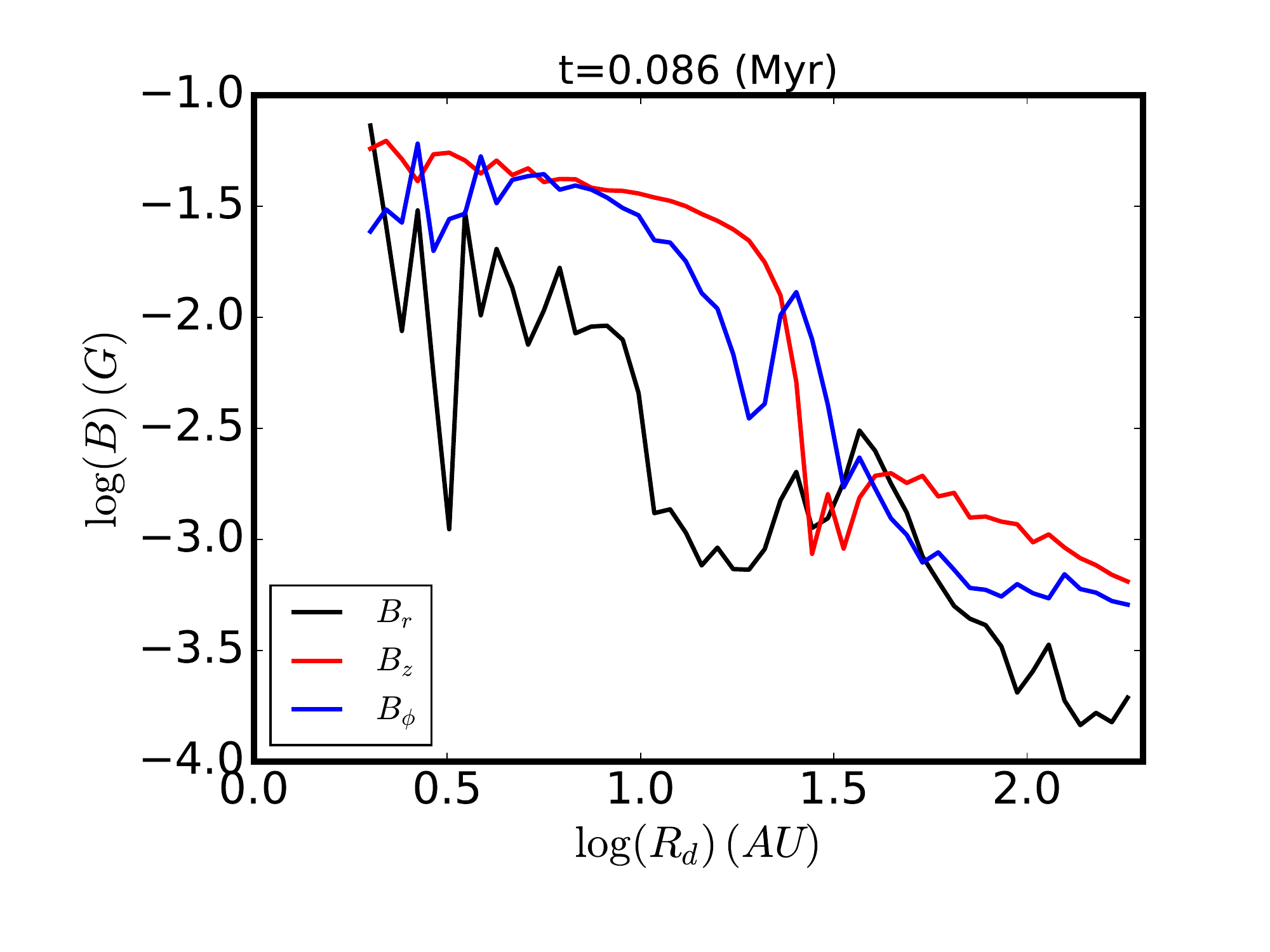}}  
\put(8,5.5){\includegraphics[width=8cm]{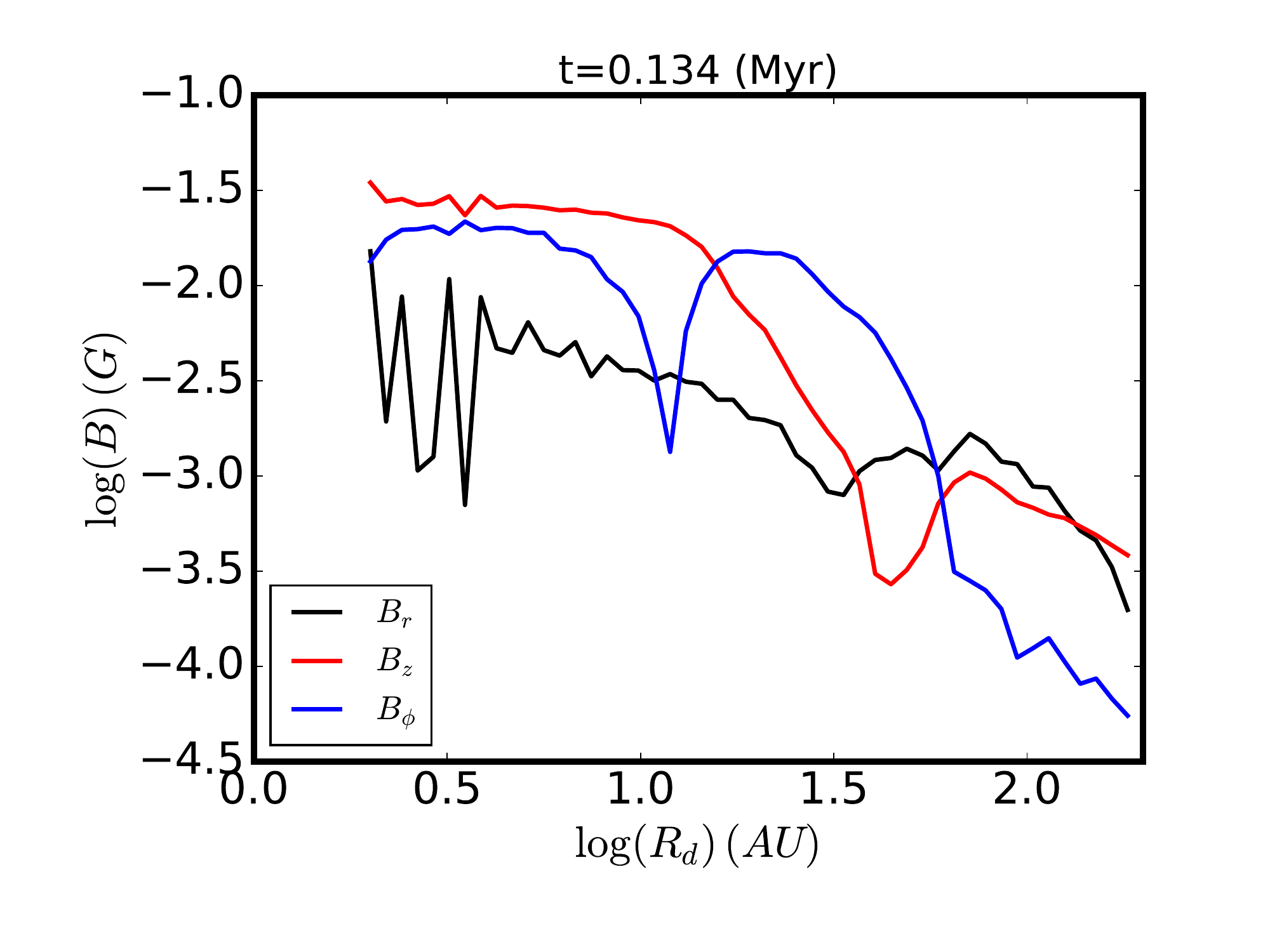}}  
\put(8,0){\includegraphics[width=8cm]{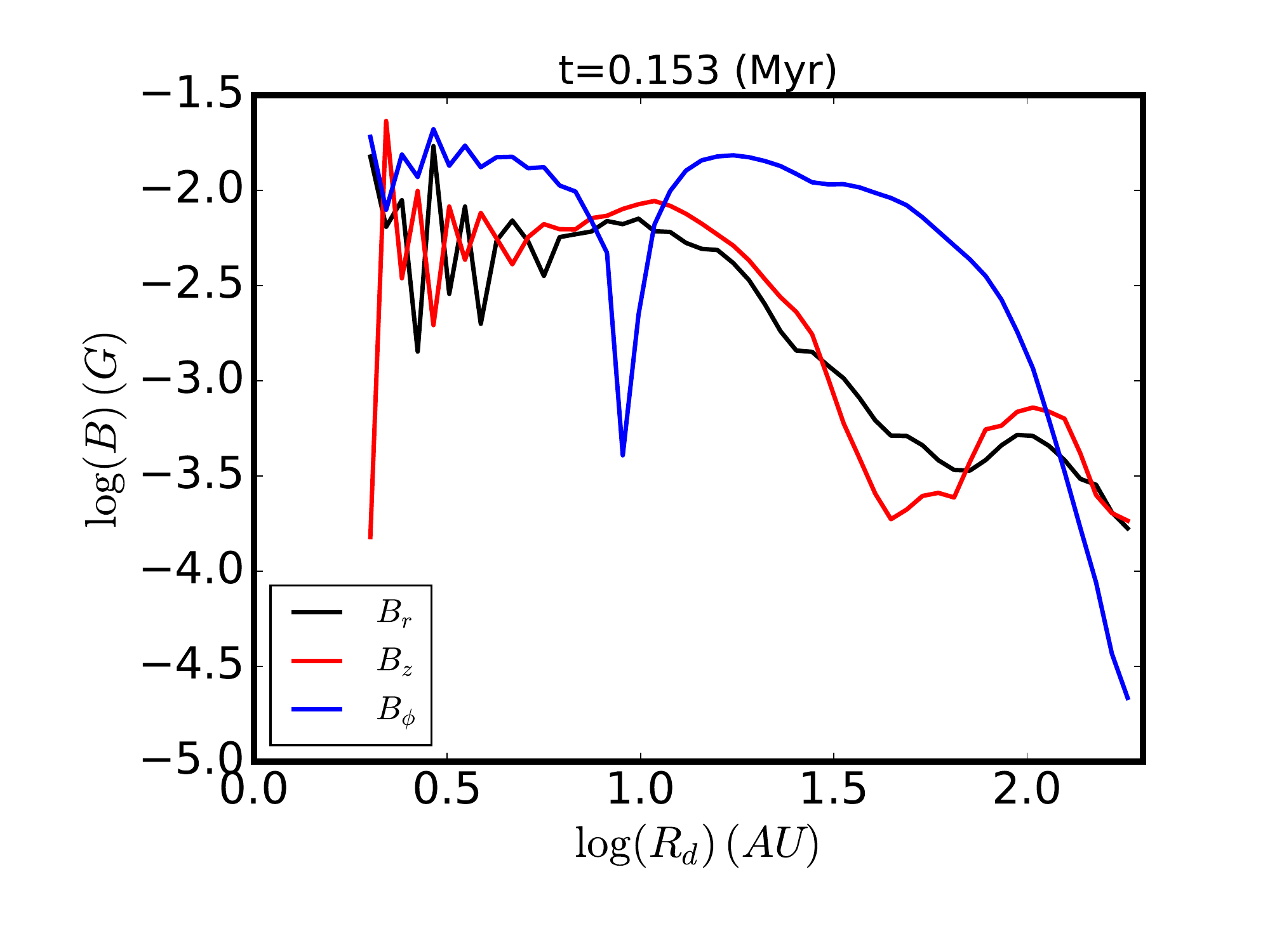}}  
\end{picture}
\caption{Radial structure of the disc and surrounding envelope. 
Top left row: mean column density of the disc of run $R2$ as a function of radius.
Second left panel: mean  density of the disc vs  radius. Third and fourth left panels: various velocities 
vs radius within the disc equatorial plane. 
Top right panel: $\beta _{\rm mag}= P_{\rm therm}/P_{\rm mag}$ as a function of radius within the disc for 
several timesteps of run $R2$. Second, third and fourth right panels display for three snapshot, the mean radial, axial and 
toroidal magnetic field as a function of radius. As can be seen while at early time, the axis field, $B_z$, is
the dominant  component, at later time it is the toroidal field which dominates.}
\label{rhov_r}
\end{figure*}

To better understand and characterize the disc, we now examine left panels of 
Fig.~\ref{rhov_r} which portray the mean radial profiles
of  the column density (top panel), the density (second panel) and several velocities at various timesteps. 
The density and the velocities are the mean values of cells located below one scale-height, $h \simeq r \times C_s / v_{\rm Kepl}$
and belonging to concentric corona, 
where $v_{\rm Kepl}$ is the Keplerian velocity and $C_s$ is the sound speed. 
In the inner part we see that the column density follow a powerlaw, $N \propto r^{-\alpha}$, 
with  $\alpha \simeq 0.5-1$. The density presents an exponent that is slightly stiffer, particularly 
within the inner 5 AU and which is about $\simeq$1-1.5.  
Extrapolating the density curves which go up to two AU (below which concentric corona are not well defined),
 we see that
the central value of the gas density
is about 10$^{13}$ cm$^{-3}$ at 1 AU, which is the threshold used for accretion onto the sink particles. 
This value slowly diminishes as time goes on. At $r$ about 20 AU, a change in the density profile
occurs and it becomes much steeper particularly at early times (i.e. $t=0.086$ and 
$t=0.113$ Myr). This corresponds to the position of the accretion shock as revealed  by the 
radial velocity profile displayed in the third panel. At later times ($t=0.134$ and $t=0.153$ Myr), the accretion shock
moves away as the disc grows in size. At time $t=0153$ Myr, it is located at about 100 AU for instance (see for example
the change of the rotation velocity at 100 AU in the left-bottom panel). We see however 
that a break in the column density and density  profile remains around 20 AU. Above this radius, the disc presents
 a steeper  profile which is broadly $\propto r^{-3}$ in column density  and $\propto r^{-4}$ in density.
Outside the disc, the envelope presents an $r^{-2}$ density profile which is close to the singular 
isothermal sphere \citep{shu77} at early times and 10-100 lower at later ones. 

The third and fourth panels display the Keplerian velocity, $v_{\rm Kepl}$, the azimuthal gas velocity, $v_\phi$,
the sound speed, $C_{\rm s}$, the Alfven speed, $v_{\rm a}$, and the radial velocity, $v_r$. 
Several important features are to be noticed:
\begin{description}
\item[-] inside the disc, $v_\phi \simeq v_{\rm Kepl}$ except below 3 AU due to the gravitational potential softening.
This confirms that the material in the disc is centrifugally supported. 
\item[-] Outside the disc, $v_\phi$ drops by a factor of about 10 below  $v_{\rm Kepl}$. 
\item[-] Inside the disc, $v_{\rm a}$ is about 10 times lower than $C_{\rm s}$ and 100 times lower than $v_{\rm Kepl}$.
\item[-] At the edge of the disc, $v_{\rm a}$ increases (steeply at early times and more gradually at later ones)
 and it dominates over the other velocities becoming comparable to the radial one, $v_r$.
\item[-] The radial velocity in the disc inner part is negative and on the order of 
$C_s/100-C_s/10$ between 3 and 10 AU (at smaller radius numerical resolution becomes insufficient).
At intermediate radius, i.e. between 10-20  AU at time 0.086 Myr and 10-80 AU at time 0.153 Myr, 
the velocity is globally positive and presents 
significant fluctuations. In the outer part, above  the accretion shock, it is negative and about 10 times 
$C_{\rm s}$.
\end{description}

Two main aspects are of particular importance. First, the value of $v_{\rm a}$ which is low inside the disc 
and high outside, indicates that the magnetic field is likely playing an important role 
in setting the disc radius most likely because strong magnetic braking occurs there (see Section~\ref{compar_analyt}).

Second we see that the mass flux is obviously not constant through the disc mid-plane since $v_r$ even becomes positive. 
Clearly to satisfy mass conservation, since the disc column density is nearly stationary in the inner part, 
 mass must flow from higher altitudes. This point is carefully addressed in Lee et al. (submitted) where 
complementary high resolution radiative calculations are presented.

\subsection{Structure of the magnetic field}

%\setlength{\unitlength}{1cm}
%\begin{figure}%[h!]
%%\centering
%\begin{picture} (0,15)
%\put(0,10){\includegraphics[width=7cm]{FIG_PAPIER_DISC_PHYS/DC_2/B_z_zoom1_01000.jpeg}}  
%\put(0,5){\includegraphics[width=7cm]{FIG_PAPIER_DISC_PHYS/DC_2/B_z_zoom1_03000.jpeg}}  
%\put(0,0){\includegraphics[width=7cm]{FIG_PAPIER_DISC_PHYS/DC_2/B_z_zoom1_04000.jpeg}}  
%\end{picture}
%\caption{Magnetic intensity and projected magnetic field direction for three snapshots for Run2. 
%The magnetic field within the disc becomes progressively more organised and dominated by the toroidal component.}
%\label{im_B}
%\end{figure}

%\setlength{\unitlength}{1cm}
%\begin{figure}%[h!]
%%\centering
%\begin{picture} (0,22)
%\put(0,16.5){\includegraphics[width=8cm]{FIG_PAPIER_DISC_PHYS/DC_2/beta_disk_r_mult.pdf}}  
%\put(0,11){\includegraphics[width=8cm]{FIG_PAPIER_DISC_PHYS/DC_2/B_disk_r_01000.pdf}}  
%\put(0,5.5){\includegraphics[width=8cm]{FIG_PAPIER_DISC_PHYS/DC_2/B_disk_r_03000.pdf}}  
%\put(0,0){\includegraphics[width=8cm]{FIG_PAPIER_DISC_PHYS/DC_2/B_disk_r_04000.pdf}}  
%\end{picture}
%\caption{Top panel: $\beta _{\rm mag}= P_{therm}/P_{mag}$ as a function of radius within the disc for 
%several timesteps of Run2. Second, third and fourth panels display for three snapshot, the mean radial, axial and 
%toroidal magnetic field as a function of radius. As can be seen while at early time, the axis field, $B_z$, is
%the dominant  component, at later time it is the toroidal field which dominates.
%}
%\label{mag_r}
%\end{figure}

Since magnetic field is believed to play an important role in the disc by inducing transport of angular momentum
through either  magneto-rotational instability or  magnetic braking, but also possibly by creating structures 
such as rings \citep{bethune2017,suriano2018}, we look more closely into its structure.
%Figure~\ref{im_B} shows the integrated field magnetic (using a simple density weigthing) in the z-direction.
%The arrows correspond to the direction of the projected field. We see that as the disc grows, 
%the structure of the magnetic field becomes progressively more organised and dominated by the toroidal component.
%There is also a trend for the field to become weaker at later times.

Right panels of Fig.~\ref{rhov_r} displays the value of $\beta _{\rm mag}$ as a function of radius for several 
timesteps as well as the mean radial profiles of $B_z$, $B_r$ and $B _\phi$ at three snapshots. The profiles 
of $\beta _{\rm mag}$ reveal that while at early times the transition between $\beta _{\rm mag} > 1$ and $\beta _{\rm mag} < 1$ occurs 
through the shock, it is not the case at later times, where the disc outer part is significantly magnetized 
and present $\beta _{\rm mag}$ values < 1. 

At time 0.086 Myr  $B_z$ and $B_\phi$ have comparable amplitudes and dominate over $B_r$. At times 
0.134 and 0.153 Myr, $B_z$ decreases and diffuses out of the disc while on the contrary the toroidal 
field grows in the disc outer part (i.e. between 20 and 100 AU). This growth of the toroidal field occurs
in a region where $\beta _{\rm mag} < 1$ and therefore where the field and the gas are well coupled. It is a natural 
consequence of the differential rotation \citep[see for example][]{ht2008}. As can be seen in 
the fourth left panel of Fig.~\ref{rhov_r}, $v_{\rm a}$ is larger than $C_{\rm s}$ in the disc outer part. This excess 
of magnetic toroidal pressure leads to inflated disc as observed in the third panel of Fig.~\ref{rho_z_Run2}.

\section{Influence of sink particles on disc evolution}

\setlength{\unitlength}{1cm}
\begin{figure}[]%[h!]
%\centering
\begin{picture} (0,10)
\put(0,5){\includegraphics[width=7cm]{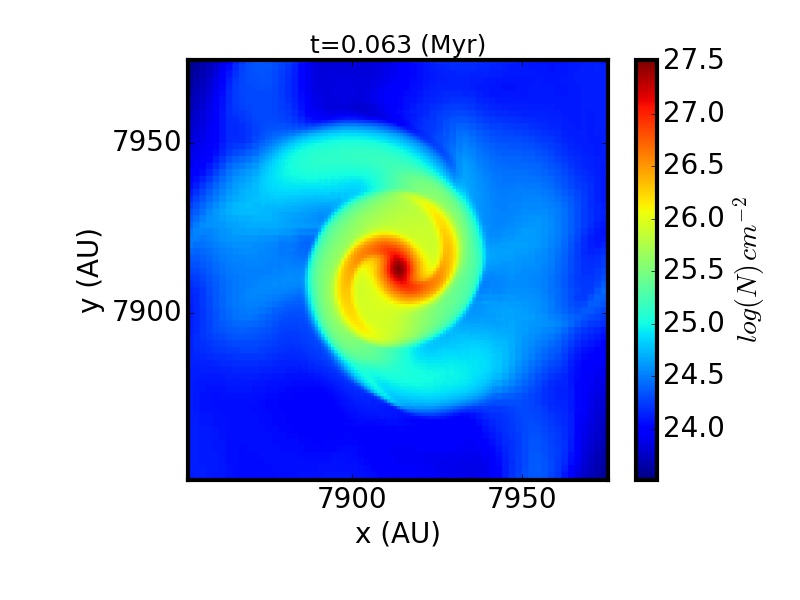}}  
\put(0,0){\includegraphics[width=7cm]{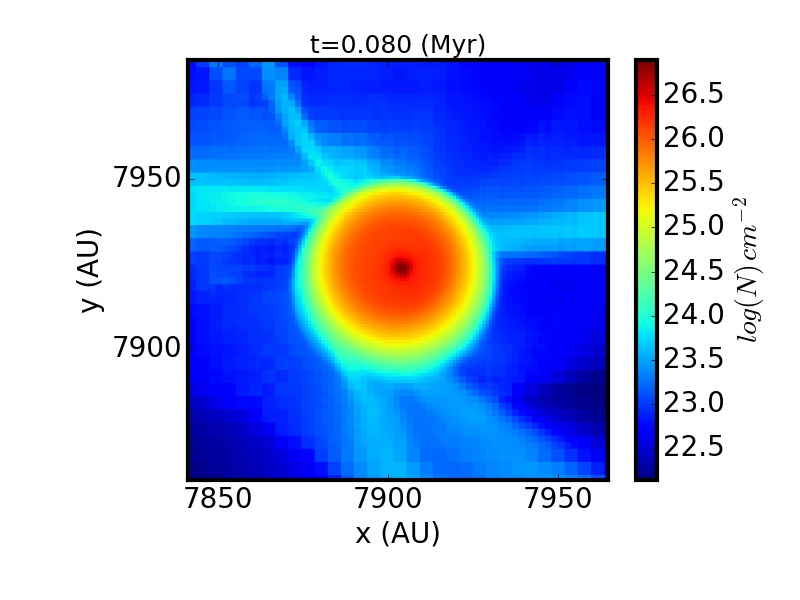}}  
\put(1.,9.8){$Rns$}
\put(1.,4.8){$Rhsink$}
\end{picture}
\caption{Column density of run $R2ns$ (top) and run $R2hsink$ (bottom). These runs are identical to the reference
run $R2$ but $R2ns$ has no sink particle while $R2hsink$ has a value $n_{\rm acc}$ which is 4 times 
higher. They should be compared with Fig.~\ref{coldens_refmod}.
In the absence of sink particles the disc is more extended due to 
prominent spiral patterns. With a higher value of $n_{\rm acc}$ the disc is also a bit bigger 
and more massive (see Fig.~\ref{time_evol_num}) than the disc in run $R2$. }
\label{coldens_nosink}
\end{figure}

\setlength{\unitlength}{1cm}
\begin{figure}%[h!]
%\centering
\begin{picture} (0,11)
\put(0,5.5){\includegraphics[width=8cm]{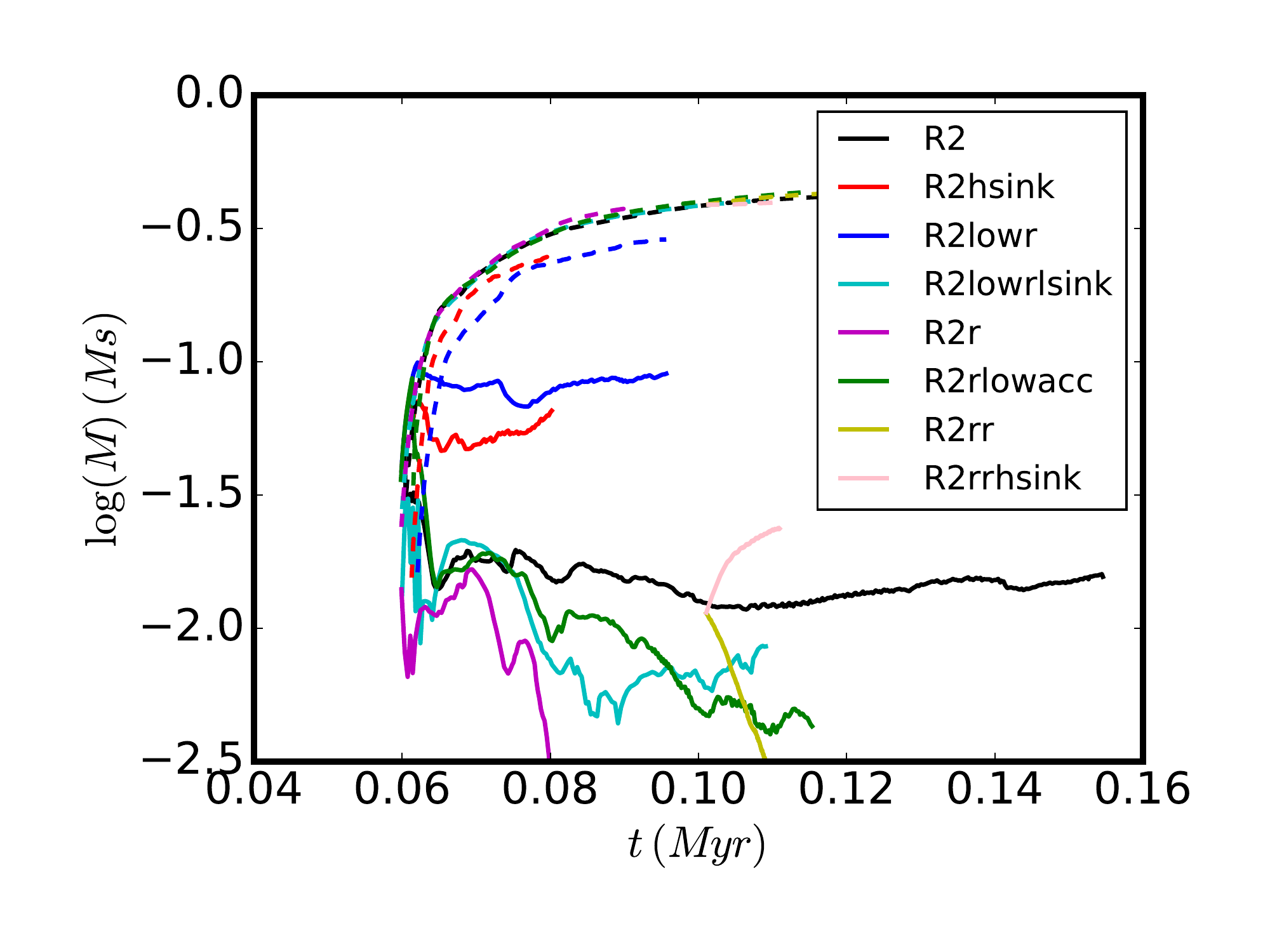}}  
\put(0,0){\includegraphics[width=8cm]{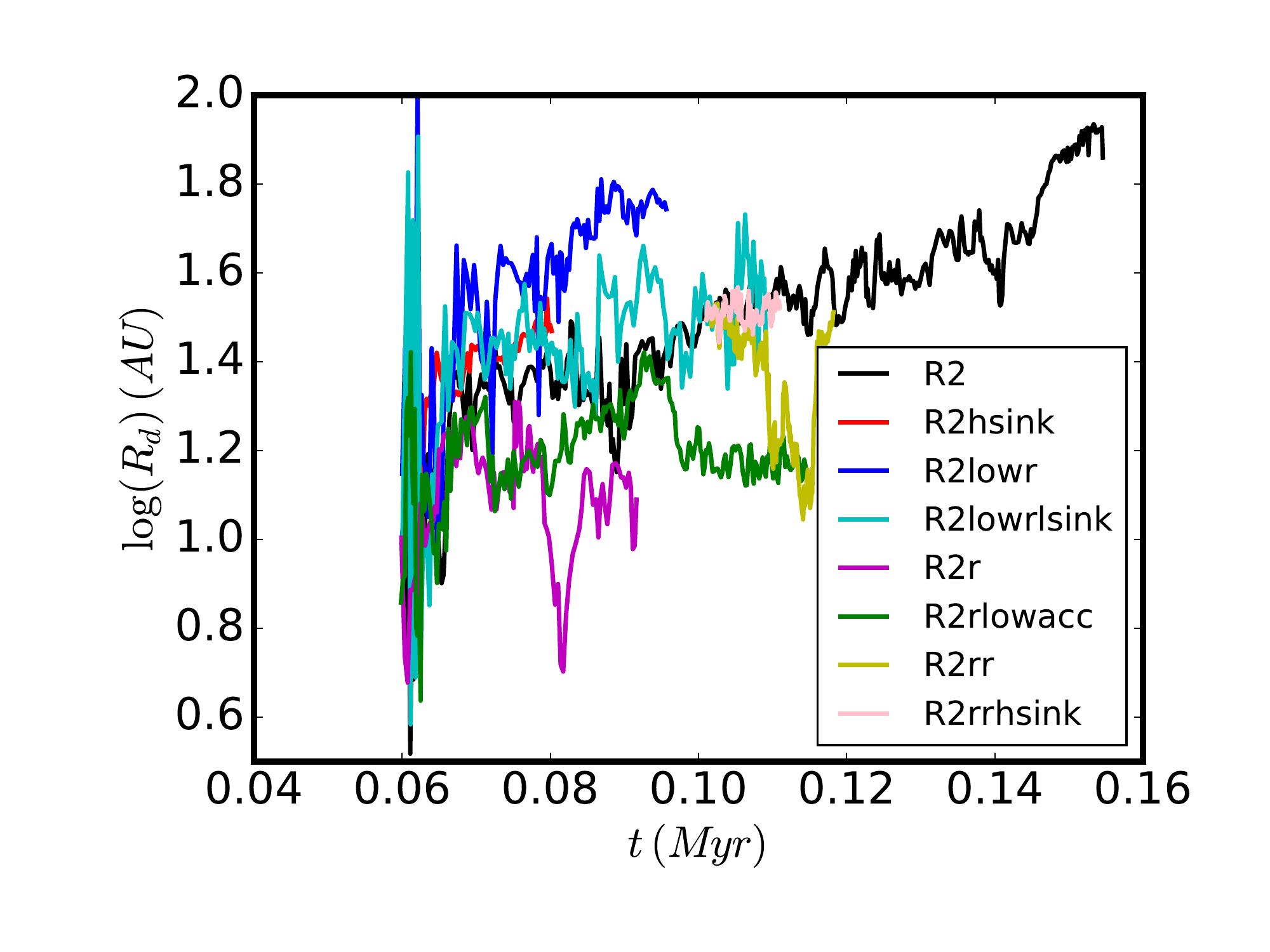}}  
\end{picture}
\caption{Comparison between disc radius and mass for the 9 numerically 
motivated runs listed in Table~\ref{table_param_num}. In top panel the dashed lines represent the 
mass of the sink particles. While the sink mass does not depend significantly on the 
detailed of the numerical algorithm (i.e. sink threshold and resolution), the disc properties 
depend very much on it. }
\label{time_evol_num}
\end{figure}

To understand disc formation, it is necessary to use sink particles. How the sink 
particles are exactly implemented remains largely arbitrary and it is 
necessary to understand the effect of the sink algorithm on the disc 
properties.

\subsection{Disc evolution without sink particle}

To see the drastic influence of sink particles, the top panel of Fig.~\ref{coldens_nosink} portrays
the column density of run $R2ns$ in which no sink particle is being used. 
The comparison with Fig.~\ref{coldens_refmod} shows drastic differences. 
Clearly the disc is larger and presents prominent spiral patterns. These later 
are largely a consequence of the fact that in the absence of sink 
particle, the gas that has accumulated in the central, thermally supported hydrostatic 
core may be transported back at larger radii and contribute to make the disc gravitationally unstable. 
Sink particles on the other hand introduce a strong irreversibility mimicking gas accretion onto stars.
Clearly runs that do not use sink particles are not sufficiently realistic to describe the 
  long time disc evolution.  As we show below, the question however of which sink parameters
should be used and how the gas is retrieved from the grid are crucial.

\subsection{Influence of sink parameters on sink mass evolution}
Top panel of Fig.~\ref{time_evol_num} shows the sink mass (dashed lines) as 
a function of time for the numerically motivated runs (Table~\ref{table_param_num}). 
In this series of runs, the numerical resolution varies by a factor 4 and the 
sink accretion threshold by a factor 40. Clearly, the sink mass evolution is 
only marginally influenced by the numerical parameters. In all runs the sinks
formed roughly at the same time and the mass evolution is very similar. 

This is an important conclusion because it confirms that the global evolution of the dense core 
is not very much affected by the numerical algorithm. In particular,  the 
accretion rate onto the sink particles is likely a consequence of the large scale of the core 
rather than the detailed small scale processes.  In other words, the traditional 
picture that the infalling gas is stored in the centrifugally supported disc and then gets
transported through it up to the star, is not supported by the present simulations, at least 
in its simplest form.

\subsection{Influence of sink parameters on disc evolution}
To understand the influence of the sink parameter,  run $R2hsink$, which 
has a value of $n_{\rm acc}$ 4 times above the one used in run $R2$, has been carried out. 
Bottom  panel of Fig.~\ref{coldens_nosink} shows the disc of run $R2hsink$ at time 
0.08 Myr. As can be seen from a comparison with  Fig.~\ref{coldens_refmod},
the disc of run $R2$ and $R2hsink$ are similar but the latter is more massive and 
slightly bigger. The reason is simply that the central disc density is higher for 
run $R2hsink$ because of the higher value of  $n_{\rm acc}$.

Top panel of Fig.~\ref{time_evol_num} portrays the disc mass (solid lines) evolution 
while bottom panel displays the disc radius for all runs listed in Table~\ref{table_param_num}. 
Very large variations are observed. 
For instance the disc mass varies by more than one order of magnitude between 
runs $R2lowr$ and runs $R2r$. This indicates that unlike for the sink mass, the disc
mass and, to a lower extent, the disc radius very much depend on the numerical scheme.
Below we analyse in details the various runs but two conclusions can already be
drawn. First of all, the disc itself does not determine the accretion onto the star
in the sense that varying the mass of the disc does not necessarily affect the accretion
onto the star.  
Second of all, the disc properties, in particular its mass, cannot be reliably predicted because they are 
controlled by numerical parameters. 

\subsubsection{Interdependence between resolution and sink particle threshold}
First, we compare runs $R2$ and $R2hsink$. The only difference is 
the value of $n_{\rm acc}$, which is 4 times larger for $R2hsink$ than for $R2$. 
The mass of the disc is about $\simeq$3 times larger in run $R2hsink$ 
while the disc radius is comparable though slightly larger. 
The reason is simply that as for run $R2$ (see Figs.~\ref{rhov_r} and~\ref{rhov_r_R2hsink}), the density at the 
edge of the sink particle is determined by $n_{\rm acc}$ and since the 
profile is similar, the mass is almost proportional to $n_{\rm acc}$. 

Second, we compare runs $R2$, $R2lowr$ and $R2r$. These three runs have same 
value of $n_{\rm acc}$ but differ in spatial resolution, which go from 
2 AU ($R2lowr$) to 0.5 AU ($R2r$).
The differences between the three discs are rather large, with a factor of $\simeq$3-5 difference between 
the disc mass of runs $R2lowr$, $R2$ and $R2r$ respectively. In terms of radius, it is a factor of 2-3. 
This clearly indicates that both the resolution and the sink density threshold, $n_{\rm acc}$, 
are important in determining the disc properties. This is confirmed by run $R2lowrlsink$ 
which has a resolution of 2 AU and a value $n_{\rm acc}= 10^{12}$ cm$^{-3}$, i.e. 10 times below the values
in the runs  $R2$, $R2lowr$ and $R2r$. In this case, the disc mass is inbetween its value in runs
$R2$ and $R2r$. 

Roughly speaking the disc mass is proportional to $dx^{-\delta} n_{\rm thres} $ with $\delta \simeq 1-2$. 
This can simply be understood as an inner boundary conditions. The density at the radius $r=dx$ is 
equal to $n_{\rm acc} = n _{thres} / 3$, 
therefore the density within the disc is about $n(r) \simeq n_{\rm acc} (dx/r)^{\delta'}$
with $\delta' \simeq 1-1.5$. Thus since the disc density decreases with resolution, the disc mass 
decreases as well. 

\subsubsection{Deciphering the role of density threshold and sink creation}
To be certain that the effects we are observing is not due to differences 
of the sink creation time, we have performed two runs, namely $R2rr$ and $R2rrhsink$,
 for which we restart from an output of run $R2$. The two runs have a resolution of 
0.5 AU and a threshold density of $n_{\rm acc}=  10^{13}$ cm$^{-3}$ and 10 times 
this value respectively. We see that the disc mass of run $R2rr$ drops by at least a factor of 3 
while the one of run $R2rrhsink$ increases by about 50$\%$ while its radius does not evolve significantly. 

This confirms that the observed differences are due to the accretion condition onto the sink particles.

\subsubsection{The amount of accreted gas}
Finally, to investigate whether the disc mass is indeed influenced by the accretion 
rate from the disc onto the star, we performed run $R2rlowacc$ which is identical to 
run $R2r$ except that the value of $C_{\rm acc}$ is 0.01 instead of 0.1. We recall that this is 
the fraction of gas above $n_{\rm thres}$ which is accreted at all timesteps. 

Figure~\ref{time_evol_num} shows that the disc of run $R2rlowacc$ is more massive (up to 3-4 times
depending of the period) and roughly 2 times larger in radii. This confirms  that the accretion 
rate at the inner boundary is a fundamental quantity that drastically  impacts the disc 
properties, in particular its mass. 

\subsection{Physical interpretation}
The results presented in Fig.~\ref{time_evol_num} demonstrate that the accretion onto the 
sink particle, which operates as  an inner boundary condition,  is a major issue for the disc. 
Similar results have been obtained by \citet{machida2014} and recently by \citet{vorobyov2019} who found that 
the accretion rate in their sink particles has a drastic impact onto the disc. 
The obvious question is then whether this choice is realistic and what physically determines the
inner boundary. We propose that this is the star-disc connection which could be limiting the accretion and 
to illustrate how it may work we propose a simple model suggesting that the choices made here are reasonable.
Note that at the T-Tauri stage, it is believed that the accretion proceeds through the 
magnetospheric accretion process \citep{bouvier2007}. However our simulations take place
far earlier than the T-Tauri phase and also unlike what is assumed in the classical models, 
we do not find, that most of the accretion proceeds through the bulk of the disc.
% although this conclusion 
%could change if we were resolving much smaller scales (particularly because at very high density
%the magnetic resistivity increases dramatically due to Ohmic resistivity \citet{tomida2013}).

We use the classical stationary $\alpha$-disc model which assumes that the effective disc viscosity is given 
by 
\begin{equation}
\label{visc}
\nu = \alpha C_s h = \alpha C_s^2 / \Omega,
\end{equation}
where $h$ is the scale height, $\Omega$ is the Keplerian rotation and $\alpha$ is a dimensionless
number on the order of $10^{-3}-10^{-2}$ depending on the disc physics. 
It is well known \citet[e.g.][]{pringle1981} that the disc surface density, $\Sigma$ is given by
\begin{equation}
\Sigma =  { \dot{M}_{\rm d} \over 3 \pi \nu} + {C \over 3 \pi \sqrt{G M_*} \nu R^{1/2} }, 
\Sigma = 2 \rho h, 
\label{sigma}
\end{equation}
where $\dot{M}_{\rm d}$ is the accretion through the disc, $M_*$ is the mass of the central star and 
$C$ is a constant, which is usually chosen in such a way that at the star radius, $R_*$, 
$\Sigma(R_*)=0$. This choice, while natural, leads to an infinite velocity at $R_*$ since 
\begin{equation}
v_{r,d} = { - \dot{M}_{\rm d} \over 2 \pi R \Sigma  },
\end{equation}
and $\dot{M}_{\rm d}$ is assumed to be constant through the disc. 

To illustrate how the connection between the star and the disc, may influence the disc
characteristics, we make very simple assumptions. First, we assume that 
the star is spherical and present a profile of density, $\rho_*$, temperature $T_*$ and 
velocity $v_*$. This latter is not zero because the star is accreting material from the 
outside at a rate $\dot{M}_*$, which we assume is nearly spherical. 
That is to say, the star is assumed to accrete both through the disc and also to receive 
a nearly spherical flux of gas.
Since we are interested 
by the outer layer at which the disc connects, we have 
\begin{equation}
v_{r,*} = { - \dot{M}_* \over 4 \pi R^2 \rho_*  }.
\end{equation}
Since the fluid velocity and density must be continuous, we have $v_{r,d} = v_{r,*} $
and  $\rho_{\rm d} = \rho_{*} $, which leads 
to
\begin{equation}
\label{acc_rap}
{ \dot{M}_{\rm d} \over     \dot{M}_* } \simeq { h \over 2 R_*} \simeq { C_s \over 2  R_* \Omega (R_*)  }.
\end{equation}
At early stage, the stellar radius is typically a few solar radii and the 
temperature of the gas on the verge to enter the star is about 
$10^4$ K \citep{masunaga2000,vaytet2018}. 
We estimate
\begin{equation}
\label{acc_rap2}
%{ \dot{M}_{\rm d} \over \dot{M}_* } \simeq  10^{-2}  \left( {T_d \over 10^4 \, K} \right)^{-1/2} 
%\left( { M \over M_*  } \right)^{-1/2}  \left( { R \over R_*  } \right)^{1/2}, 
{ \dot{M}_{\rm d} \over \dot{M}_* } \simeq  10^{-2}  \left( {T_d \over 10^4 \, K} \right)^{1/2} 
\left( { M_* \over M_\odot  } \right)^{-1/2}  \left( { R_* \over R_\odot  } \right)^{1/2}, 
\end{equation}
which shows that under these circumstances, the accretion rate through the disc is a few percent
of the accretion onto the star. In the context of an $\alpha$-disc, this also 
fixes the column density profile. 

We stress that the present model is essentially illustrative and that a more 
detailed analysis should be performed. Other assumptions will lead to different disc properties. 
It however supports the idea that the accretion 
rate through the disc and its density profile, at large scale, may be influenced by the small scale star-disc connection 
at the early stage of the star formation process. Unfortunately little is presently 
known on this phase as studies which have addressed the second collapse and the formation 
of the protostar itself \citep[e.g.][]{machida2006,tomida2013,vaytet2018} are very limited in time
due to the small Courant conditions.

\subsection{A criteria for sink accretion threshold}
The obvious question is therefore what  value of the accretion threshold should be used.
As discussed in the previous section, this is not an easy question as it may request a 
detailed knowledge of the disc structure at small radii may be up to the star itself. 
To proceed we again use the standard $\alpha$-disc model. Using Eqs.~(\ref{visc}-\ref{sigma})
we get
 
\begin{eqnarray}
\rho \simeq {\Sigma \over 2 h} =  {\dot{M}_{d} \over 6 \pi \alpha C_s h^2 } = {\dot{M}_{d} \Omega^2 \over 6 \pi \alpha C_s^3},
\end{eqnarray}
Now with Eq.~(\ref{eq_full_eos}), we have  when $n \gg n_{\ad}:$
\begin{eqnarray}
C_s \simeq C_s^0 n ^{2/10} n_{ad}^{-2/6} n_{ad,2}^{5/6-7/10},
\end{eqnarray}
where $C _s ^0$ is the sound speed in the isothermal envelope and is about 0.2 km s$^{-1}$.
and this leads for the gas density to
\begin{eqnarray}
n (R) \simeq  \left( {G \dot{M}_{d}  M_* \over 6 \pi \alpha m_p} (C_s^0)^{-3} n_{ad} n_{ad,2}^{-2/5}  R^{-3} \right)^{5/8},
\end{eqnarray}
where $m_p$ is the mean mass per particle and $n = \rho / m_p$.
We therefore obtain
\begin{eqnarray}
\label{n_acc}
n (R) &\simeq& 5.5 \times 10^{13} \rm{cm}^{-3} \\
& & \left( {\dot{M}_{d} \over 10^{-7} M_\odot \, \rm{yr}^{-1}}  \right)^{5/8} \left(  { M_* \over 1 M _\odot} \right)^{5/8} 
\left( {\alpha \over 0.01} \right)^{-5/8}  \left(  {R \over 1 \rm{AU} } \right)^{-15/8}.
\nonumber
\end{eqnarray}
% 10^-7 Ms / yr = 6.6e18 g / s
%>>> print ( (6.6e18) /6. / 3.1416 / 0.01 / 8. / 1.e12 / 4.e-24 * (3.e-14/4.e-24)**1. * (9.e-13/4.e-24)**(-0.4) *6.8e-8 * 2.e33 / (1.5e13)**3 )**(5./8.)
%5.45243822365e+13

Several points must be discussed. First the typical value of $n$ at 1 AU is expected to 
be around $10^{13}$ cm$^{-3}$ if $M_* = 0.3   M_\odot$ as it is the case in the present study. 
Second this density depends on the accretion rate through the disc as well as the effective 
viscosity. Both are heavily uncertain and their estimate requires a careful and specific 
investigation.  The reference value $\dot{M}_d = 10^{-7}$ M$_\odot$ yr$^{-1}$, is simply based on 
Eq.~(\ref{acc_rap2}) and on the measured accretion ratio in the sink which is on the order of
$10^{-6}-10^{-5}$  M$_\odot$ yr$^{-1}$. 

When  using the sink particles, one must make sure that the density at which the gas is accreted 
within the sink is close to the value stated by Eq.~(\ref{n_acc}). 
There is however another source of uncertainty
which comes from the radius at which this should be evaluated. Typically, it could be 
either the sink radius, $4 dx$, or the resolution, $dx$. As revealed by
Fig.~\ref{rhov_r}, the density reached $n_{\rm acc}$ only in the very center. Therefore estimating 
$n_{\rm acc}$ at $dx$ seems to be a better choice. 
In any case, there is, as described in the present section, a dependence of the accretion threshold 
$n_{\rm acc}$
into the resolution. Typically from Eq.~(\ref{n_acc}), it broadly scales as $dx^{-2}$, which is 
compatible with what as been inferred from the numerical experiments presented above. 
Finally, we stress that in reality the value of $n_{\rm acc}$ should be time dependent as it depends 
on M$_*$ and $\dot{M}_d$, which both evolve with time. It is also possibly physics dependent since
it depends on $\alpha$, which in turn may  depend on the magnetic intensity for instance.

\section{Influence of initial conditions on disc formation and evolution}
 We now investigate the influence that initial conditions have onto the disc evolution. We focus 
on the global disc properties, i.e. its mass and size. As discussed in the previous section, the 
accretion threshold $n_{\rm acc}$ is of primary importance while highly uncertain. We adopt the value 
$n_{\rm acc} = 10^{13}$ cm$^{-3}$ as it is close to our analytical estimate. While it is likely that 
better,  time-dependent,
 estimates will have to be used in the future, it is nevertheless important to investigate 
the influence of initial conditions because they are also playing an important role in the disc evolution.
In particular, the resulting discs can be compared to understand the consequences 
of the  various processes. From the previous section, we saw that the mass of the disc is 
nearly proportional to $n_{\rm acc}$, as long as this parameter varies from a factor of a few. So
in principle, the results presented below should remain valid for other values 
of $n_{\rm acc}$ by simply multiplying the disc mass by $n_{\rm acc} / 10^{13}$ cm$^{-3}$.

\subsection{The impact of rotation}

\setlength{\unitlength}{1cm}
\begin{figure}%[h!]
%\centering
\begin{picture} (0,11)
\put(0,5.5){\includegraphics[width=8cm]{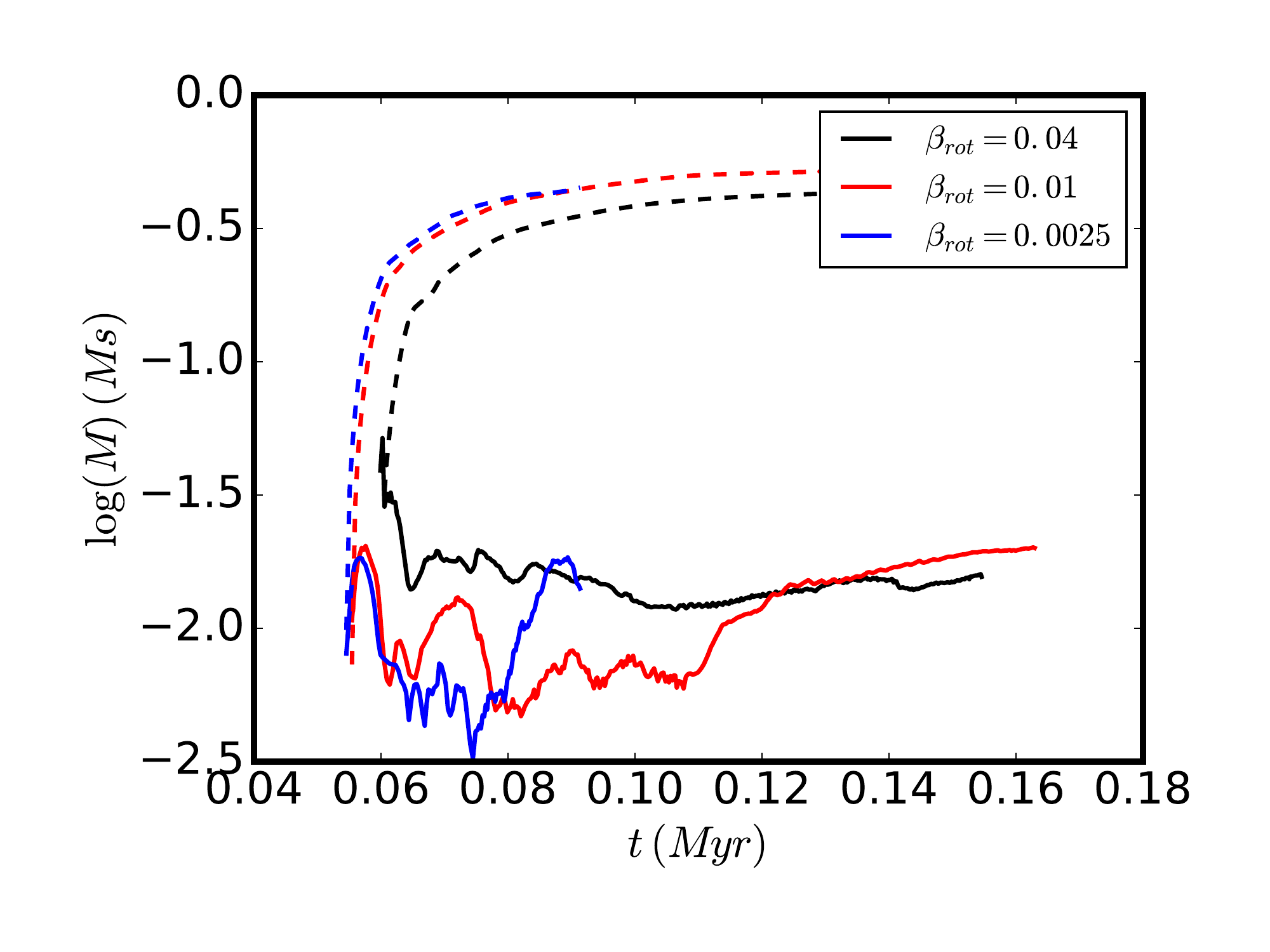}}  
\put(0,0){\includegraphics[width=8cm]{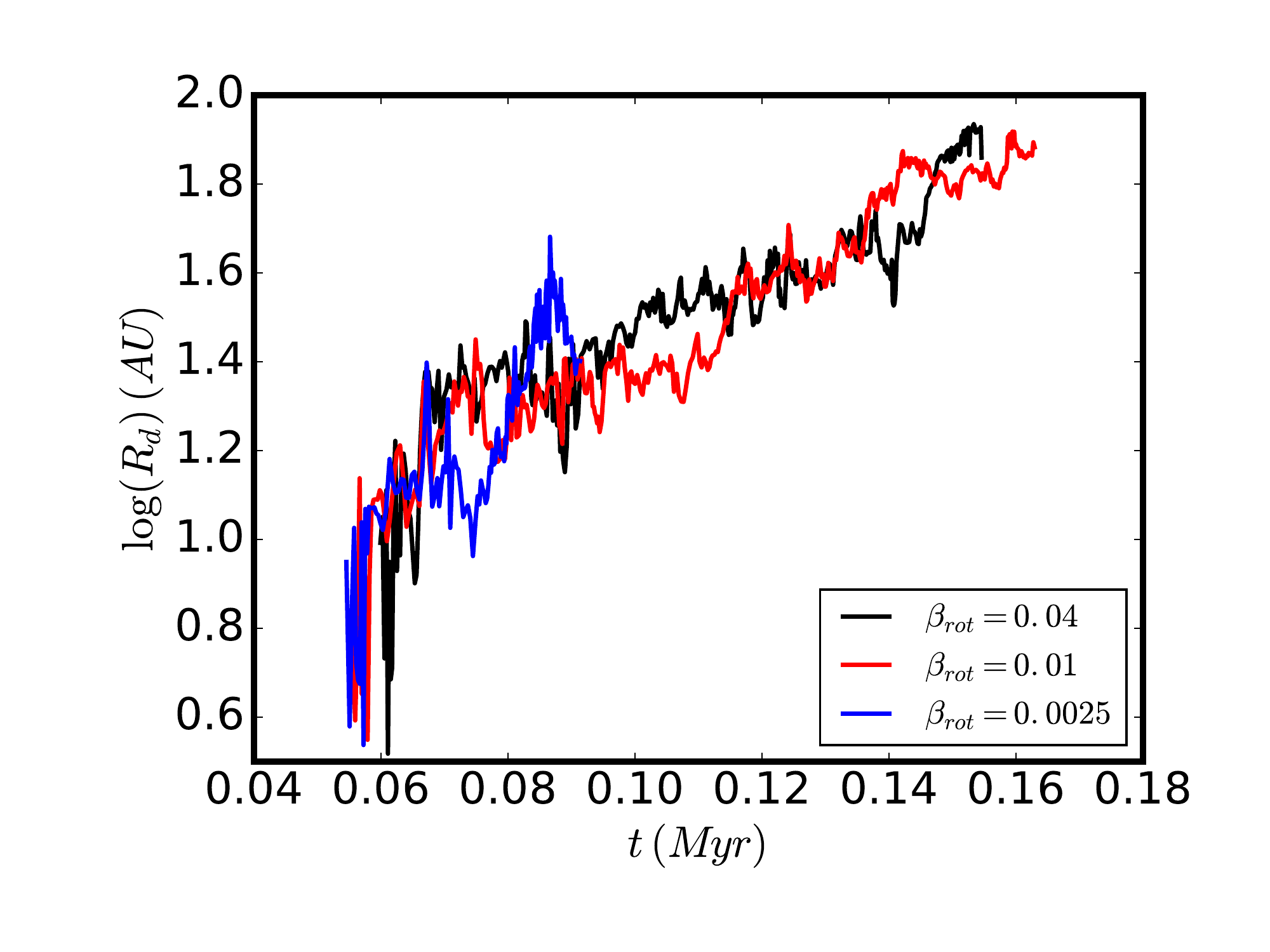}}  
\end{picture}
\caption{Comparison between disc mass and radius for runs $R1$, $R2$ $R10$, i.e. having 3 levels of 
rotation corresponding to $\beta_{\rm rot}=0.01$, 0.04, and 0.0025 respectively.}
\label{time_evol_rot}
\end{figure}

Figure~\ref{time_evol_rot} displays the disc mass and the disc radius as a function of time for 
runs $R1$, $R2$ and $R10$ (i.e. $\beta_{\rm rot}=0.01$, 0.04, and 0.0025 respectively). During the first 50 kyr, there is a clear 
trend for the disc of run $R1$ to be about 2 times less massive on average that the disc of run $R2$
 but at time $\simeq$0.11 Myr, the 
two runs present discs of comparable masses. The radius of  the two discs are  close but there 
is a clear trend for the disc of run $R1$ to be slightly smaller by a few tens of percents. 
With run $R10$ the situation is a bit more confusing. There is a short phase, between 0.07 and 0.08 Myr
during which the disc of run $R10$ is about 50$\%$ smaller and less massive than the disc of run $R1$. However, 
passed this phase, the disc of run $R10$ becomes even slightly bigger and more massive than the disc of run $R2$. 
A more detailed analysis reveals that the disc in run $R10$ becomes very elliptical.
Altogether rotation appears to have only a modest influence on the disc characteristics, though obviously very 
small rotation would certainly lead to major differences. 

At first sight this is surprising since the disc radius is proportional to $j^2$ where $j$
is the angular momentum. Therefore increasing $\beta_{\rm rot}$ by a factor of 4 may lead
to a disc 4 times bigger. This however is not the case because magnetic braking largely 
controls the formation of disc and is responsible for redistributing the angular momentum 
through the core. Indeed the  analytical expression of the disc radius (Eq.~(13) of \citet{hetal2016})
does not entail any dependence on $\beta_{\rm rot}$ (contrarily to Eq.~(14)  of \citet{hetal2016}). 
This is a consequence of magnetic braking and magnetic field generation. 
A larger rotation leads to a stronger $B_\phi$, which 
in turn leads to more efficient magnetic braking.

\subsection{The impact of magnetic intensity}

\setlength{\unitlength}{1cm}
\begin{figure*}%[h!]
%\centering
\begin{picture} (0,15)
\put(0,10){\includegraphics[width=7cm]{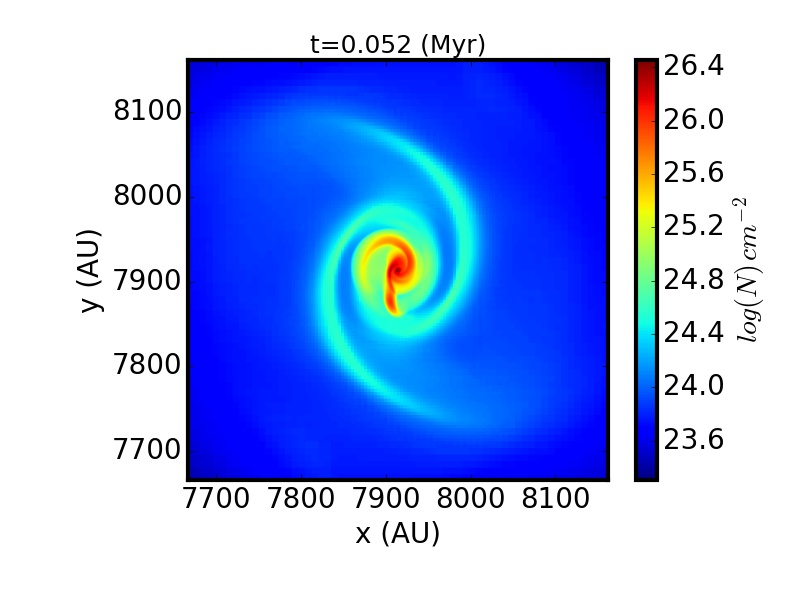}}  
\put(0,5){\includegraphics[width=7cm]{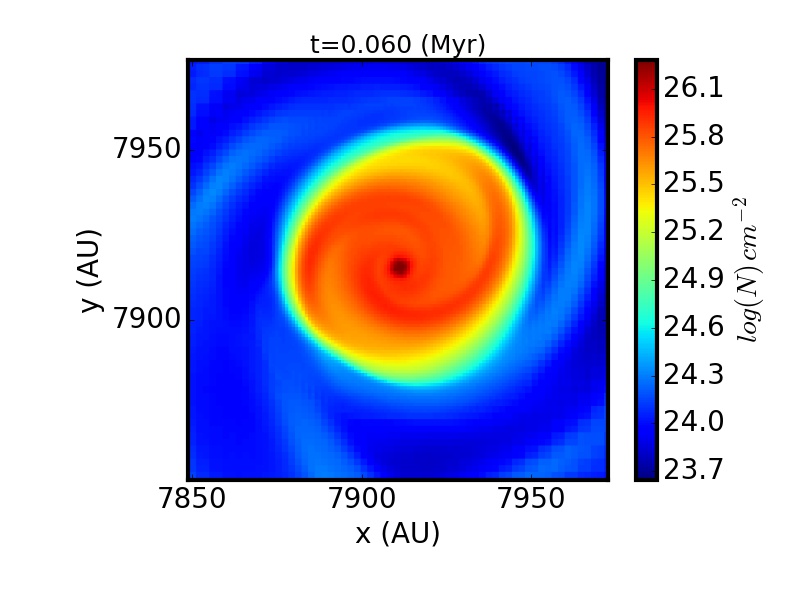}}  
\put(0,0){\includegraphics[width=7cm]{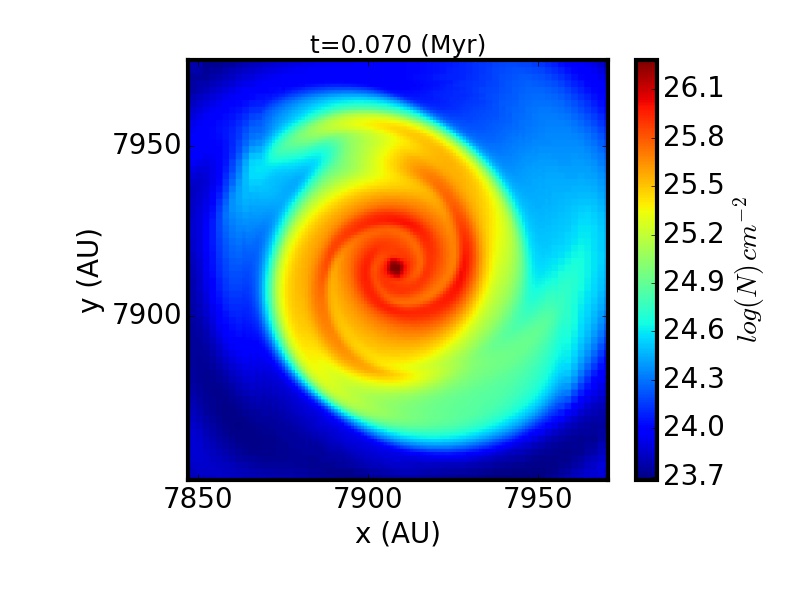}}  
\put(7,10){\includegraphics[width=7cm]{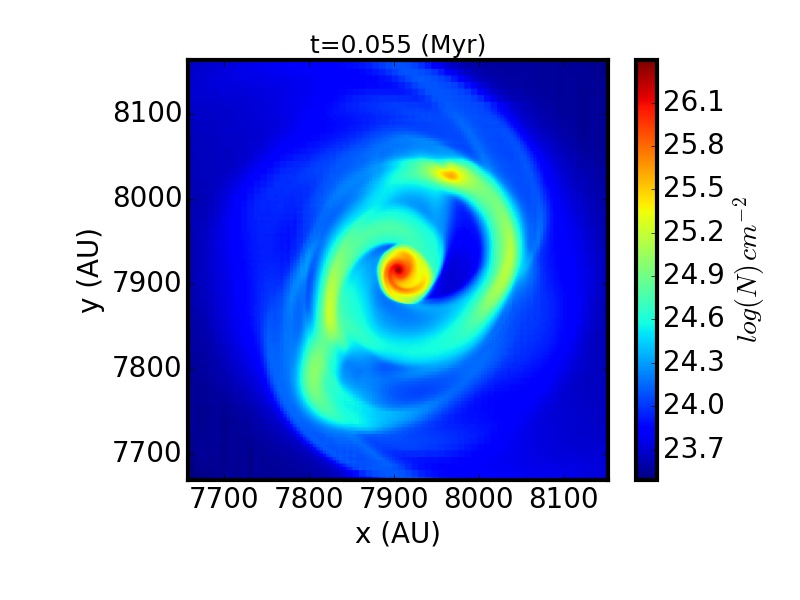}} 
\put(7,5){\includegraphics[width=7cm]{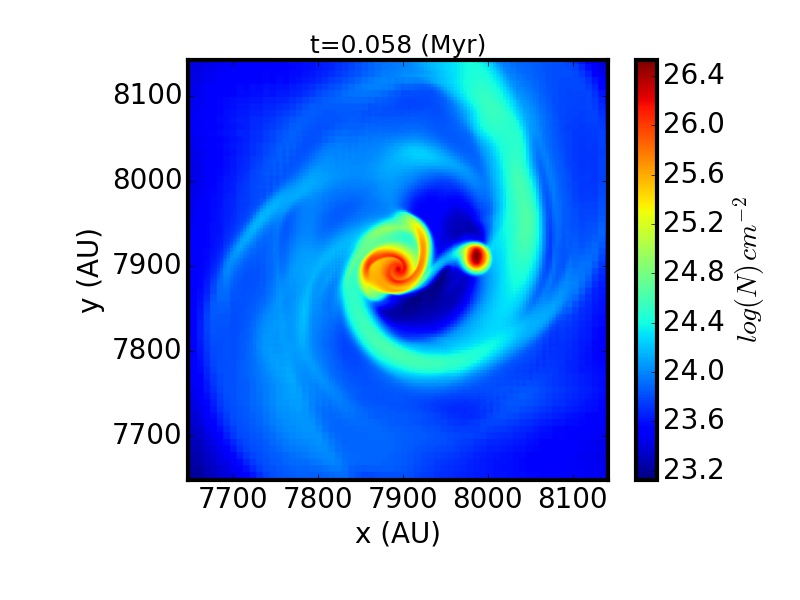}}   
\put(7,0){\includegraphics[width=7cm]{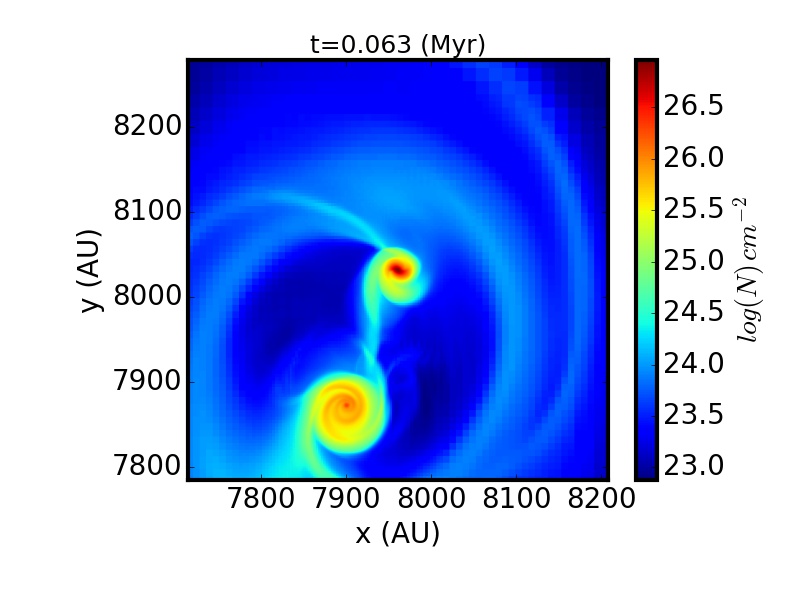}}  
\put(1.,14.8){$R3$}
\put(8.,14.8){$R5$}
\end{picture}
\caption{Column density of runs $R3$ (left) and $R5$ (right), i.e. weakly magnetized runs, for two  snapshots.
For run $R3$ while the magnetic intensity is 3 times weaker than for run $R2$, the disc is only marginally 
larger. Two noticeable differences however are that the disc presents strong and persistent spiral arms and 
that  large scale spiral arms outside the disc also remain.  As can be seen, run $R5$ (which has a magnetic intensity about 4 times lower than run $R2$) fragments in 2 objects.}
\label{coldens_bigdisc}
\end{figure*}

\setlength{\unitlength}{1cm}
\begin{figure}%[h!]
%\centering
\begin{picture} (0,11)
\put(0,5.5){\includegraphics[width=8cm]{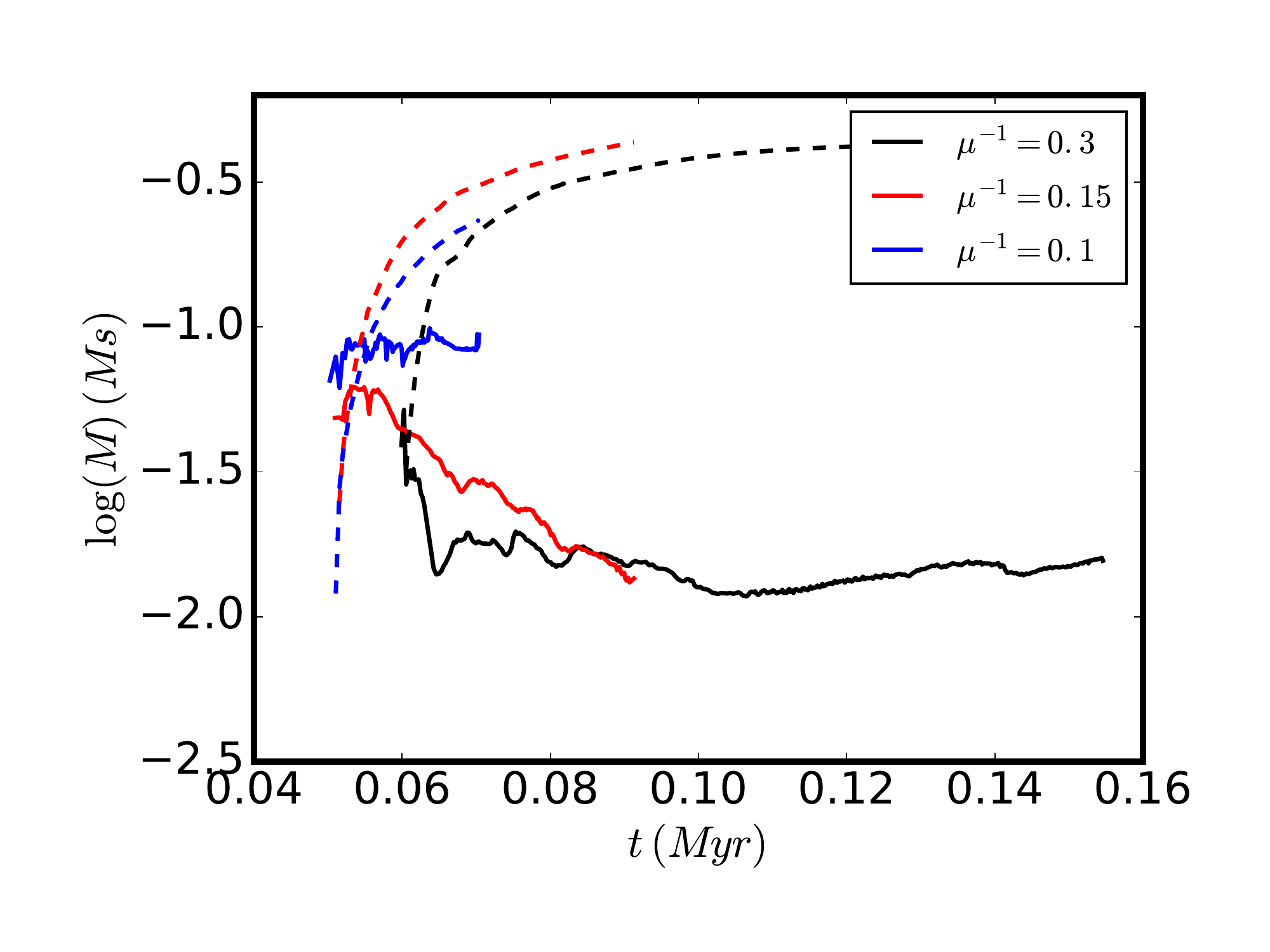}}  
\put(0,0){\includegraphics[width=8cm]{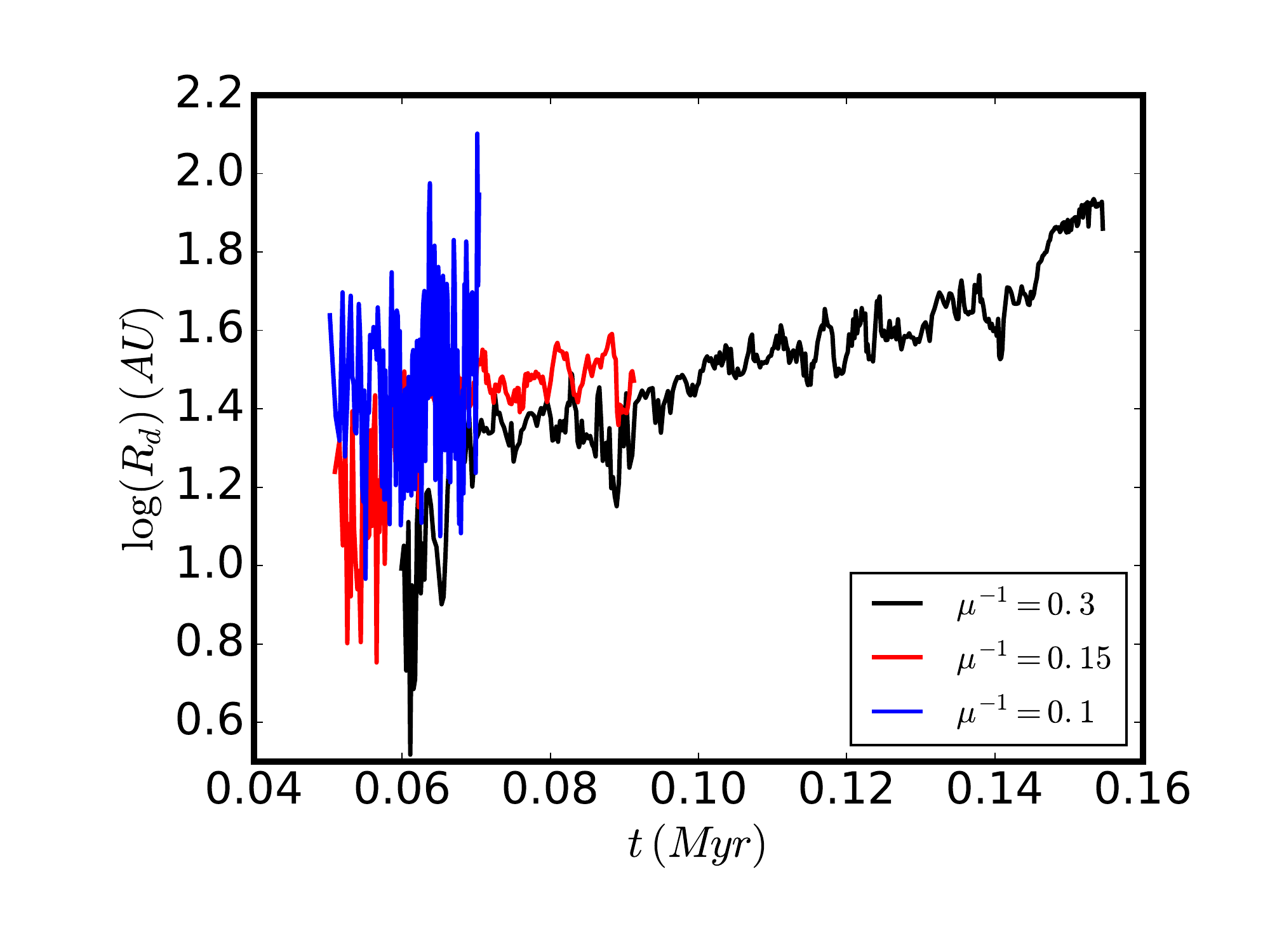}}  
\end{picture}
\caption{Comparison between disc mass and radius for runs $R2$, $R3$ and $R4$, i.e. having 3 levels of 
magnetization corresponding to $\mu^{-1}=0.3$, 0.1 and 0.15 respectively.}
\label{time_evol_mag}
\end{figure}

The magnetic intensity is expected to be a fundamental parameter regarding disc formation and here 
we investigate four values of $\mu$, going from significantly magnetized (i.e. $\mu^{-1} = 0.3$) to
weakly magnetized ($\mu^{-1} = 0.07$). 
Figure~\ref{coldens_bigdisc} shows the disc at 3 snapshots for runs $R3$ (left, $\mu^{-1} =0.1$) and $R5$ (right, $\mu^{-1} \simeq 0.07$). 
The disc in run $R3$ is a bit larger than the disc of run $R2$ and is surrounded by spiral arms whose size is about 200-400 AU, 
that persist and even grow with time (not shown here for conciseness). There are also prominent spiral arms within the 
disc itself as shown by the middle and bottom left panel. This is very similar to the disc shown by \citet{tomida2017}.

The disc formed early in run $R5$ (top right panel) is visibly both massive and large with prominent spiral arms.
Further fragments develop through gravitational instability within the arms and a new object forms as seen in 
bottom-right panel. Two smaller discs form around each of the two stars with radii on the order of 20-40 AU. Around the 
binary a larger scale rotating structure is seen, which may give raise to a circumbinary disc.

Figure~\ref{time_evol_mag} portrays the disc mass and disc radius as a function of time for 
runs $R2$, $R3$ and $R4$ ($\mu^{-1}$ = 0.3, 0.1, and 0.15 respectively, but not run $R5$ since it fragments). First let us note that 
the sinks form about 10 kyr later in run $R2$ than in runs $R3$ and $R4$ because of the magnetic support 
which delays the collapse by about 25\% in time. The disc in run $R3$ is about 5-6 times more massive than the disc 
of run $R2$ and it is about 2 times larger (the large fluctuations are due to the spiral arms which continuously 
evolve). The disc of run $R4$ is initially a little more massive than the one in run $R2$ but 
over time $t=0.08$ Myr, the mass of the discs of runs $R2$ and $R4$ are almost identical. In terms of radius, the 
difference is on the order of $\simeq 50 \%$ until $t=0.09$ Myr where the two radius become very close. 

Altogether these results show that for magnetic fields such that $\mu^{-1} > 0.15$, the magnetic intensity does not 
influence very strongly the disc properties. For weaker field, typically corresponding to $\mu^{-1} = 0.07-0.1$, we 
see a stronger dependence leading to bigger discs that can possibly fragment. As explained in 
\citet{hetal2016} 
 this is largely a consequence of the ambipolar diffusion that tends to regulate the 
magnetic intensity at high density. Note that more variability is reported in \citet{masson2016}
than what is inferred here particularly regarding the cases $\mu=3$ and $\mu \simeq 6$. The difference 
comes from the sink particles, which are not used in  \citet{masson2016}.
As shown in Fig.~\ref{coldens_nosink},  when they are not used 
strong spiral patterns develop because the gas that has piles up in the centre can expand back forming 
more massive disc than it should.

\subsection{The impact of misalignment}

\setlength{\unitlength}{1cm}
\begin{figure*}%[h!]
%\centering
\begin{picture} (0,10)
\put(0,5){\includegraphics[width=7cm]{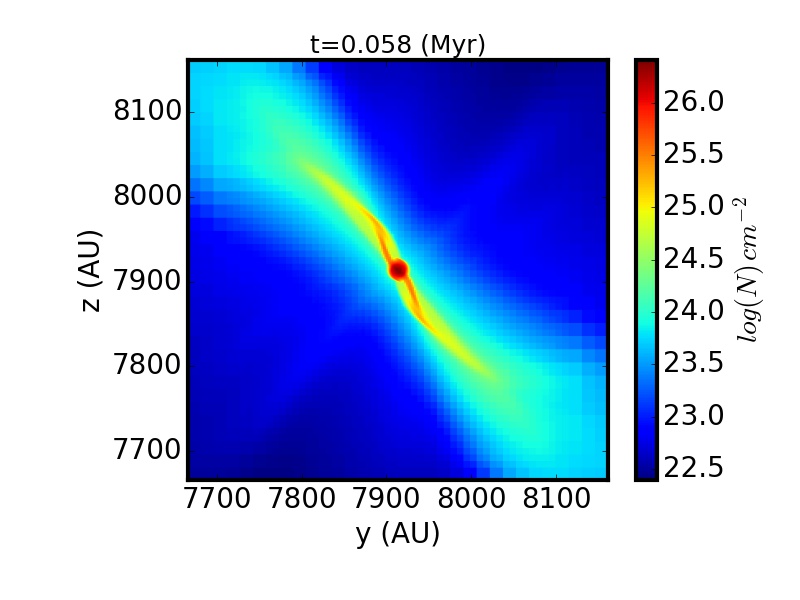}}  
\put(0,0){\includegraphics[width=7cm]{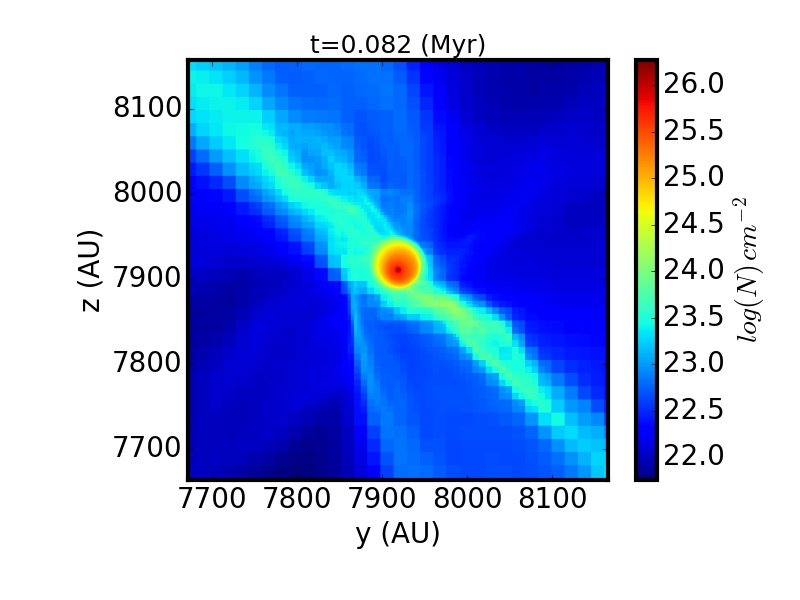}}  
\put(7,5){\includegraphics[width=7cm]{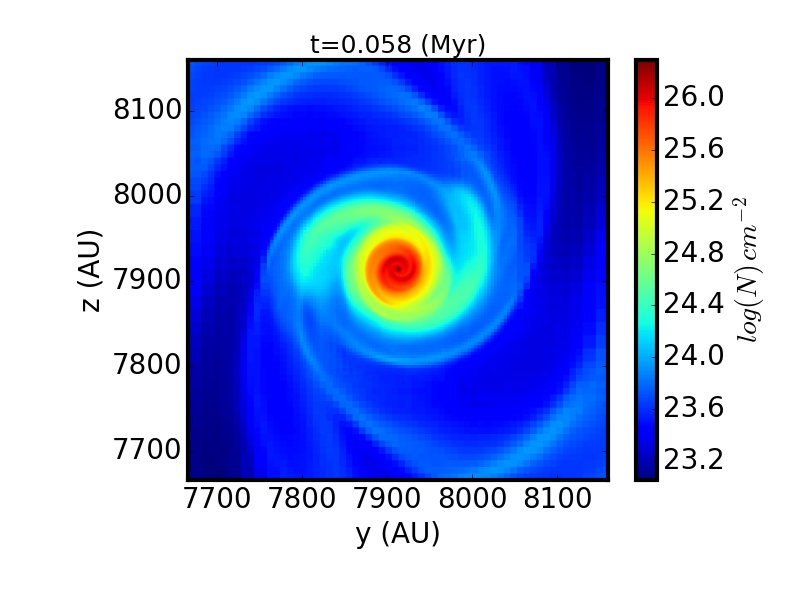}}  
\put(7,0){\includegraphics[width=7cm]{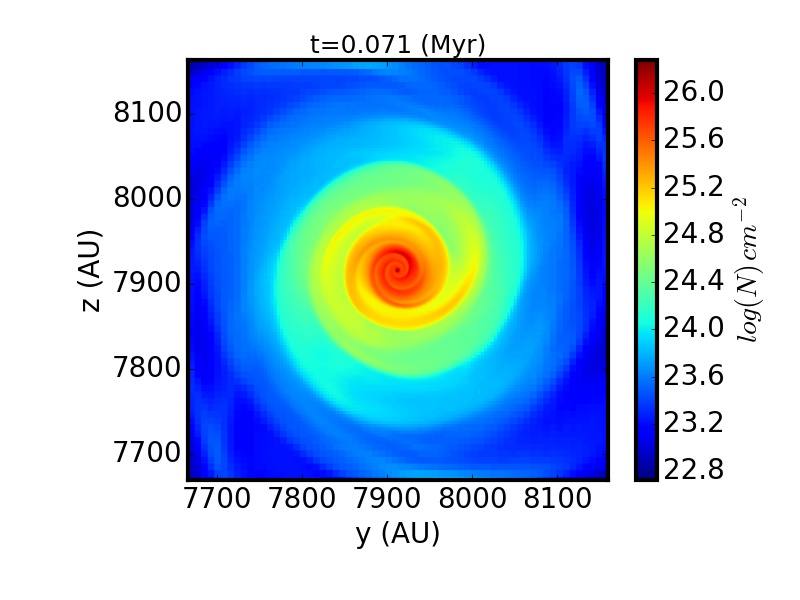}}  
\put(1.,9.8){$R7$}
\put(8.,9.8){$R9$}
\end{picture}
\caption{Column density of runs $R7$ (left) and $R9$ (right) for two snapshots. Both runs present initially a misalignment 
between the magnetic and rotation axis of 90$^\circ$. Run $R7$ which is has the same magnetic intensity than 
run $R2$, presents a disc
that is rather similar to it. Run $R9$ which is 3 times less magnetized than run $R2$ (and have the same magnetization than run $R3$), presents 
a massive and large disc. }
\label{coldens_angle}
\end{figure*}

\setlength{\unitlength}{1cm}
\begin{figure}%[h!]
%\centering
\begin{picture} (0,11)
\put(0,5.5){\includegraphics[width=8cm]{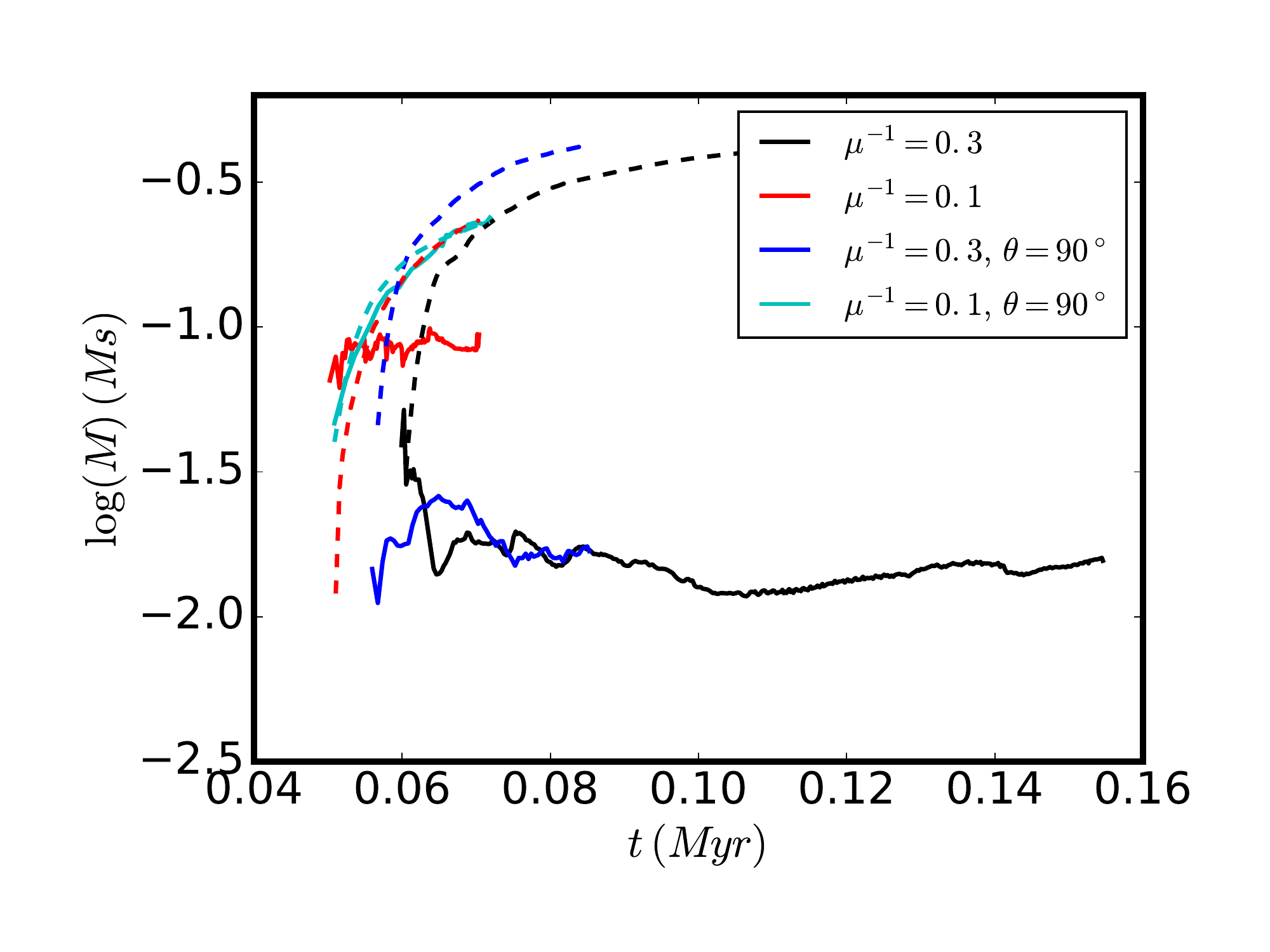}}  
\put(0,0){\includegraphics[width=8cm]{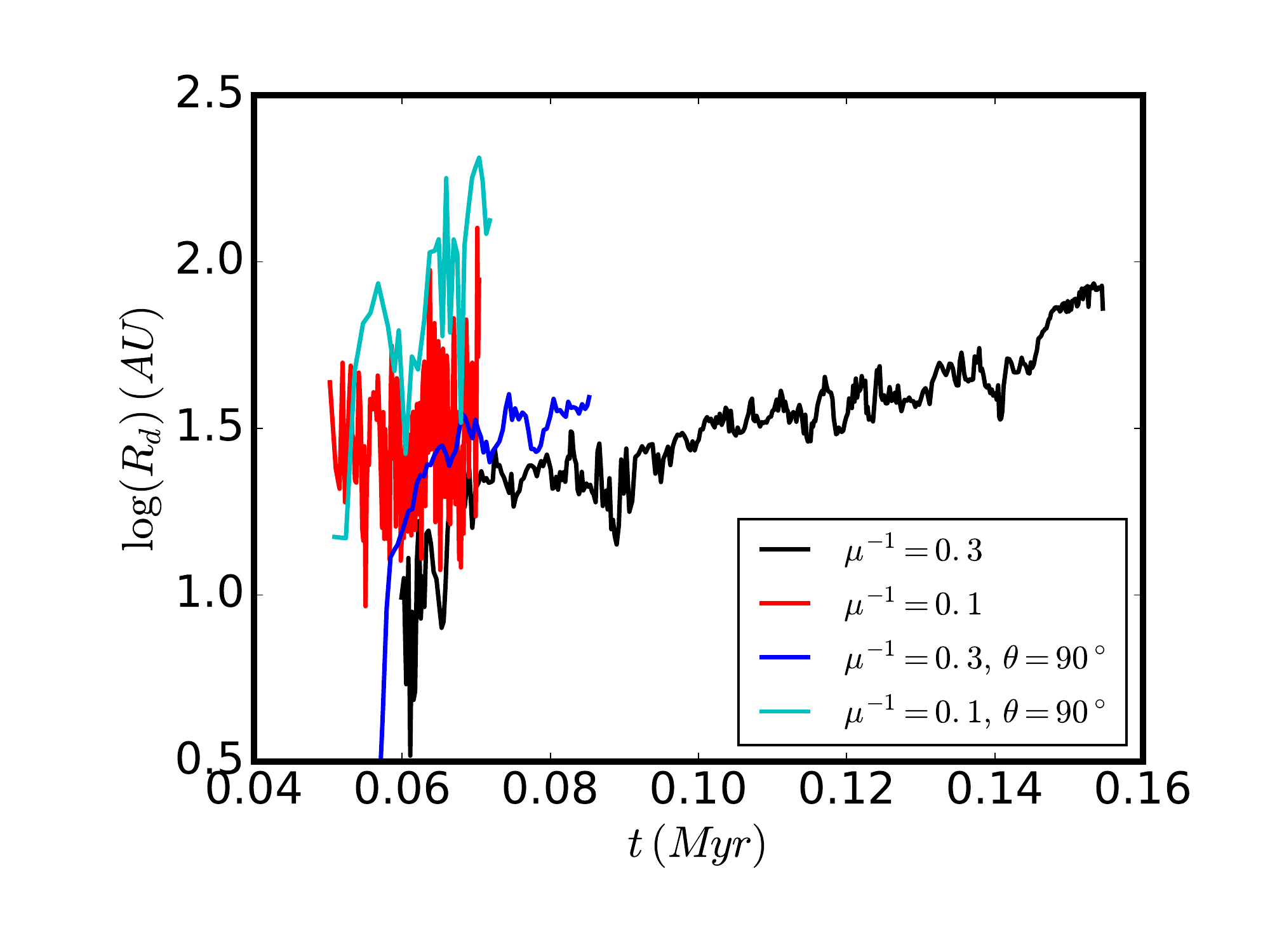}}  
\end{picture}
\caption{Comparison between disc radius and mass for runs $R2$, $R3$,  $R7$ and $R9$, i.e. having 2 levels of 
magnetization corresponding to $\mu^{-1}=0.3$ and 0.1 as well as two angles between the rotation and magnetic 
field axis initially, namely $30^\circ$ and $90^\circ$.}
\label{time_evol_angle}
\end{figure}

Misalignment has been identified to be playing a significant role in the process of disc formation 
particularly for values of $\mu > 5$ \citep{hciardi2009,joos2012,li2013,gray2018,hirano2019}. Typically 
it has been found that large discs form more easily in the perpendicular configuration 
 although 
\citet{tsukamoto2018} concluded based on an analysis of the amount of angular momentum that this 
may not be the case when non-ideal MHD is considered. This is however based on the amount of angular 
momentum available in the dense gas as they do not report about the discs themselves.

Figure~\ref{coldens_angle} shows the disc for runs $R7$ (left panels) and $R9$ (right panels) which both have an initial 
angle between magnetic and rotation axis equal to 90$^\circ$ and a magnetisation 
of $\mu^{-1}=0.3$ and $\mu^{-1}=0.1$, respectively. Both discs have their axis along 
the x-axis i.e. along the initial rotation axis. 

The disc of run $R7$ is small in size and comparable to the one of run $R2$. This is confirmed by 
Fig.~\ref{time_evol_angle}, which shows that both the mass and the radius are very similar 
to the one of run $R2$ with the radius of the latter being a few tens of percent smaller than the former.

The case of larger $\mu$, i.e. runs $R3$ and $R9$ is different. From Fig.~\ref{coldens_angle}, it is clear 
that the disc of run $R9$ is rather large and typically 3 times larger than the disc of run $R3$. 
This is also clearly shown by Fig.~\ref{time_evol_angle} where it is seen that at the end of the 
integration, both the mass and the radius of run $R9$  disc are almost 3 times the ones of run $R3$ disc. 
Note that while run $R9$ is both massive and large, it does not seem to show any sign of fragmentation 
so far. The reason is likely because it has a stronger field than run $R5$ for instance. 

The conclusion is therefore broadly similar to the one inferred from ideal MHD calculations 
\citep{joos2012}, that is to say the orientation does not have a drastic 
impact for moderate to strong field (i.e. $\mu < 3$), it can make a significant difference 
for weak magnetisation (in the present case $\mu =10$). 
 We note nevertheless that in ideal MHD calculations, the influence of the misalignment 
appears to be important for values  as low as $\mu=3$.

\subsection{The impact of turbulence}

\setlength{\unitlength}{1cm}
\begin{figure}%[h!]
%\centering
\begin{picture} (0,11)
\put(0,5.5){\includegraphics[width=8cm]{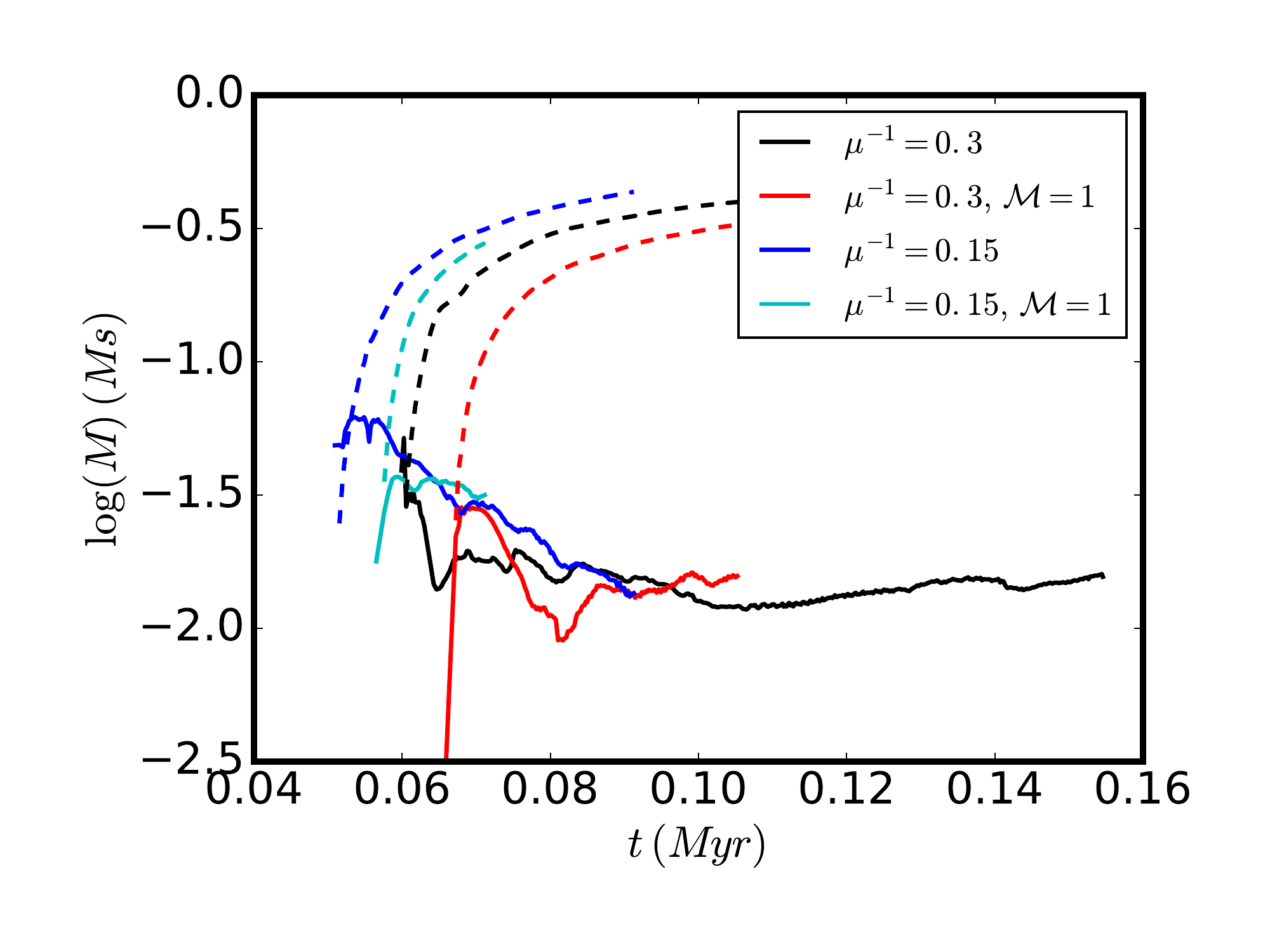}}  
\put(0,0){\includegraphics[width=8cm]{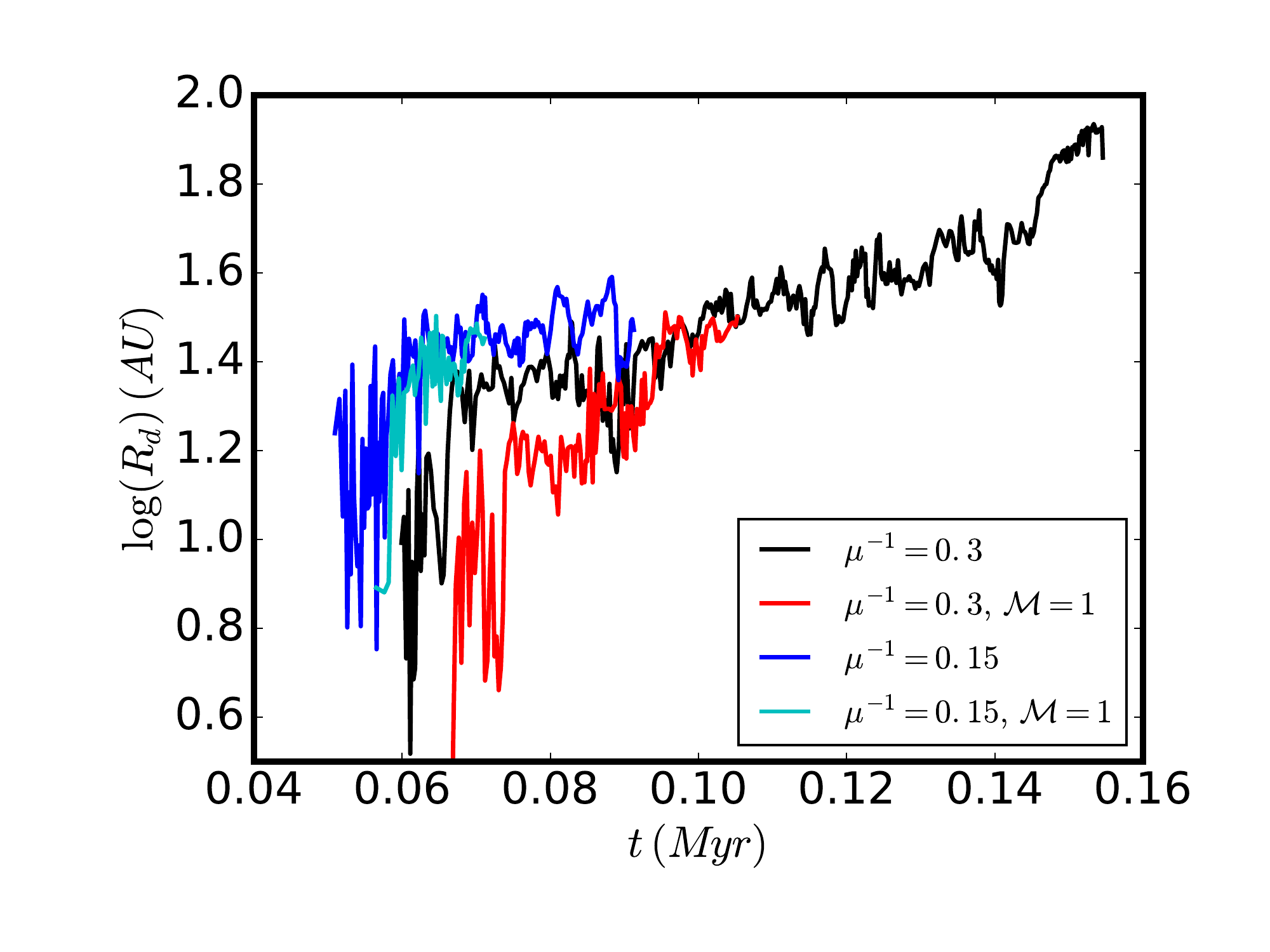}}  
%\put(0,5.5){\includegraphics{FIG_PAPIER_DISC_PHYS/DC_8/mass_disk_serie_t.pdf}}  
%\put(0,0){\includegraphics{FIG_PAPIER_DISC_PHYS/DC_8/rad_disk_serie_t.pdf}}  
\end{picture}
\caption{Comparison between disc radius and mass for runs $R2$, $R4$,  $R6$ and $R8$, i.e. having 2 levels of 
magnetization corresponding to $\mu^{-1}=0.3$ and 0.15 as well as two values of initial turbulent Mach number initially, 
namely 0 and 1.}
\label{time_evol_turb}
\end{figure}

The impact of turbulence on disc formation  has been studied in the context of ideal MHD and 
it has been found \citep{santos2012,joos2013} that it reduces the magnetic flux in the 
core inner part therefore favouring disc formation. It has also been proposed \citep{seifrid2013}  that 
 it is the disorganised structure of the field, induced by turbulence, which reduces 
the effect of magnetic braking while
\citet{gray2018} concluded that the most important effect of turbulence is the misalignment 
it induces \citep[see also][]{joos2013}. 

Cases that included both turbulence and ambipolar diffusion have been presented in \citet{hetal2016} and 
it has been found that the discs were not particularly different from the cases where 
no turbulence was initially present. Both runs $R6$ and $R8$ (which are identical apart for their magnetisation) 
confirm this conclusion as revealed by Fig.~\ref{time_evol_turb} where the disc and radius 
mass are portrayed for runs $R2$ ($\mu^{-1}=0.3$ and no turbulence) and $R6$ ($\mu^{-1}=0.3$ 
with an rms Mach number of 
1 initially) as well as runs $R4$ ($\mu^{-1}=0.15$ and no turbulence) and $R8$ ($\mu^{-1}=0.15$ 
with an rms Mach number of 
1 initially). At early time,   disc of run $R6$ is 50$\%$ smaller than for $R2$  but this may be 
in part due to the slightly delayed collapse of run $R6$ with respect to run $R2$ (a consequence of turbulence) and also 
to the lower rotation of run $R6$.  After 20 kyr, the discs of runs $R2$ and $R6$ have indistinguishable sizes and masses.

We conclude that in the presence of ambipolar diffusion, turbulence does not have a significant impact on the disc. 
This is likely because ambipolar diffusion has an effect on magnetic field that dominates over the diffusion 
induced by turbulence. Another complementary effect is the dissipative nature of ambipolar diffusion, which 
likely tends to reduce turbulence by enhancing its dissipation.

\subsection{Magnetic intensity vs density}

\setlength{\unitlength}{1cm}
\begin{figure}%[h!]
%\centering
\begin{picture} (0,5.5)
\put(0,0){\includegraphics[width=8cm]{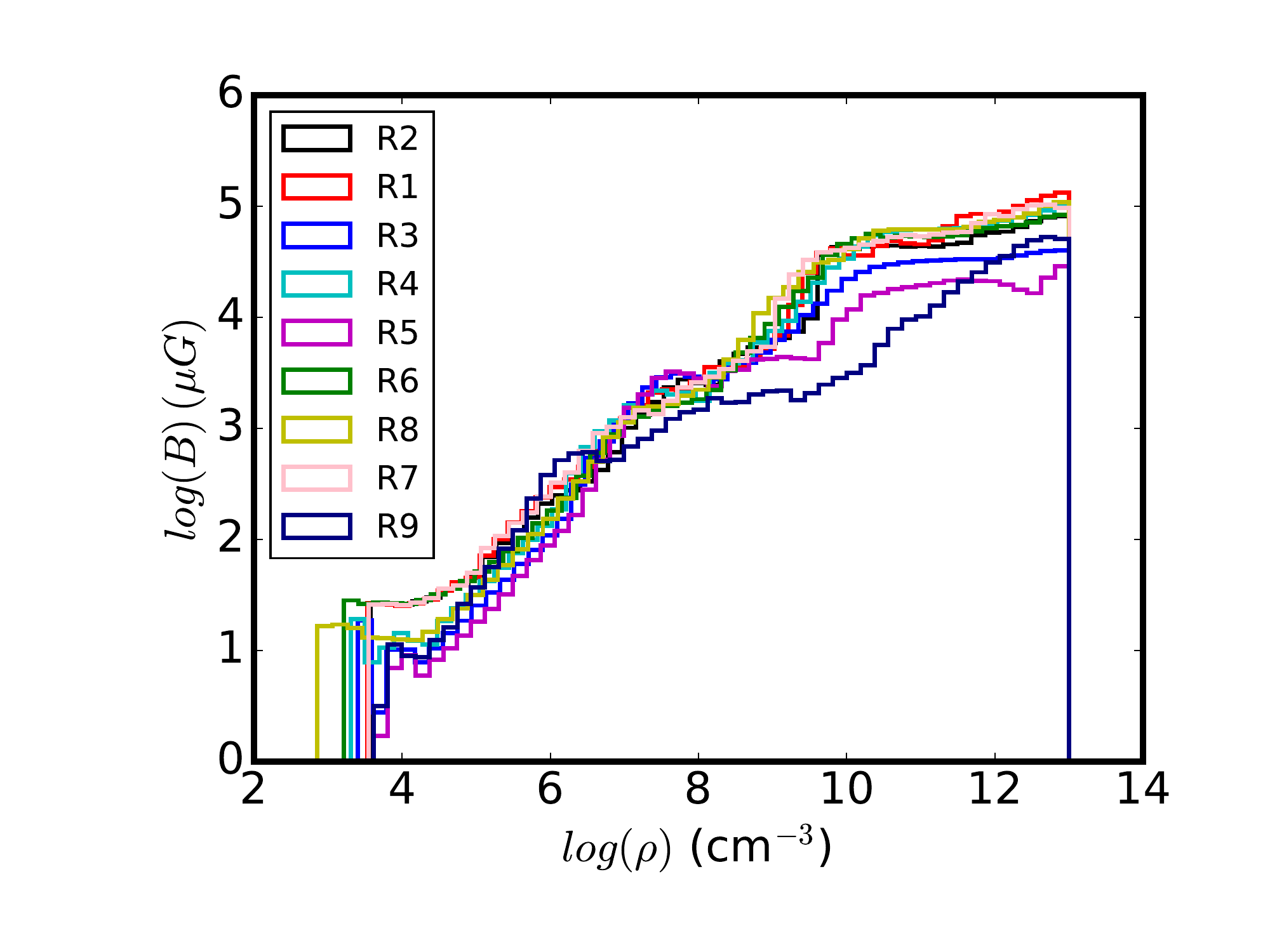}}  
\end{picture}
\caption{Mean magnetic intensity as a function of density within the 9 runs. Apart for runs $R3$, $R5$ and $R9$
which all have $\mu^{-1} < 0.1$, i.e. a very low field initially, the distribution of 
the magnetic intensity is remarkably similar for all runs. }
\label{Bvsrho}
\end{figure}

To interpret the results inferred in this section, the magnetic
intensity distribution is of fundamental importance since magnetic 
braking is largely controlling disc formation. 
Figure~\ref{Bvsrho} portrays the mean magnetic intensity, $B$, per logarithmic 
bin of densities for all runs  at time 0.06 Myr. We note that the magnetic field intensity
as a function of density does not evolve very significantly with time so the particular choice 
made here is not consequential.

In the envelope for densities between 10$^4$ and 10$^8$ cm$^{-3}$, 
$B$ increases with the density. Some discrepancies 
are seen between the runs. The most magnetized ones 
present a more shallow profiles (typically with  powerlaw exponents of about 1/2) while 
for the less magnetized ones, the profiles are stiffer (and more compatible with 
 powerlaw exponents close to 2/3). 
At higher densities, i.e. 10$^8$-10$^{10}$ cm$^{-3}$ depending of the runs, 
a plateau of magnetic intensities  formed. Its value is nearly the same 
for all runs, except for runs $R3$, $R5$ and $R9$ which present lower intensities. 
The reason of this behaviour \citep[see][]{hetal2016} is that the
ideal MHD holds in the low density envelope while at high density, ambipolar diffusion is dominant and 
an equilibrium settles between the inwards flux of magnetic field dragged by the 
collapsing motion and the ambipolar diffusion.

The distribution for run $R9$ is particularly intriguing. It has the same initial magnetisation as run $R3$,  
which is more magnetised than the run $R5$, but 
the magnetisation at high density in run $R9$ is the weakest. On the other hand, we see that it presents an excess compared
to the other runs of magnetic intensity around $n \simeq 10^6$ cm$^{-3}$. This strongly suggests that in 
run $R9$ efficient 
diffusion of the magnetic field occurred. This is probably a consequence of the perpendicular configuration as 
well as the weak intensity which lead to a strongly wrapped field.

Importantly, we find an obvious qualitative correlation between the size of the disc and the field intensity 
in the core inner part (i.e. $n > 10^8$ cm$^{-3}$). 

\subsection{Comparison between simulation and analytical theory of disc radii}
\label{compar_analyt}

\setlength{\unitlength}{1cm}
\begin{figure}%[h!]
%\centering
\begin{picture} (0,5.5)
\put(0,0){\includegraphics[width=8cm]{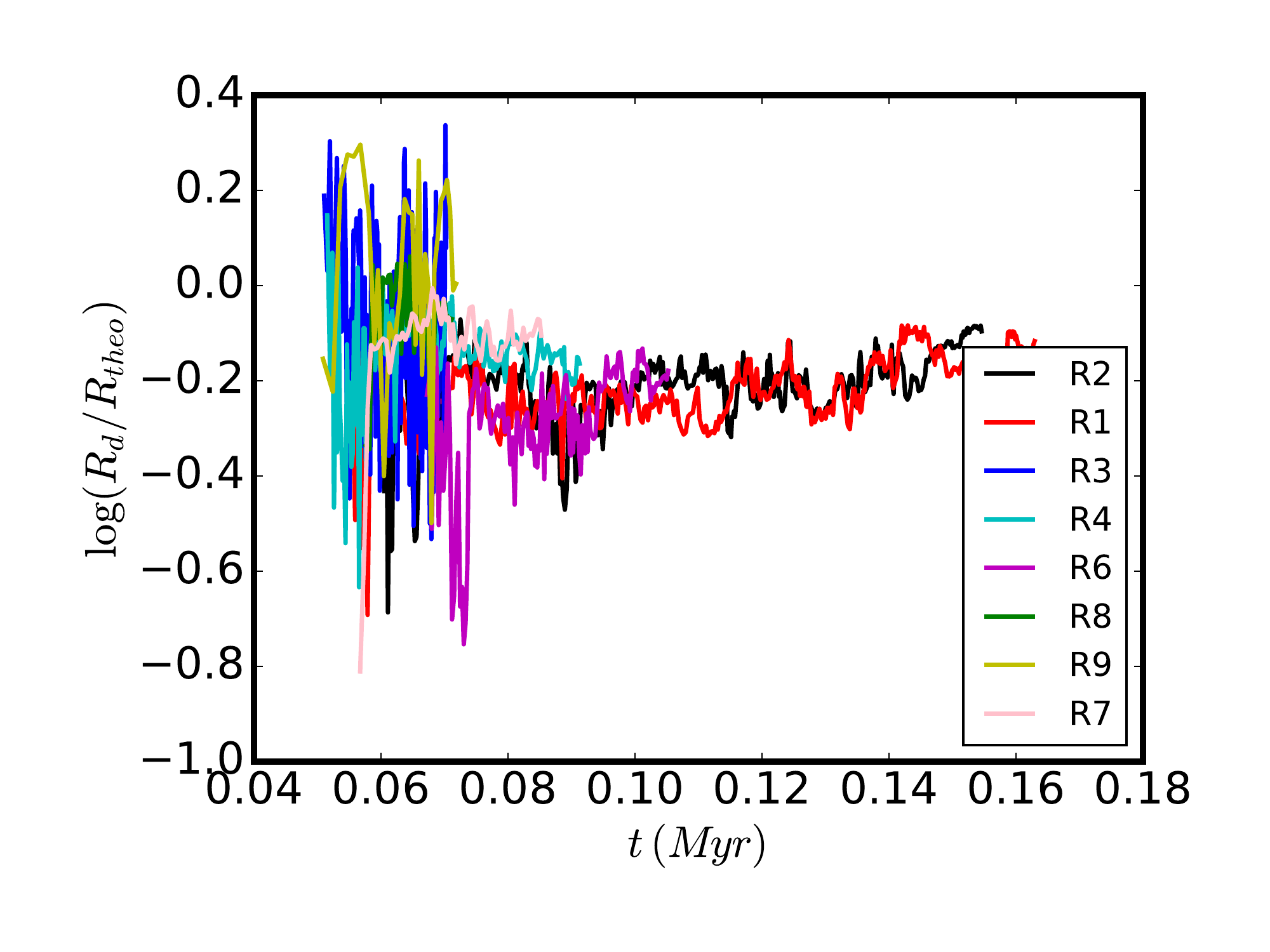}}  
\end{picture}
\caption{Ratio of  disc radius and theoretical prediction from \citet{hetal2016} for runs $R1$-$R9$. As can be 
seen the analytical model leads to reasonable agreement with the simulations within a factor of about 2.}
\label{comp_theo}
\end{figure}

To better understand the physics that governs the disc evolution, we now compare the 
numerical results presented above with the analytical modelling presented in 
\citet{hetal2016}. Let us remind that this expression is obtained by simply 
requiring the equality between on one hand the magnetic braking and rotation time and 
on the other hand, the magnetic diffusion and generation  of the toroidal 
magnetic field at the disc edge. The radius expression is

\begin{eqnarray}
%\nonumber
% R_{\rm ad}  \simeq  18 \, {\rm AU}  \;   & \times & \\ 
%\delta ^{2/9 }\left( {   \eta_\mathrm{AD} \over 0.1 \, {\rm s}   } \right)^{2/9} &&
%\left( {B_z \over 0.1\,{\rm G} } \right) ^{-4/9}   \left( {  M_* \over 0.1 M_\odot} \right) ^{1/3}. \\
 R_{\rm theo}  \simeq  18 \, {\rm AU}  \;    \times \;
\delta ^{2/9 }\left( {   \eta_\mathrm{AD} \over 0.1 \, {\rm s}   } \right)^{2/9}
\; \left( {B_z \over 0.1\,{\rm G} } \right) ^{-4/9}   \left( {  M_* \over 0.1 M_\odot} \right) ^{1/3}.
\label{radius_ad}
\end{eqnarray}

In this expression, 
$\eta_{AD}$ is the magnetic resistivity, 
$B_z$ is the vertical component of the magnetic field at the edge of the disc and $M_*$ the mass of the star.
$\delta$ is a coefficient such that $\rho = \delta \rho_{SIS} (1 + v_\phi^2 / C_s^2 / 2)$, 
where $\rho_{SIS}$
is the density of the singular isothermal sphere. 
Note that while this expression is reasonable to describe the equatorial density of 
a rotating hydrodynamical collapsing core, it is not obviously the case in a misaligned 
magnetized dense core as the densest part of the collapsing envelope, the pseudo-disc, is generally
not lying in the  plane of the disc (as is particularly obvious from Figs.~\ref{rho_z_Run2} and~\ref{rho_z_Run3}).
Moreover as it is clear from Fig.~\ref{rhov_r}, the envelope and the disc are connected through a shock.
This suggests that using for the density at the disc edge the expression
$\rho = \delta \rho_{SIS} {\mathcal M}^2$ would be more appropriate. However, since 
${\mathcal M} = v_r/C_s \simeq v_\phi / C_s$, this leads to exactly the same dependence and 
leaves Eq.~(\ref{radius_ad}) essentially unchanged. 

To test this formula, we have measured the mean magnetic intensity in the  disc as a function of 
time. It gives a reasonable proxy for the magnetic field at the edge. The values of 
$\eta_{AD}$ and $\delta$ are not easy to estimate because the density changes abruptly at the edge 
of the disc. Fortunately, $\eta_{AD}$ in this range of parameter does not change too steeply and 
both parameters come with an exponent $2/9$ in Eq.~(\ref{radius_ad}). 
For the sake of simplicity, we therefore keep them constant. 

Figure~\ref{comp_theo} shows the ratio of the measured disc radius to the theoretical expression for the 
8 runs (run $R5$ is not considered because it fragments) as a function of time.  
Altogether we see that within a factor of $\simeq 2$, we indeed have   $R_d \simeq  R_{\rm theo}$.
Apart 
for run $R9$ and maybe $R3$, there is a 
general trend for $R_{\rm theo}$ to be about 50$\%$ larger, which is probably a consequence of $\eta_{AD} \simeq 0.1$ 
to be  a little too high and more generally to the difficulty in inferring a single value of $B_z$ and of $\eta$.
For most runs, apart from the large fluctuations at the beginning where 
the disc is not well settled, $R/R_{\rm theo}$ varies within a factor of about 2. 
 This is similar to what has been inferred in 
\citet{hetal2016} although since sink particles are used here, the agreement is even better.
Compared with the disc radius as a function of time and for different runs (see for instance
Fig.~\ref{time_evol_angle}), it is clear that the radio $R_d / R_{\rm theo}$ presents much less variability 
which also illustrates the reasonable validity of  Eq.~(\ref{radius_ad}). In particular, we see that 
$R_d / R_{\rm theo}$ does not vary appreciably with time, which indicate that $R_d \propto M^{1/3}$ is a good 
approximation.

We concluded that altogether the theoretical expression provides the correct answer 
within a factor of $\simeq$2-3.

%__________________________________________________________________

\section{Conclusion}

In this paper we have presented two series  of  numerical simulations aiming 
at describing the collapse of a magnetized dense core in which 
non-ideal  MHD operates, namely the ambipolar diffusion. 
The first series of runs explore the influence of the numerical 
parameters such as the numerical resolution and the sink particle accretion scheme.
The  goal of the second series of runs is to understand the impact of
initial conditions on the disc properties.
 In particular, we vary the initial 
magnetic intensity, the rotation speed, the inclination of the 
field with respect to the rotation axis and the level of 
turbulence. 

We found that while the mass of the central object is insensitive to the 
numerical scheme, the disc mass on the contrary depends significantly on 
the accretion scheme onto the sink particle.  Essentially at a given spatial resolution, 
the disc density increases with $n_{\rm acc}$, the density above which the sink would accrete. 
 This in particular implies that the disc mass increases with $n_{\rm acc}$. 
For a given value of $n_{\rm acc}$, the disc mass decreases if the resolution increases. 
This is because the sink particles act as an inner boundary and what matters is the 
density imposed at this boundary.
We suggest that in reality, it is the small scale accretion, perhaps the 
star-disc interaction that determines, together with the disc transport properties, 
the disc density profile and therefore its mass distribution. 

Exploring a large sets of initial conditions, we found that the most critical parameter is the 
magnetic intensity particularly when it becomes small, i.e. 
for $\mu^{-1} < 0.1$, for which the disc radius increases with $\mu$. For $\mu^{-1} < 0.07$, we found 
that the disc fragments, leading to a binary. 
 For  magnetisation  larger than $\mu^{-1} > 0.15$ the disc properties 
do not change strongly with it. By changing the rotation speed
from a $\beta_{\rm rot}=0.01$ to $\beta_{\rm rot}=0.04$, we found that at least in this range of 
parameter, it does not modify significantly the size of the disc for instance. 
The reason of this somewhat counter-intuitive result is that magnetic 
braking really controls the process of disc formation and more rotation also leads 
to larger toroidal field and more braking.
The misalignment between rotation axis and magnetic field direction is found 
to play a minor role when the field is substantial ($\mu^{-1}=0.3$) but has a major 
impact when the field is weak ($\mu^{-1}=0.1$) in which case a  large disc of 
about 100 AU in radius forms.
Finally, the two runs that include turbulence present discs which are nearly identical 
to the discs formed in the absence of turbulence. 

We conclude that the disc physical characteristics,  and in particular its mass, are to a large extent 
determined by the local physical processes that is to say the collapse, 
the magnetic braking, the field diffusion, as well as an inner boundary condition
possibly determined by the star-disc interaction. 
Only when the magnetic field is 
relatively weak, i.e. $\mu^{-1} < 0.15$, we do find that the resulting 
discs present significantly different properties.

The standard disc that forms in our simulations has a mass of 
about $10^{-2}$ $M_\odot$ (which we stress depends on the accretion condition onto the central sink), 
a radius of 20-30 AU and is weakly magnetized 
in the inner part ($\beta_{\rm mag}$>100) and strongly magnetized in the outer 
part ($\beta_{\rm mag}$< 0.1). At later time, as the envelope 
disappears, the inner part of the disc remains barely changed by the external part as the disc 
develops. It is characterized by a weaker $\beta_{\rm mag}$ and a stiff density 
profile (we estimate it to be broadly$\propto r^{-4}$).

%__________________________________________________________________

\begin{acknowledgements}
We thank the anonymous referee for a detailed and helpful report which has improved the paper. 
This work was granted access to HPC
   resources of CINES and CCRT under the allocation x2014047023 made by GENCI (Grand
   Equipement National de Calcul Intensif). 
   This research has received funding from the European Research Council under
   the European Community's Seventh Framework Programme (FP7/2007-2013 Grant  Agreement no. 306483). Y.-N. Lee acknowledges the financial support of the UnivEarthS Labex program at Sorbonne Paris Cit\'e (ANR-10-LABX-0023 and ANR-11-IDEX-0005-02). We thank the programme national de physique stellaire, the programme 
national de plan\'etologie  and the 
programme national de physique et chimie du milieu interstellaire for their supports.
\end{acknowledgements}

%This research has received funding from the European Research Council under the European
% Community's Seventh Framework Programme (FP7/2007-2013 Grant Agreement no. 306483).
%\end{acknowledgements}

% for the bibliography
\bibliography{lars}{}
\bibliographystyle{aa} % style aa.bst

\appendix

\section{Disc profiles for runs $R2hsink$ and $R3$}

\setlength{\unitlength}{1cm}
\begin{figure}%[h!]
%\centering
\begin{picture} (0,22)
\put(0,16.5){\includegraphics[width=8cm]{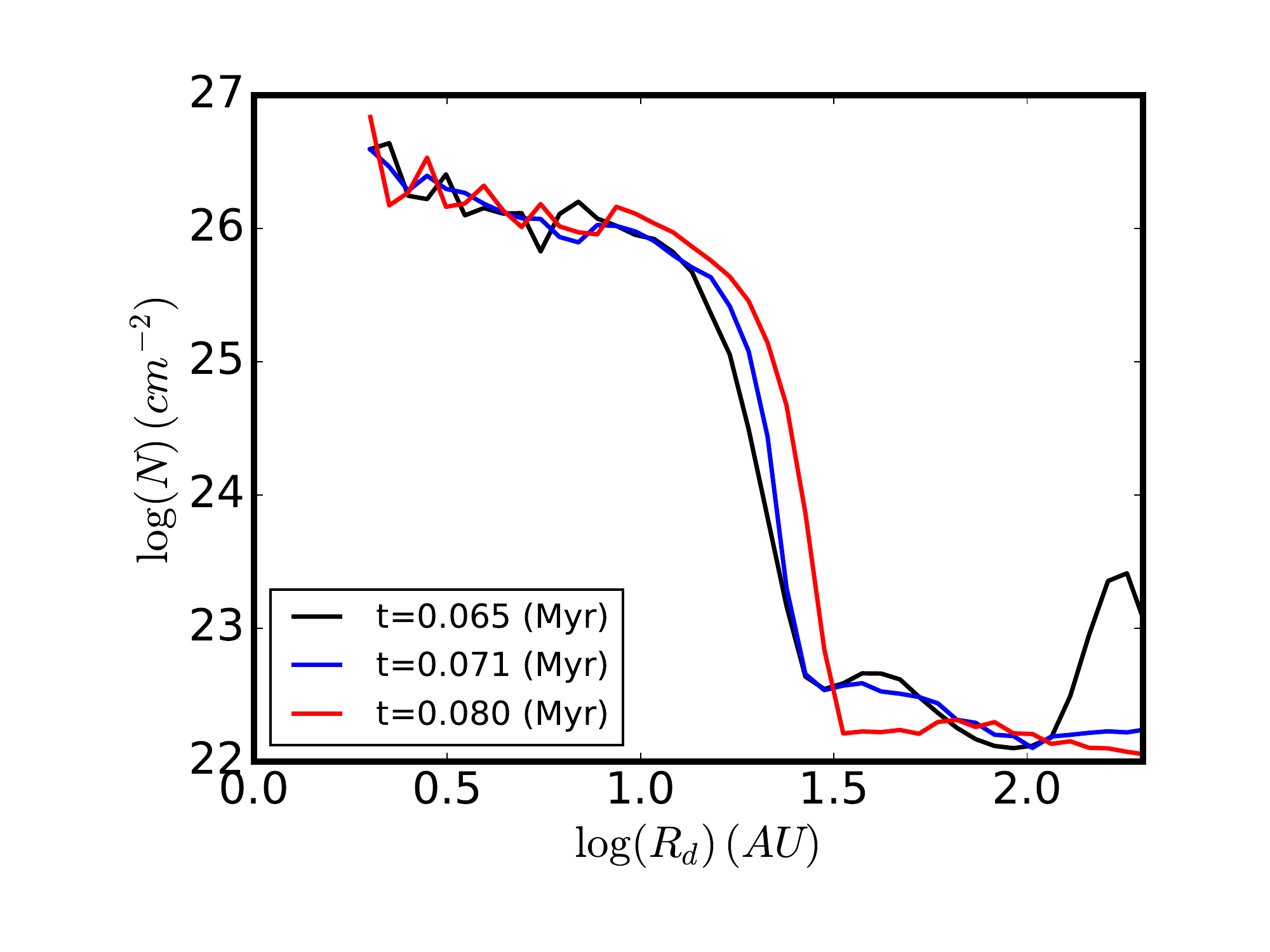}}  
\put(0,11){\includegraphics[width=8cm]{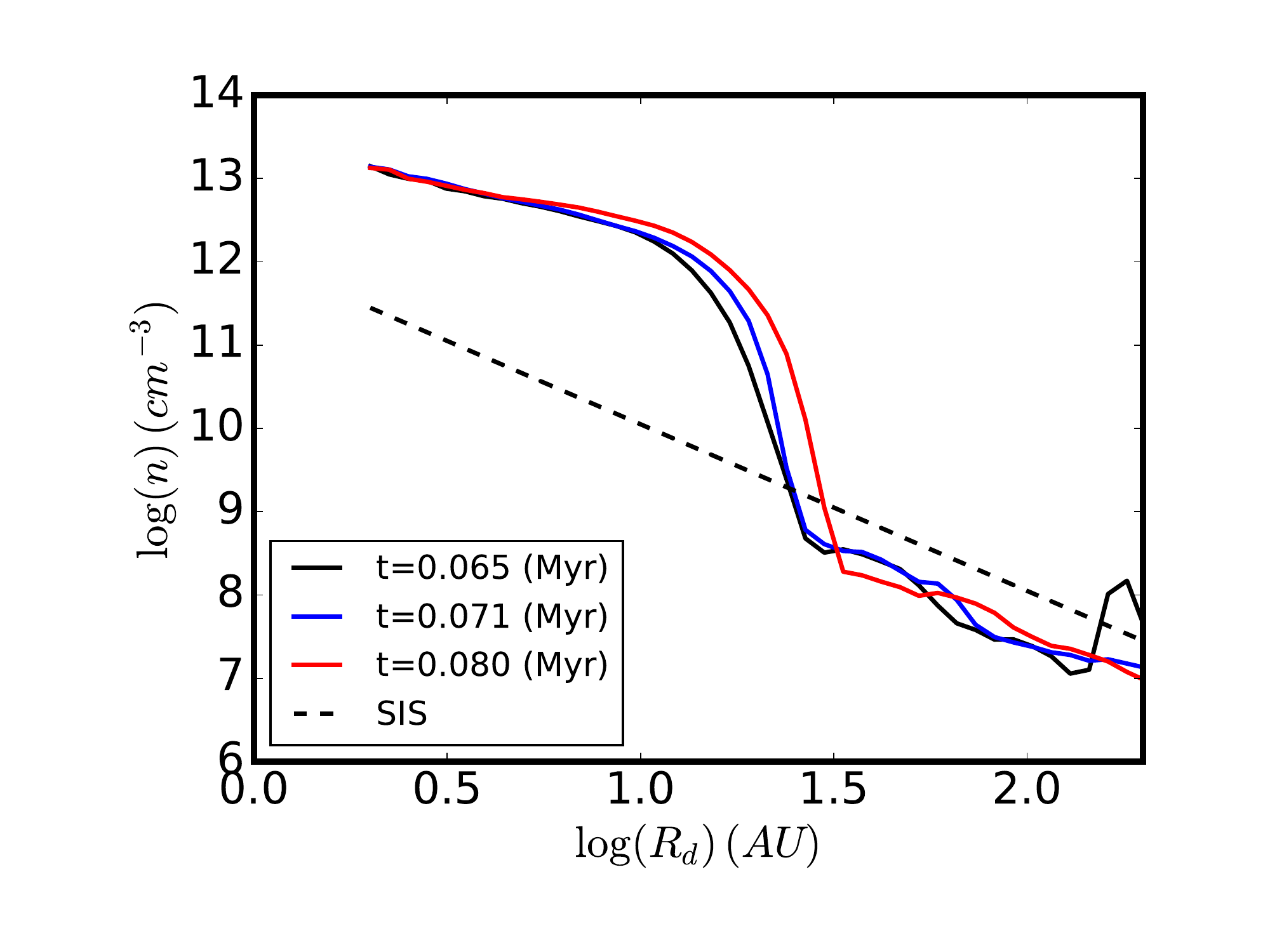}}  
\put(0,5.5){\includegraphics[width=8cm]{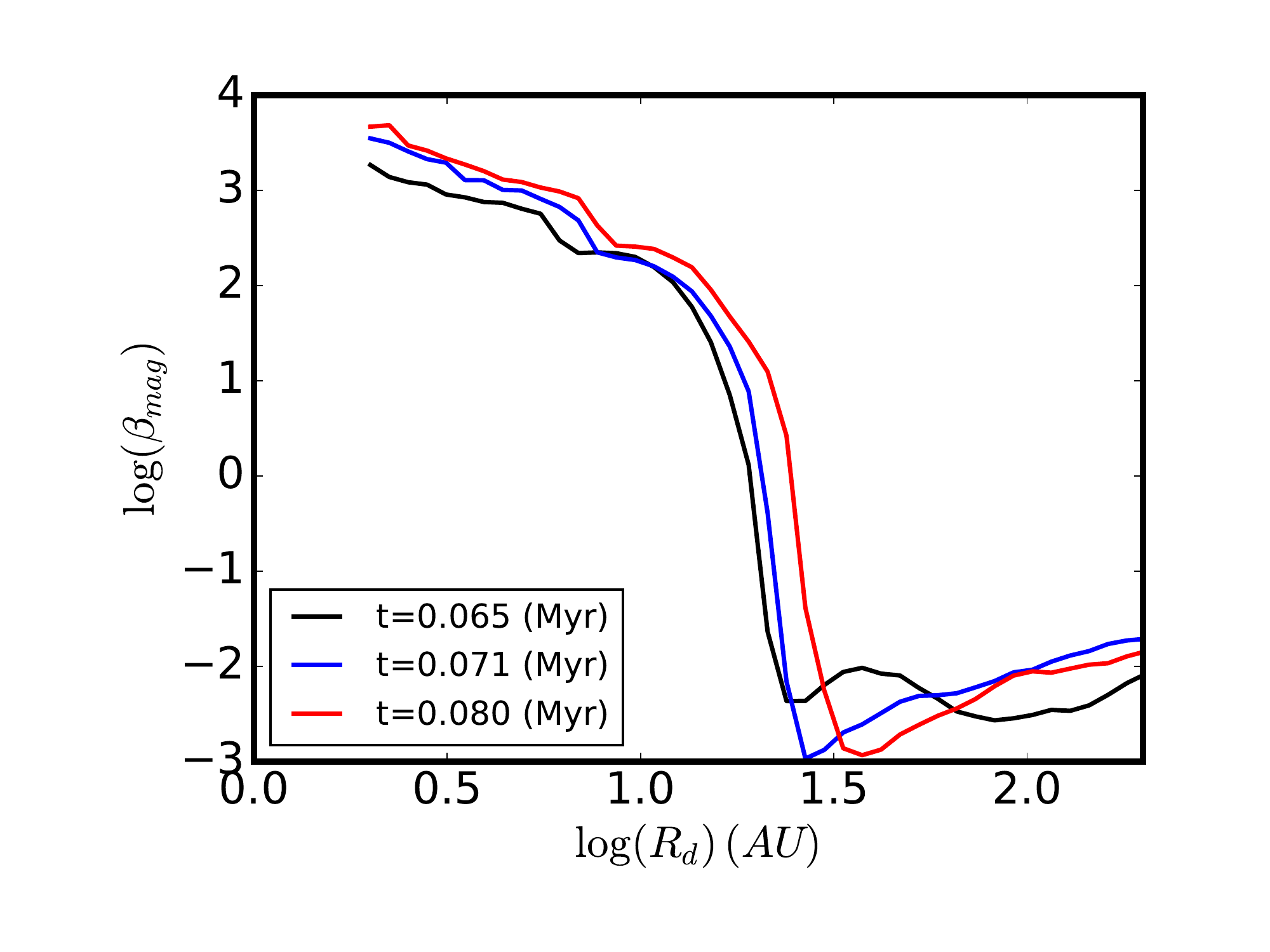}}  
\put(0,0){\includegraphics[width=8cm]{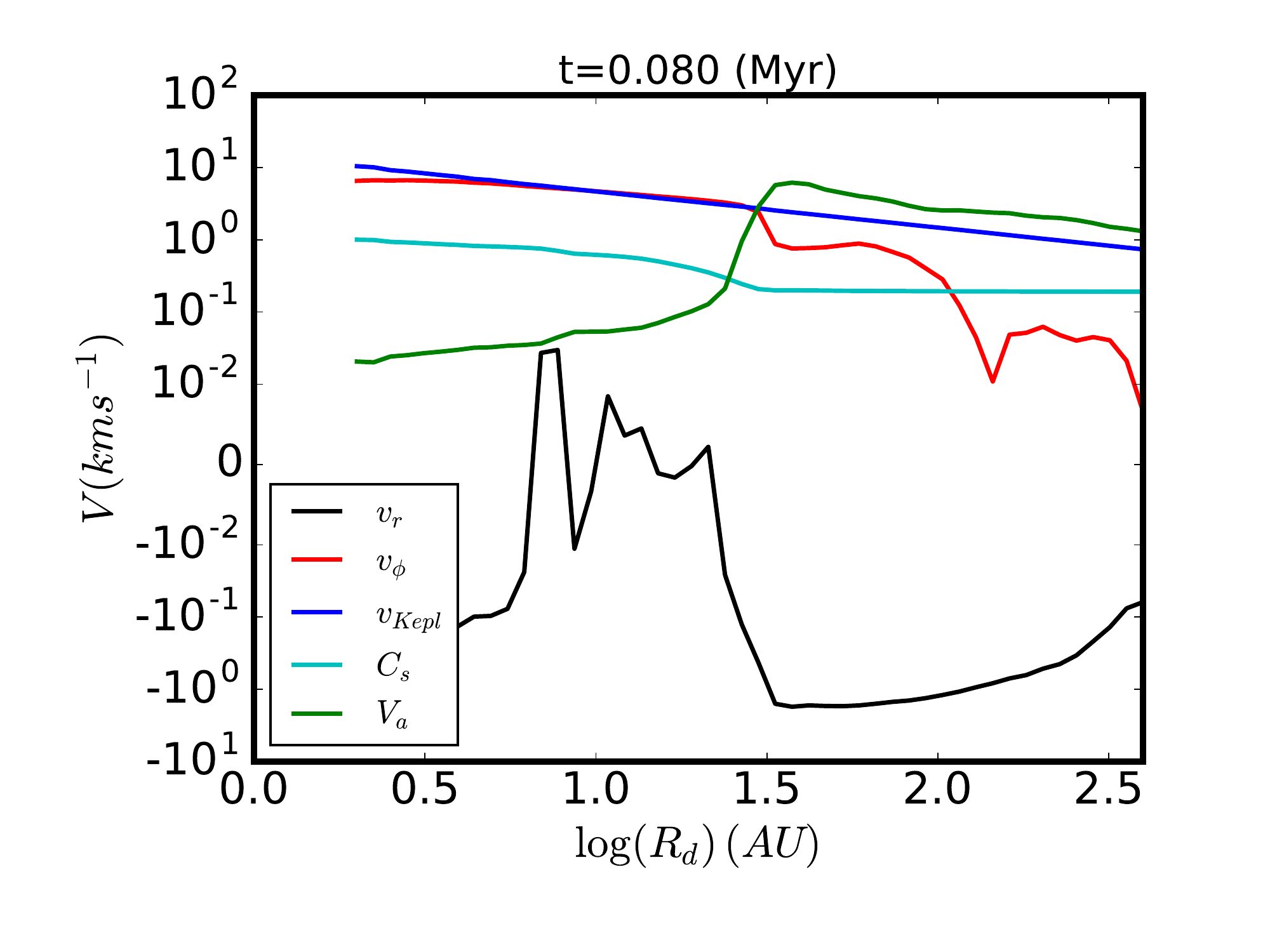}}  
\end{picture}
\caption{Radial structure of the disc of run $R2hsink$ (i.e. $n_{\rm thres}= 1.2 \, 10^{14}$ cm$^{-3}$). Top  row: mean column density of the disc  as a function of radius.
Second  panel: mean  density of the disc vs  radius. Third panel: $\beta _{\rm mag}= P_{\rm therm}/P_{\rm mag}$ as a function of radius.
Fourth  panel: various velocities 
vs radius within the disc equatorial plane. }
\label{rhov_r_R2hsink}
\end{figure}

\setlength{\unitlength}{1cm}
\begin{figure}%[h!]
%\centering
\begin{picture} (0,22)
\put(0,16.5){\includegraphics[width=8cm]{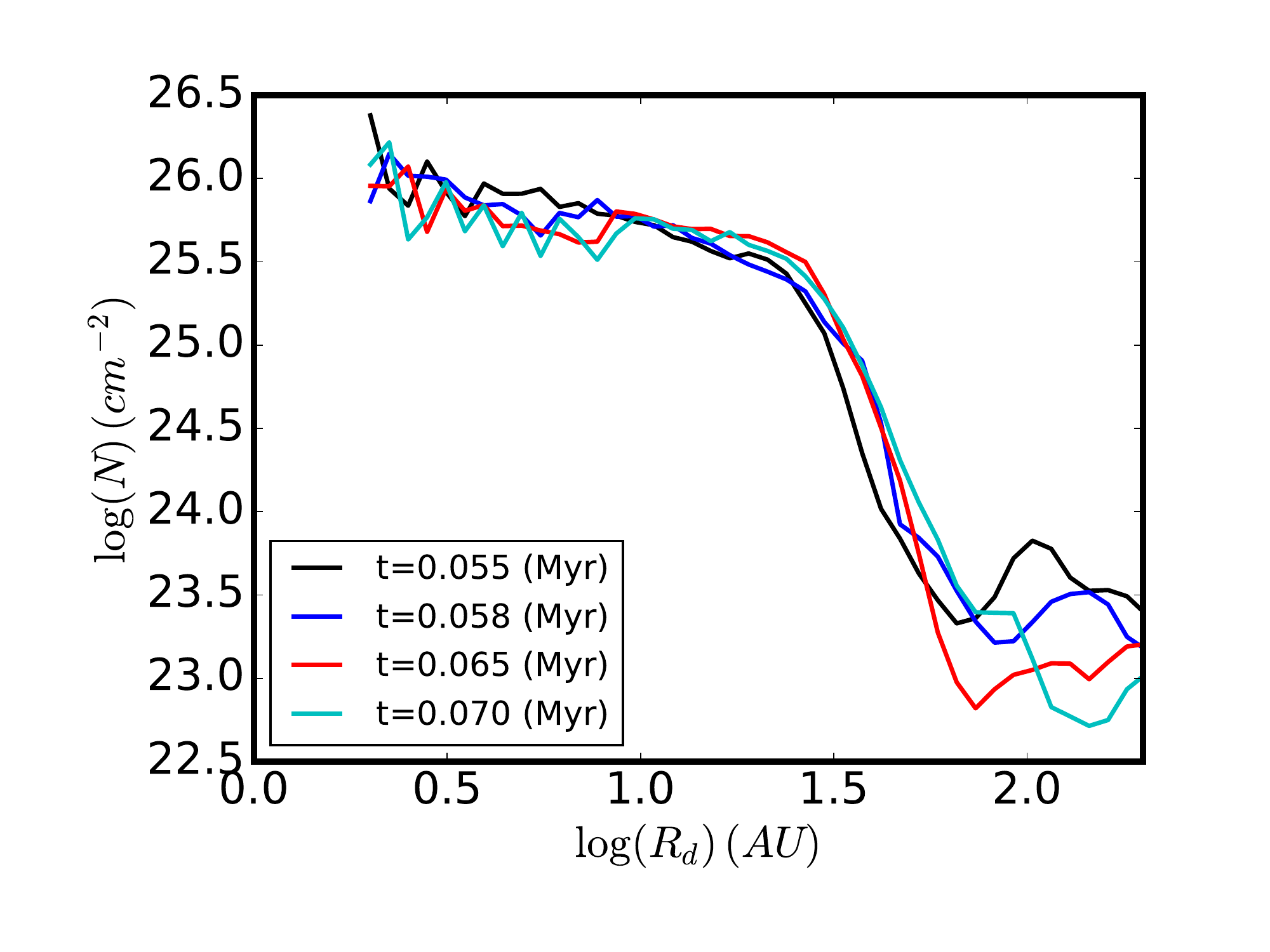}}  
\put(0,11){\includegraphics[width=8cm]{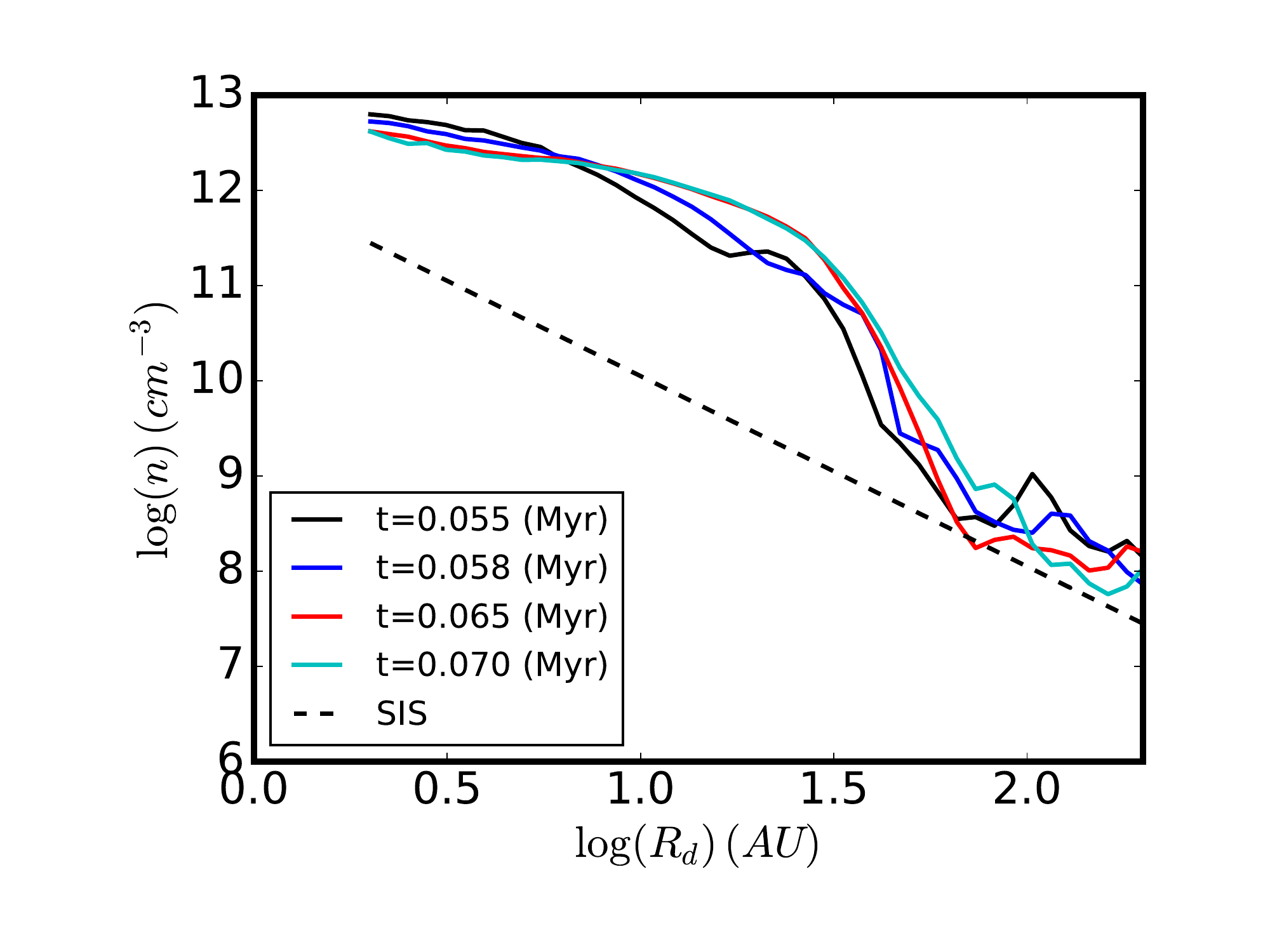}}  
\put(0,5.5){\includegraphics[width=8cm]{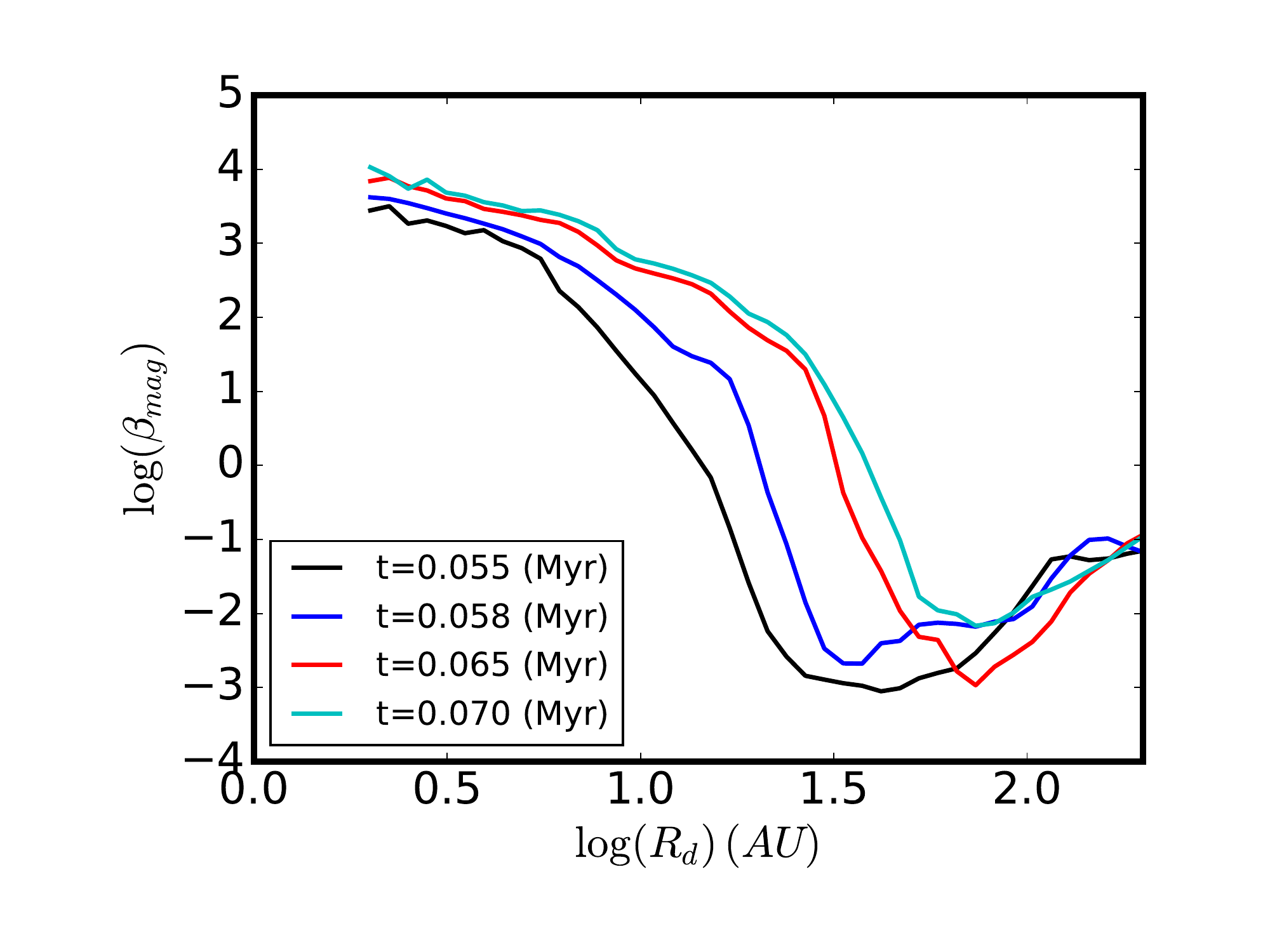}}  
\put(0,0){\includegraphics[width=8cm]{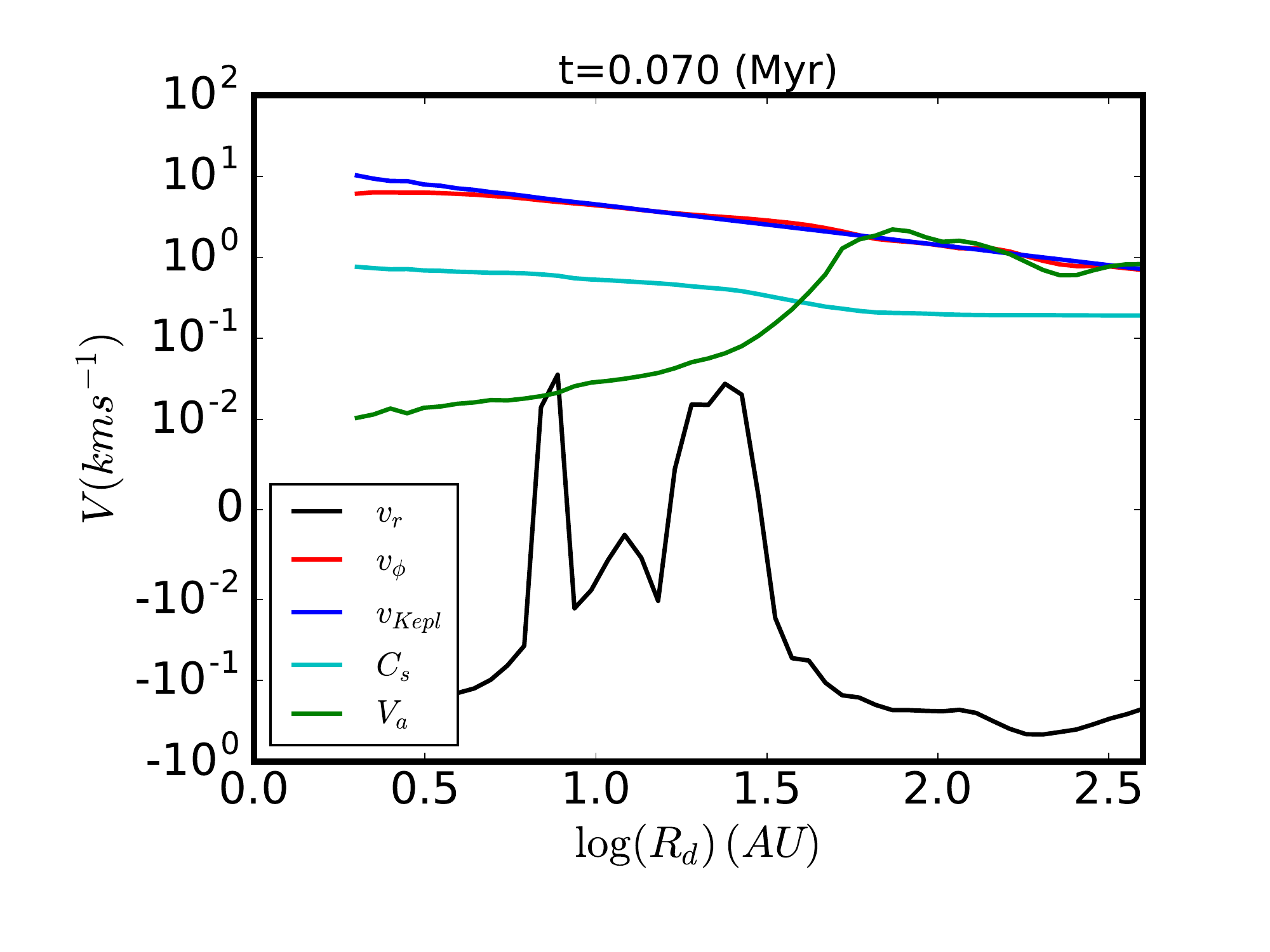}}  
\end{picture}
\caption{Radial structure of the disc of run $R3$ (i.e. $\mu^{-1}=0.1$). Top  row: mean column density of the disc  as a function of radius.
Second  panel: mean  density of the disc vs  radius. Third panel: $\beta _{\rm mag}= P_{\rm therm}/P_{\rm mag}$ as a function of radius.
Fourth  panel: various velocities 
vs radius within the disc equatorial plane. }
\label{rhov_r_R3}
\end{figure}

In this section we give more details, on two of the runs namely Run2hsink and $R3$ that we recall 
differ from our reference run $R2$ by a higher value of $n_{\rm thres}$ (i.e. the value at which the sink accretes) and 
a weaker magnetisation.  The two discs are larger than the disc of $R2$ and also more massive 
(\ref{time_evol_num} and \ref{time_evol_mag}).

Figures~\ref{rhov_r_R2hsink} and~\ref{rhov_r_R3} display the column density, the density in the equatorial plan, 
the $\beta_{\rm mag}$ as a function of radius for several timesteps. It also shows several velocities at 
a given snapshot. These plots can be compared with Fig.~\ref{rhov_r}.

For run $R2hsink$, the most important difference comes from the central density that is 3-4 times 
higher, which is a direct consequence of the higher value of $n_{\rm thres}$. 
The value of $\beta_{\rm mag}$ is also slightly higher because of higher density. 

For run $R3$, the mean density profile is not very different from $R2$ (the spiral structure 
is not seen on the 1D profile). The value of $\beta_{\rm mag}$ is about 2-4 times lower. 

Apart from these differences, the other profiles tend to be remarkably similar.

\section{Outflows}
As it is the case in most MHD collapse calculations \citep[e.g.][]{banerjee2004,machida2006,ciardi2010,tomida2010,hirano2019}, 
prominent outflows due to the magneto-centrifugal mechanism \citep{frank2014} also develop in the present simulations and for the sake of completeness 
we present them here. 

\subsection{Outflow qualitative description}

\setlength{\unitlength}{1cm}
\begin{figure}%[h!]
%%%\centering
\begin{picture} (0,15)
\put(0,10){\includegraphics[width=7cm]{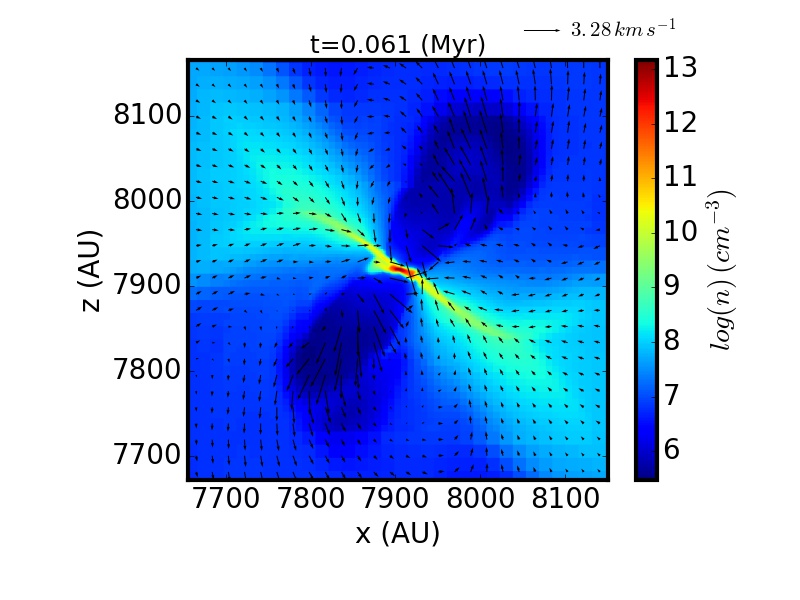}}  
\put(0,5){\includegraphics[width=7cm]{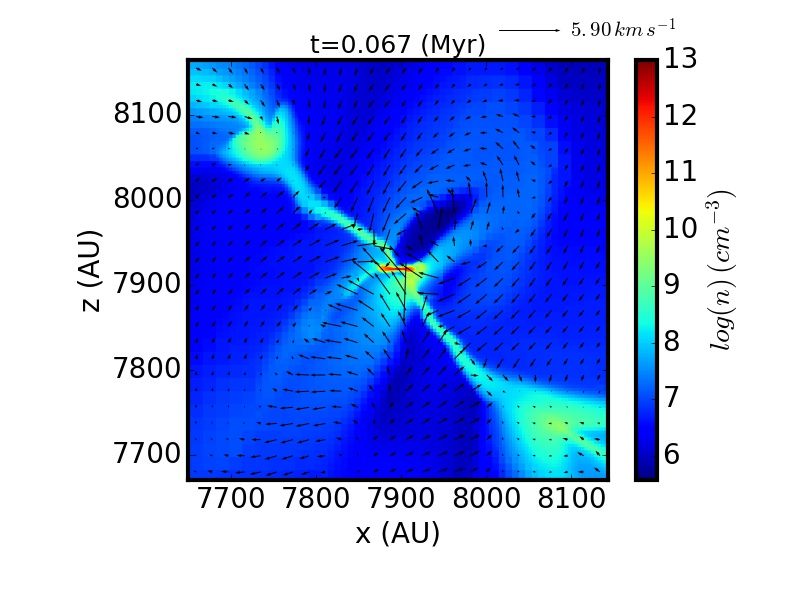}}  
\put(0,0){\includegraphics[width=7cm]{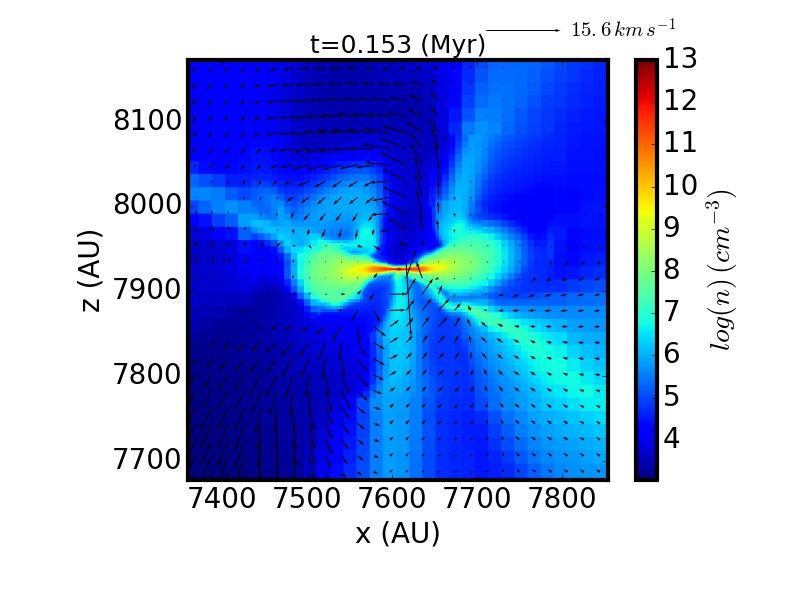}}  
\end{picture}
\caption{Density and velocity field at three snapshots for run $R2$.}
\label{rho_z_Run2}
\end{figure}

\setlength{\unitlength}{1cm}
\begin{figure}%[h!]
%%\centering
\begin{picture} (0,15)
\put(0,10){\includegraphics[width=7cm]{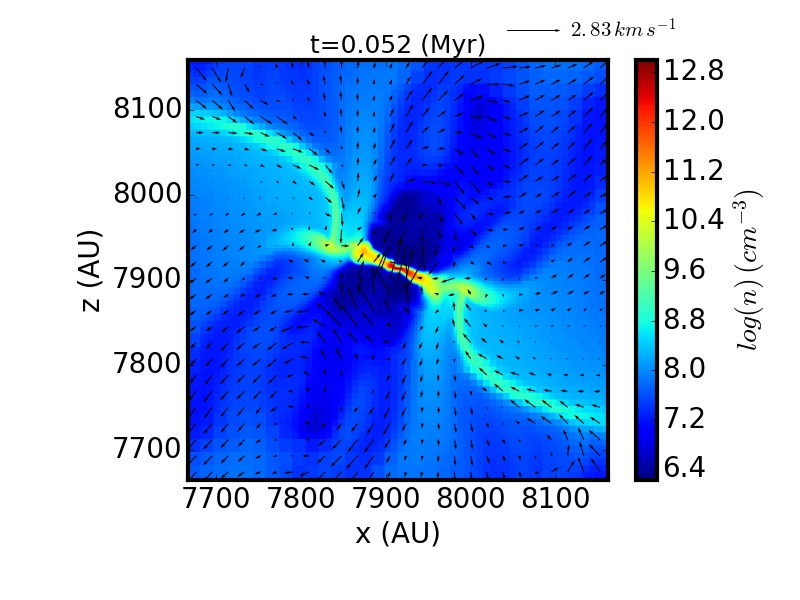}}  
\put(0,5){\includegraphics[width=7cm]{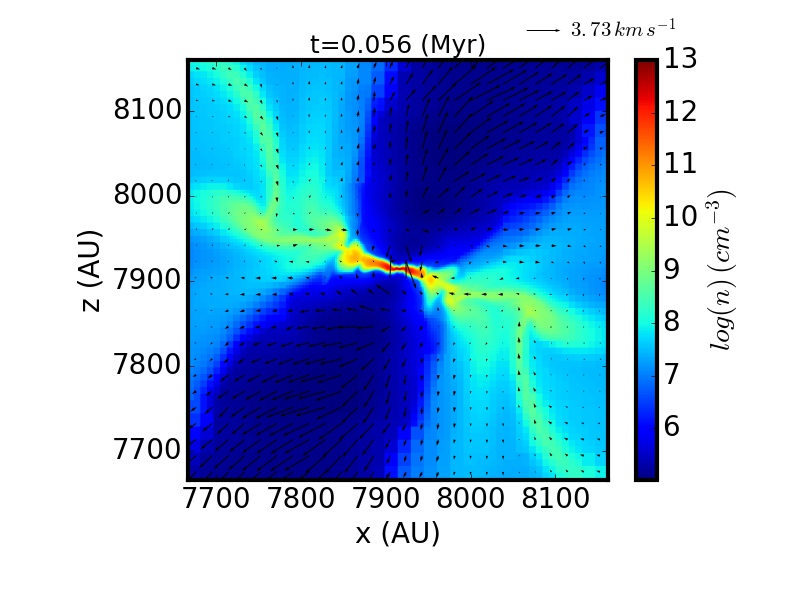}}  
\put(0,0){\includegraphics[width=7cm]{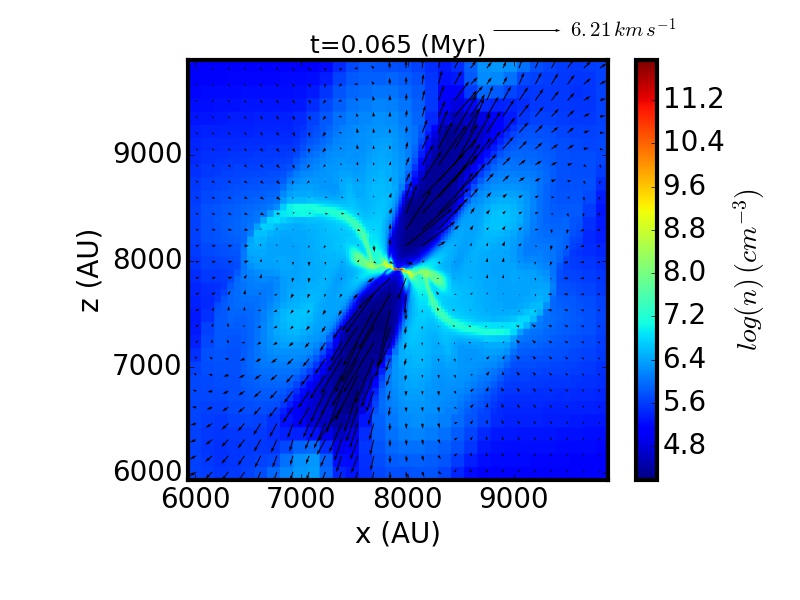}}  
\end{picture}
\caption{Density and velocity field at three snapshots for run $R3$.}
\label{rho_z_Run3}
\end{figure}

Figure~\ref{rho_z_Run2} portrays  density slices on which the velocity 
field is overplotted at 3 different snapshots. From the top and middle panels, we see
that a magnetized cavity with expanding motion starts developing early.
Indeed the corresponding time is very close to the moment when the sink 
particle is being introduced in run $R2$. Interestingly the gas 
close to the sink and disc is infalling and it is only at a distance of a few 
tens of AU that it expands. The cavity is broadly aligned with the direction of the mean 
magnetic field. 
Bottom panel shows that the outflow persists at later time and becomes 
relatively faster. The cavities are asymmetrical and do not occupy 
the whole disc surface as accretion along the disc axis is still 
occurring.

Figure~\ref{rho_z_Run3} is identical to Fig.~\ref{rho_z_Run2} but 
shows run $R3$ (which has $\mu^{-1}=0.1$).
Top panel shows again that the outflow starts early and 
is concomitant to the sink creation. Again the velocity close to the disc 
and sink reveals that the gas is infalling. 
Middle panel shows that the cavities develop in a rather symmetric way and occupy all the 
surface within and above the disc. 
Bottom panel shows a larger scale view about 10 kyr later. The outflow
stays symmetrical an well collimated. 

\subsection{Outflow velocity and mass}

\setlength{\unitlength}{1cm}
\begin{figure}%[h!]
%%%\centering
\begin{picture} (0,11)
\put(0,5.5){\includegraphics[width=8cm]{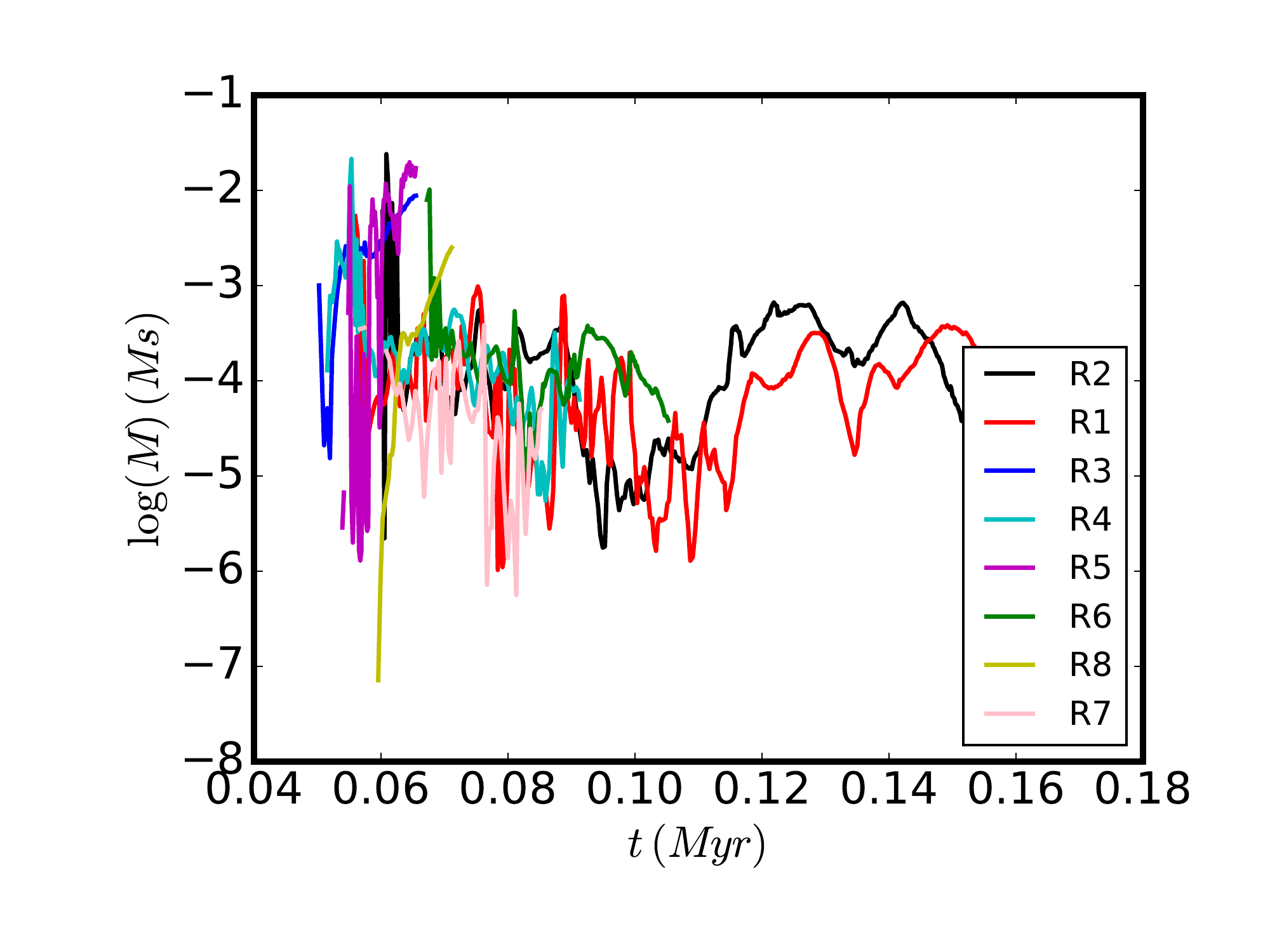}}  
\put(0,0){\includegraphics[width=8cm]{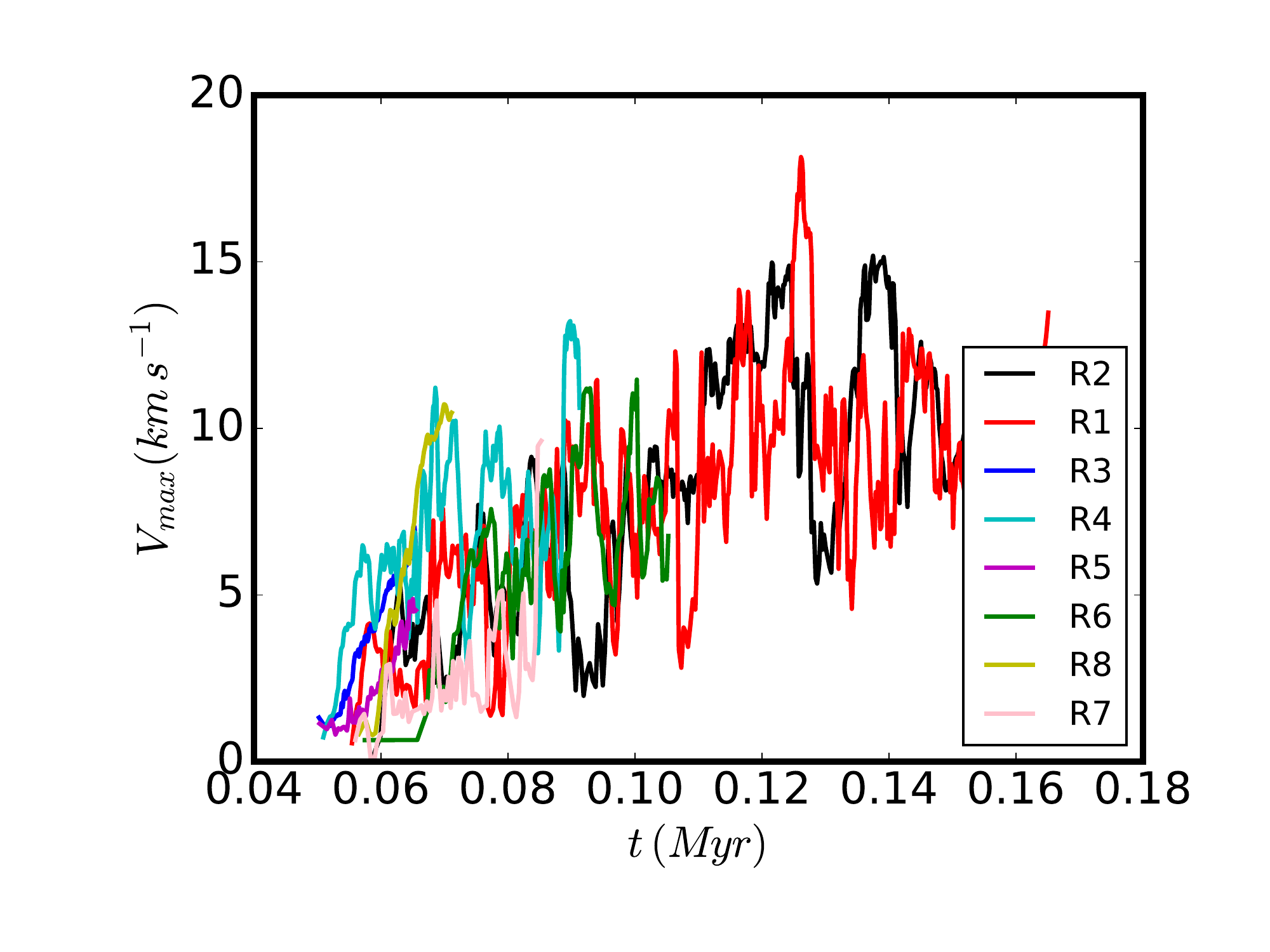}}  
\end{picture}
\caption{Top panel and bottom panel show respectively the mass within and the largest velocity of,
 the outflow defined as gas having a positive radial velocity larger than 
1 km s$^{-1}$ for all runs  (except run $R9$ which presents no outflow). } 
\label{vmax_flow}
\end{figure}

To quantify further the outflows in our simulations, we have selected all computational cells which have 
a radial velocity that is positive (therefore expanding) and larger than 1 km s$^{-1}$.
Figure~\ref{vmax_flow} displays the total mass within the flow as well as its maximum velocity for 
all the runs listed in Table~\ref{table_param} except for run $R9$ which do not present outflow. 
The total mass remains below 1$\%$ of the sink mass, except initially when the sink has a low mass. 
It should be stressed however that after a few kyr, the outflows reach the boundary of the computational domain
and therefore the mass is not properly counted. The velocity threshold used to define the flow is 
obviously critical. Choosing smaller values, obviously leads to higher estimate but there is 
then a confusion with the expanding envelope which surrounds the core since the medium 
outside the core has initially a low pressure. 
The largest velocity within the flow tends to be similar in all runs and goes from about 5 km s$^{-1}$ at 
early times to 15 km s$^{-1}$ at later ones.

%\subsection{Outflow location, density and magnetisation}

%\setlength{\unitlength}{1cm}
%\begin{figure}%[h!]
%%%%\centering
%\begin{picture} (0,22)
%\put(0,16.5){\includegraphics[width=8cm]{FIG_PAPIER_DISC_PHYS/DC_2/mass_flow_t.pdf}}  
%\put(0,11){\includegraphics[width=8cm]{FIG_PAPIER_DISC_PHYS/DC_2/nmean_flow_t.pdf}}  
%\put(0,5.5){\includegraphics[width=8cm]{FIG_PAPIER_DISC_PHYS/DC_2/size_flow_t.pdf}}  
%\put(0,0){\includegraphics[width=8cm]{FIG_PAPIER_DISC_PHYS/DC_2/beta_flow_t.pdf}}  
%\end{picture}
%\caption{Top pannel and bottom panel show respectively tha mass within and the largest velocity of,
% the outflow defined as gas having a positive radial velocity larger than 
%1 km s$^{-1}$ for all runs  (except run R9 which presents no outflow). }
%\label{run2_flow}
%\end{figure}

\end{document}